\documentclass[11pt,a4paper]{article}
\usepackage{graphicx}
\usepackage{float}
\usepackage{afterpage}
\usepackage{epsfig,cite}
\usepackage{amssymb}
\usepackage{amsmath}
\usepackage{dsfont}
\usepackage{multirow}
\usepackage{url,hyperref}
\usepackage{booktabs}
\usepackage{tabularx}
\usepackage{xcolor}
\usepackage{bm}
\usepackage{textcomp}
\usepackage{caption}
\usepackage{slashed}
\usepackage{changes}
\usepackage{url}
\usepackage{hyperref}

\usepackage[normalem]{ulem}

\usepackage{url}
\usepackage{hyperref}

\begin{document}

\vspace{-2.0cm}

\begin{center} 

  {\Large \bf 
Perturbative RGE systematics in precision observables}
  \vspace{.7cm}

V.~Bertone,$^1$ G.~Bozzi,$^{2,3}$ F.~Hautmann$^{4,5}$

\vspace{.3cm}
$^1${\it IRFU, CEA, Universit\'e Paris-Saclay, F-91191 Gif-sur-Yvette, France,}\\
$^2${\it Dipartimento di Fisica, Universit\`a di Cagliari, Cittadella Universitaria, I-09042 Monserrato,}\\
$^3${\it INFN, Sezione di Cagliari, Cittadella Universitaria, I-09042 Monserrato,}\\
$^4${\it Elementaire Deeltjes Fysica, Universiteit Antwerpen, B 2020 Antwerpen,}\\
$^5${\it Theoretical Physics Department, University of Oxford, Oxford OX1 3PU.}\\
\end{center}  

\vspace{0.1cm}

\begin{center}
  {\bf \large Abstract}\\
\end{center}

QCD calculations for collider physics make use of perturbative
solutions of renormalisation group equations (RGEs). Ambiguities
related to these solutions can contribute significantly to systematic
uncertainties of theoretical predictions for physical observables. We
propose a general method to estimate these systematic effects using
techniques inspired by soft-gluon and transverse-momentum resummation
approaches. We first discuss the cases of the evolution of strong
coupling $\alpha_s$, collinear parton-distribution functions (PDFs),
and transverse-momentum-dependent distributions (TMDs). We then study
the implications for precision observables in hadron-collider
processes, such as the deep-inelastic scattering structure functions
and the transverse-momentum distribution of the lepton pair in
Drell-Yan production.
\vspace{1cm}


\newpage

\tableofcontents

\section{Introduction}
\label{Intro}

Many applications of Quantum Chromodynamics (QCD) to physics at
high-energy colliders involve the solution of renormalisation group
equations (RGEs).  With the dramatic increase in precision of
perturbative calculations for hard-scattering cross sections over the
past few years, theoretical systematic uncertainties from RGE
solutions become a potentially important factor in determining the
overall accuracy of theoretical predictions for collider processes.

In our recent work, Ref.~\cite{Bertone:2022sso}, we studied these
systematics in the case of QCD coupling $\alpha_s$ and collinear
parton distribution functions
(PDFs)~\cite{Gribov:1972ri,Altarelli:1977zs,Dokshitzer:1977sg},
estimating the perturbative uncertainty associated with the solution
of the corresponding RGEs. These effects were found to be
non-negligible in kinematic regions which are phenomenologically
important for PDF determinations from global fits to experimental
data~\cite{Ball:2022hsh,Ball:2021leu,NNPDF:2019ubu,Bailey:2020ooq,Hou:2019efy,Alekhin:2017kpj,Abramowicz:2015mha},
as well as for extractions from future collider
experiments~\cite{Azzi:2019yne,LHeC:2020van,FCC:2018byv,Proceedings:2020eah}. As
argued in Ref.~\cite{Bertone:2022ope}, the fact that modern PDF fits
do not account for these systematics is likely to lead to an
underestimate of their
uncertainties. Refs.~\cite{Bertone:2022sso,Bertone:2024uyf} illustrate
this by studying the impact of RGE-related uncertainties on the
deep-inelastic scattering (DIS) structure functions $F_2$ and $F_L$ at
next-to-leading order (NLO) and next-to-next-to-leading order (NNLO)
accuracy in perturbative QCD.  Further evidence is provided by the
presence of logarithmically-enhanced higher-order contributions to the
RGE kernels for PDF evolution at small values of Bjorken's
$x_{\rm B}$~\cite{Jaroszewicz:1982gr,Catani:1993rn,Catani:1994sq}
which are known to give rise to significant effects in $F_2$. While
these effects can be tamed by means of dedicated resummation
techniques~\cite{Catani:1994sq,Kwiecinski:1985cq,Ellis:1993rb,Ellis:1995gv,Forshaw:1995ga,Ball:1995vc,Kwiecinski:1997ee,Ciafaloni:2007gf,Altarelli:2008aj,White:2006yh,Ball:2017otu,xFitterDevelopersTeam:2018hym},
a reliable estimate of RGE-related uncertainties is fundamental to
assess the applicability region of fixed-order calculations. The
relevance of such effects has also been noted in
Refs.~\cite{Ebert:2021aoo,Billis:2019evv} for analytic Sudakov
resummation calculations, and in Ref.~\cite{Hautmann:2019biw} for
parton branching calculations.

In this paper, we provide more details to the analysis of the
evolution of $\alpha_s$ and PDFs given in Ref.~\cite{Bertone:2022sso},
and extend it to the evolution of transverse-momentum-dependent
distributions
(TMDs)~\cite{Collins:1984kg,Collins:2011zzd,Angeles-Martinez:2015sea,Boussarie:2023izj}.
We present a general framework aimed at evaluating theoretical
systematic uncertainties related to the solution of RGEs on these
quantities. We then study the impact of these effects on predictions
for the inclusive neutral-current DIS structure functions ($F_2$,
$F_L$, and $F_3$) and for the transverse-momentum ($q_T$) distribution
of the lepton pair in inclusive Drell-Yan (DY) production. The case of
DIS is primarily relevant to current and forthcoming extractions of
PDFs, and to precision QCD studies at future lepton-hadron
experiments~\cite{LHeC:2020van,Proceedings:2020eah,AbdulKhalek:2022hcn}.
The case of DY is instead relevant to extractions of
TMDs~\cite{Scimemi:2019cmh,Bacchetta:2019sam,Hautmann:2020cyp,Bacchetta:2022awv,Bury:2022czx,Moos:2023yfa,Bacchetta:2024qre},
as well as to precise determinations of the electroweak parameters of
the Standard Model, such as the mass of the $W$
boson~\cite{CDF:2022hxs, ATLAS:2024erm}. These studies often require
the matching of the small-$q_T$ calculation, which can be written in
terms of TMDs, to the large-$q_T$ calculation, which is instead
expressed in terms of (collinear)
PDFs~\cite{Camarda:2019zyx,Camarda:2021ict,Camarda:2022qdg,Camarda:2023dqn,Bizon:2018foh,Bizon:2019zgf,Chen:2022cgv,BermudezMartinez:2019anj,BermudezMartinez:2021lxz,BermudezMartinez:2022bpj,Bubanja:2023nrd,Isaacson:2023iui,Becher:2020ugp,Neumann:2022lft,Campbell:2023lcy,Ebert:2016gcn,Ebert:2020dfc,Ju:2021lah
}. We carry out the matching keeping track of the theoretical
uncertainties stemming from both small- and large-$q_T$ regimes,
finally providing an estimate over the entire spectrum which
consistently incorporates all uncertainties.

We organise the presentation of our study as follows. In
Sec.~\ref{sec:emergent}, we discuss in general terms the emergence of
RGE-related perturbative uncertainties and devise the strategy to
estimate them by introducing the resummation scales. In
Secs.~\ref{sec:runningcoupling} and~\ref{sec:coll}, we examine the
cases of the strong coupling and PDFs by reviewing and extending the
results of Ref.~\cite{Bertone:2022sso}.  In Sec.~\ref{sec:sud}, we
analyse the case of TMDs. We express their evolution in terms of a
Sudakov form factor and present quantitative results on the associated
RGE systematic uncertainties. In Sec.~\ref{sec:PhysicalObservables},
we discuss uncertainty estimates for specific physical observables:
the DIS structure functions and the transverse-momentum distribution
of the lepton pair in DY production. In
Sec.~\ref{sec:numapplications}, we perform numerical calculations for
the neutral-current structure functions $F_2$, $F_L$ and $F_3$, and
for the DY differential cross $d \sigma / d q_T$, illustrating the
corresponding RGE theoretical uncertainties in comparison with
renormalisation- and factorisation-scale uncertainties. We give our
conclusions in Sec.~\ref{sec:concl}. The lengthy expressions for the
$g$-functions that govern the evolution of the quantities considered
in this paper, \textit{i.e.}, $\alpha_s$, PDFs, and TMDs, are
collected in the App.~\ref{app:formulas}.

\section{Emergence of resummation scales}
\label{sec:emergent}

In this section, we discuss in general terms how resummation scales
emerge when solving RGEs whose anomalous dimensions are known to some
fixed order in perturbation theory. To carry out this investigation,
we start by considering RGEs of the general form
\begin{equation}
  \label{eq:RGEproto}
  \frac{d}{d\ln \mu} \ln R (\mu)= \gamma(\alpha_s(\mu))\,,
\end{equation}
where $R$ is a renormalised quantity, function of the renormalisation
scale $\mu$, and $\gamma$ is the appropriate anomalous dimension
computable in perturbation theory as a power series in the coupling
$\alpha_s$. The perturbative expansion of the anomalous dimension
$\gamma$ truncated at order $k$ is given by
\begin{equation}
  \label{eq:gamma-expand}
  \gamma(\alpha_s(\mu)) =
  \sum_{n=0}^k a_s^{n+1}(\mu)\gamma_{n}\,,
\end{equation}
with $a_s(\mu) \equiv \alpha_s(\mu)/(4\pi)$. 

At any finite truncation order $k$, Eq.~(\ref{eq:RGEproto}) can be
solved either analytically (i.e., expressing $R(\mu)$ in terms of the
boundary condition $R(\mu_0)$ in a closed analytic form) or
numerically. We stress that both kind of solutions, based on an
anomalous dimension truncated at the same order, are equally good and
equivalent from the viewpoint of perturbative accuracy. The purpose of
this work is to devise a strategy to estimate the truncation
uncertainty in both cases.

In the context of analytic solutions, this can be achieved using
techniques borrowed from soft-gluon and transverse-momentum
resummation~\cite{Catani:1996yz,Catani:2003zt,Bozzi:2005wk}. Indeed,
such solutions have the advantage of explicitly exposing terms
proportional to $L_1=a_s(\mu_0)\ln(\mu/\mu_0)$. Since $\mu_0$ and
$\mu$ are potentially far apart, $L_1$ can be large in spite of the
presence of $a_s(\mu_0)$ and thus it cannot be treated as a
perturbative parameter. This implies the need to sum powers of $L_1$
to all orders in perturbation theory. Once the analytic solution of
the RGE has been obtained, one can decompose all instances of $L_1$ as
follows,
\begin{equation}
  L_1=L_{\kappa}- a_s(\mu_0)\ln\kappa\,,\quad\mbox{with}\quad L_{\kappa}=a_s(\mu_0)\ln\left(\frac{\kappa \mu}{\mu_0}\right)\,,
  \label{eq:SplitResScale}
\end{equation}
where the scale $\kappa\mu$ is dubbed \textit{resummation scale}.
Provided that $\kappa$ is of order ${\cal O} (1)$, the second term in
the decomposition of $L_1$, \textit{i.e.}, $a_s(\mu_0)\ln\kappa$, is
perturbatively small. This allows one to expand the analytic solution
of the RGE around $L_1=L_{\kappa}$, retaining only terms within
perturbative accuracy. The net result is an analytic solution that
sums all powers of $L_{\kappa}$, while retaining powers of
$a_s(\mu_0)\ln\kappa$ only up to the nominal accuracy. Variations of
$\kappa$ in the vicinity of $\kappa = $ 1 generate subleading terms
that provide an estimate of the perturbative uncertainty associated
with the analytic solution of the RGE.

A similar method can be employed in the context of numerical
solutions.  One could introduce arbitrary subleading corrections to
the anomalous dimension $\gamma$ and then solve the corresponding RGE
numerically.  Differences at the level of numerical solutions obtained
with different subleading terms in the anomalous dimension would then
provide an estimate of the perturbative uncertainty of the solution
itself. A way to generate subleading corrections in fixed-order
quantities is through scale variations. To be more specific, let us
consider the anomalous dimension $\gamma$ truncated at NLO accuracy,
\begin{equation}
  \gamma = a_s(\mu)\gamma_0+a_s^2(\mu)\gamma_1\,.
  \label{eq:NLOanomalousdimension}
\end{equation}
Then we use the expanded solution for the running of the strong
coupling to express $a_s(\mu)$ in terms of $a_s(\xi\mu)$, where $\xi$
is a parameter of order ${\cal O} (1)$,\footnote{The expansion in
  Eq.~(\ref{eq:alphasNLOexp}) comes from the solution to the RGE
  \begin{equation}
    \frac{d\ln a_s(\mu)}{d\ln \mu} = a_s(\mu)\beta_0+\mathcal{O}(\alpha_s^2)\,.
  \end{equation}}
\begin{equation}
a_s(\mu) = a_s(\xi\mu)-a_s^2(\xi\mu)\beta_0\ln\xi +\mathcal{O}(\alpha_s^3)\,,
\label{eq:alphasNLOexp}
\end{equation}
that plugged into Eq.~(\ref{eq:NLOanomalousdimension}) gives
\begin{equation}
  \gamma = a_s(\xi\mu)\gamma_0+a_s^2(\xi\mu)\left[\gamma_1-\gamma_0\beta_0\ln\xi\right]+\mathcal{O}(\alpha_s^3)\,,
  \label{eq:NLOanomalousdimensionShifted}
\end{equation}
where subleading terms of order $\mathcal{O}(\alpha_s^3)$ have been
neglected. From a perturbative perspective, solving
Eq.~(\ref{eq:RGEproto}) using the anomalous dimension in
Eq.~(\ref{eq:NLOanomalousdimension}) or that in
Eq.~(\ref{eq:NLOanomalousdimensionShifted}) is equivalent. However,
the two solutions will be numerically different. As argued above,
their difference can be interpreted as an estimate of the perturbative
uncertainty associated to the solution of the RGE obtained with
anomalous dimension truncated at NLO accuracy. Of course, this
procedure can be extended to any finite truncation order. Therefore,
variations of the parameter $\xi$ in
Eq.~(\ref{eq:NLOanomalousdimensionShifted}) when solving
Eq.~(\ref{eq:RGEproto}) numerically are analogous to variations of the
parameter $\kappa$ introduced in Eq.~(\ref{eq:SplitResScale}) when the
solution is analytic. In this sense, also the scale $\xi\mu$ is a
resummation scale. We will explicitly show below, considering the RGE
of the strong coupling at NLO accuracy, that there is indeed a
one-to-one correspondence between $\kappa$ and $\xi$.  Therefore,
variations by the same amount of $\kappa$ (in analytic solutions) and
$\xi$ (in numerical solutions) are expected to provide comparable
estimates of the same truncation uncertainty.

In the following, we will implement the strategy outlined above for
the cases of strong coupling $\alpha_s$, PDFs, and TMDs
(Secs.~\ref{sec:runningcoupling}, \ref{sec:coll}, and~\ref{sec:sud},
respectively). In all cases, we will derive analytic solutions to the
RGEs and compare them to the respective numerical solutions. In doing
so, we will also vary the resummation-scale parameters $\kappa$ and
$\xi$ and compare the resulting uncertainty bands. On a case-by-case
basis, we will also define the accuracy of the results.  We will
consider the \textit{logarithmic accuracy} (leading-logarithm (LL),
next-to-leading-logarithm (NLL), and so on, as opposed to fixed-order
accuracy used for perturbative calculations truncated at some finite
order in $\alpha_s$) for ``single-logarithmic'' resummation, as in the
cases of $\alpha_s$ and PDFs, and for ``double-logarithmic''
resummation, as in the case of TMDs.

After the discussion on the solution of RGEs, we will move to
examining the implications of RGE systematic uncertainties on the
perturbative computation of phenomenologically relevant physical
observables (Secs.~\ref{sec:PhysicalObservables}
and~\ref{sec:numapplications}). The role of these uncertainties in
single-logarithmic resummation will be studied in the cases of
inclusive neutral-current DIS and large-$q_T$ DY production. The case
of double-logarithmic resummation will be studied by considering DY
production at small $q_T$. In these studies, we will also investigate
the interplay between RGE systematic uncertainties, parameterised by
resummation scales, and the more common perturbative uncertainties
obtained by variations of renormalisation and factorisation scales.

\section{Strong coupling}
\label{sec:runningcoupling}

We start our analysis by considering the evolution of the strong
coupling $\alpha_s$. In Eqs.~(\ref{eq:RGEproto})
and~(\ref{eq:gamma-expand}), we set $R= a_s$ and $\gamma=\beta$, where
$\beta$ is the QCD $\beta$-function, and obtain\footnote{Note that
  often the $\beta$-function is defined through the equation (see
  \textit{e.g.}  Ref.~\cite{Ellis:1996mzs})
  \begin{equation}
    \frac{d \alpha_s}{d\ln \mu^2} = -\alpha_s(\mu)\sum_{n=0}^k a_s^{n+1}(\mu)\beta_n\,.
  \end{equation}
  Using this form, the anomalous dimension $\gamma$ in
  Eq.~(\ref{eq:RGEproto}) should be identified with
  $\gamma= - 2 \beta / a_s$.}
\begin{equation}
\frac{d}{d\ln \mu} \ln a_s (\mu)=
\beta(\alpha_s(\mu))=\sum_{n=0}^{k}a_s^{n+1}(\mu)\beta_n\,.
\label{eq:RGEalphas}
\end{equation}
Truncating the series at order $k$ in Eq.~(\ref{eq:RGEalphas}),
\textit{i.e.},  using a N$^k$LO accurate $\beta$-function, produces 
solutions that are N$^k$LL accurate. The corresponding analytic
solution can be written as~\cite{Bertone:2022sso}
\begin{equation}
  \label{eq:alphasgs}
  a_s^{{\rm N}^{k}{\rm LL}}(\mu) = \sum_{l=0}^{k}a_s^{l+1}(\mu_0) g_{l+1}^{(\beta)}(\lambda , \kappa )\,,
\end{equation}
where we have defined 
\begin{equation}
  \label{eq:lambdadefalphas}
  \lambda= a_s(\mu_0)\beta_{0}\ln\left(\frac{\kappa\mu}{\mu_0}\right)\,,
\end{equation}
with the parameter $\kappa$ introduced in
Eq.~(\ref{eq:SplitResScale}), and the functions $g_{i}^{(\beta)}$
collected in App.~\ref{App} up to $i=4$, necessary to achieve N$^3$LL
accuracy. As explained above, variations of $\kappa$ around
$\kappa = 1$ by a moderate factor (typically of 2) allow one to
estimate the perturbative uncertainty that affects the running of the
coupling.

Alternatively, Eq.~(\ref{eq:RGEalphas}) can be solved numerically. As
argued in Sec.~\ref{sec:emergent}, an estimate of the theoretical
uncertainty associated to the numerical solution can be achieved by
performing scale variations at the level of the
$\beta$-function. Displacing the scale $\mu$ by a factor $\xi$ in the
r.h.s.~of Eq.~(\ref{eq:RGEalphas}) with $k=4$ and using the expanded
evolution of $\alpha_s$ truncated at $\mathcal{O}(\alpha_s^4)$, we
obtain
\begin{eqnarray}
  \label{eq:betaxi}
  \beta(\alpha_s(\mu))&=&\displaystyle a_s(\xi\mu) \beta_{0}\Big[1
                                 +a_s (\xi\mu) \left(b_1-2 \beta_{0}
                                 \ln \xi\right)+ a_s^2 (\xi\mu) \left(-5 b_1 \beta_{0} \ln \xi+3 \beta_{0}^2 \ln ^2\xi+b_2\right)\nonumber\\
                             &+&\displaystyle a_s^3 (\xi\mu) \Big(13 b_1 \beta_{0}^2 \ln^2\xi-3 b_1^2 \beta_{0} \ln \xi -6 b_2 \beta_{0} \ln \xi-4 \beta_{0}^3 \ln ^3\xi+b_3\Big)\Big]+\mathcal{O}(\alpha_s^5),
\end{eqnarray}
where $b_n=\beta_n/\beta_0$. Eq.~(\ref{eq:betaxi}) defines a new
${\beta}$-function that differs from the original one by subleading
corrections. This ${\beta}$-function can be used in numerical
solutions to estimate the corresponding theoretical uncertainty by
varying the parameter $\xi$.

An inconvenient feature of the ${\beta}$-function in
Eq.~(\ref{eq:betaxi}) is that it relates the derivative of the strong
coupling $\alpha_s$ computed at $\mu$ (l.h.s.~of
Eq.~(\ref{eq:RGEalphas})) with $\alpha_s$ computed at $\xi \mu$. This
gives rise to a ``retarded'' or an ``advanced'' differential equation
whose solution is more complicated than that of a standard
differential equation. To overcome this difficulty, one can employ the
analytic solution in Eq.~(\ref{eq:alphasgs}) to express $a_s(\xi\mu)$
in terms of $a_s(\mu)$ so that, retaining only the relevant orders,
one obtains
\begin{eqnarray}
  \label{eq:betabscale}
  \beta(\alpha_s(\mu))&=&\displaystyle a_s(\mu) \beta _0\Big[
                          (g_1^{(\beta)})^2+a_s (\mu) \left(b_1 (g_1^{(\beta)})^3+2 g_1^{(\beta)} \left(g_2^{(\beta)}-
                          (g_1^{(\beta)})^2 \beta _0\ln \xi\right)\right) \nonumber\\
                      &+&\displaystyle a_s ^2 (\mu) \big(-5 b_1 (g_1^{(\beta)})^4 \beta _0\ln \xi+b_2 (g_1^{(\beta)})^4+3 b_1
                          g_2^{(\beta)} (g_1^{(\beta)})^2+3 (g_1^{(\beta)})^4 \beta
                          _0^2\ln ^2\xi \nonumber\\
                      &&\qquad\quad -6 g_2^{(\beta)} (g_1^{(\beta)})^2 \beta _0\ln
                         \xi+2 g_3^{(\beta)} g_1^{(\beta)}+(g_2^{(\beta)})^2\big) \nonumber\\
                      &+&\displaystyle a_s ^3(\mu) \big(13 b_1 (g_1^{(\beta)})^5 \beta _0^2\ln^2\xi-3 b_1^2 (g_1^{(\beta)})^5 \beta _0\ln \xi-6 b_2 (g_1^{(\beta)})^5 \beta _0\ln
                          \xi \\
                      &&\qquad\quad -20 b_1 g_2^{(\beta)} (g_1^{(\beta)})^3 \beta _0\ln \xi+b_3 (g_1^{(\beta)})^5+4 b_2 g_2^{(\beta)}
                         (g_1^{(\beta)})^3+3 b_1 g_3^{(\beta)} (g_1^{(\beta)})^2 \nonumber\\
                      &&\qquad\quad +3 b_1 (g_2^{(\beta)})^2 g_1^{(\beta)}-4
                         (g_1^{(\beta)})^5 \beta _0^3\ln ^3\xi+12 g_2^{(\beta)}
                         (g_1^{(\beta)})^3 \beta _0^2\ln ^2\xi\nonumber\\
                      &&\qquad\quad-6 g_3^{(\beta)} (g_1^{(\beta)})^2 \beta _0\ln \xi-6 (g_2^{(\beta)})^2 g_1^{(\beta)} \beta _0\ln \xi+2 g_4^{(\beta)} g_1^{(\beta)}+2 g_2^{(\beta)} g_3^{(\beta)}\big)
                         \Big]+\mathcal{O}(\alpha_s^5)\nonumber ,
\end{eqnarray}
where the $g^{(\beta)}$-functions are to be computed at
$\lambda_\xi=a_s(\mu)\beta_0\ln\xi$ and $\kappa=1$. We note that,
setting $\xi=1$ so that $\lambda_\xi=0$, one finds that
$g_1^{(\beta)}=1$ while $g_i^{(\beta)}=0$ for $i>1$, which
consistently reduces Eq.~(\ref{eq:betabscale}) to the usual
$\beta$-function. Thus, Eq.~(\ref{eq:betabscale}) allows for estimates
of theoretical uncertainties by variations of $\xi$, and produces a
differential equation which can be solved by standard methods.

We now show explicitly that, at NLL accuracy, the analytic solution in
Eq.~(\ref{eq:alphasgs}) for a generic value of $\kappa$ corresponds,
up to subleading terms, to the numerical solution obtained using the
$\beta$-function in Eq.~(\ref{eq:betabscale}) with $\xi=\kappa$. To do
so, we observe that the NLL analytic solution obtained using
Eq.~(\ref{eq:alphasgs}) can be written as
\begin{equation}
  a_s^{\rm NLL}(\mu) = a_s^{\rm NLL}(\kappa\mu,1) -
  \left(a_s^{\rm LL}(\kappa\mu,1)\right)^2\beta_{0}\ln\kappa \,.
\end{equation}
Taking the derivative with respect to $\ln\mu$, one gets
\begin{equation}
  \label{eq:kappa-xi-id}
  \frac{d \ln a_s^{\rm NLL}(\mu,\kappa)}{d\ln\mu} = a_s^{\rm
    NLL}(\kappa\mu,1)\beta_{0} + \left(a_s^{\rm
      LL}(\kappa\mu,1)\right)^2\left[\beta_{1} -
    2\left(\beta_{0}\right)^2\ln\kappa \right]+\mathcal{O}(\alpha_s^3)\,,
\end{equation}
which coincides with the r.h.s.~of Eq.~(\ref{eq:betabscale}) truncated
at $\mathcal{O}(\alpha_s^2)$ if one identifies $\kappa$ with $\xi$. It
can be shown by similar methods that the same equivalence between
$\kappa$ and $\xi$ holds also at higher orders in perturbation theory.

\begin{figure}[h]
  \begin{centering}
    \includegraphics[width=0.49\textwidth]{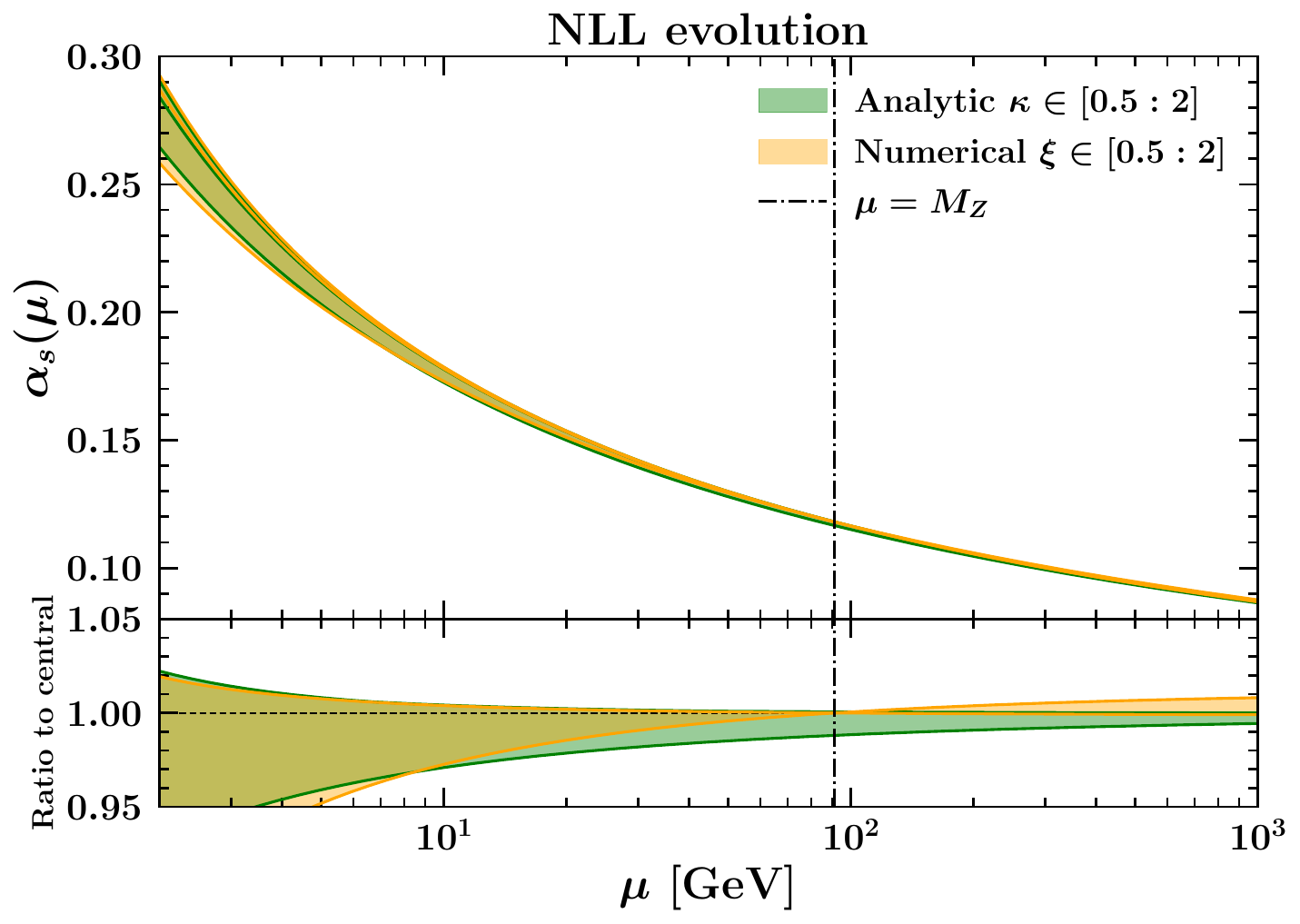}
    \includegraphics[width=0.49\textwidth]{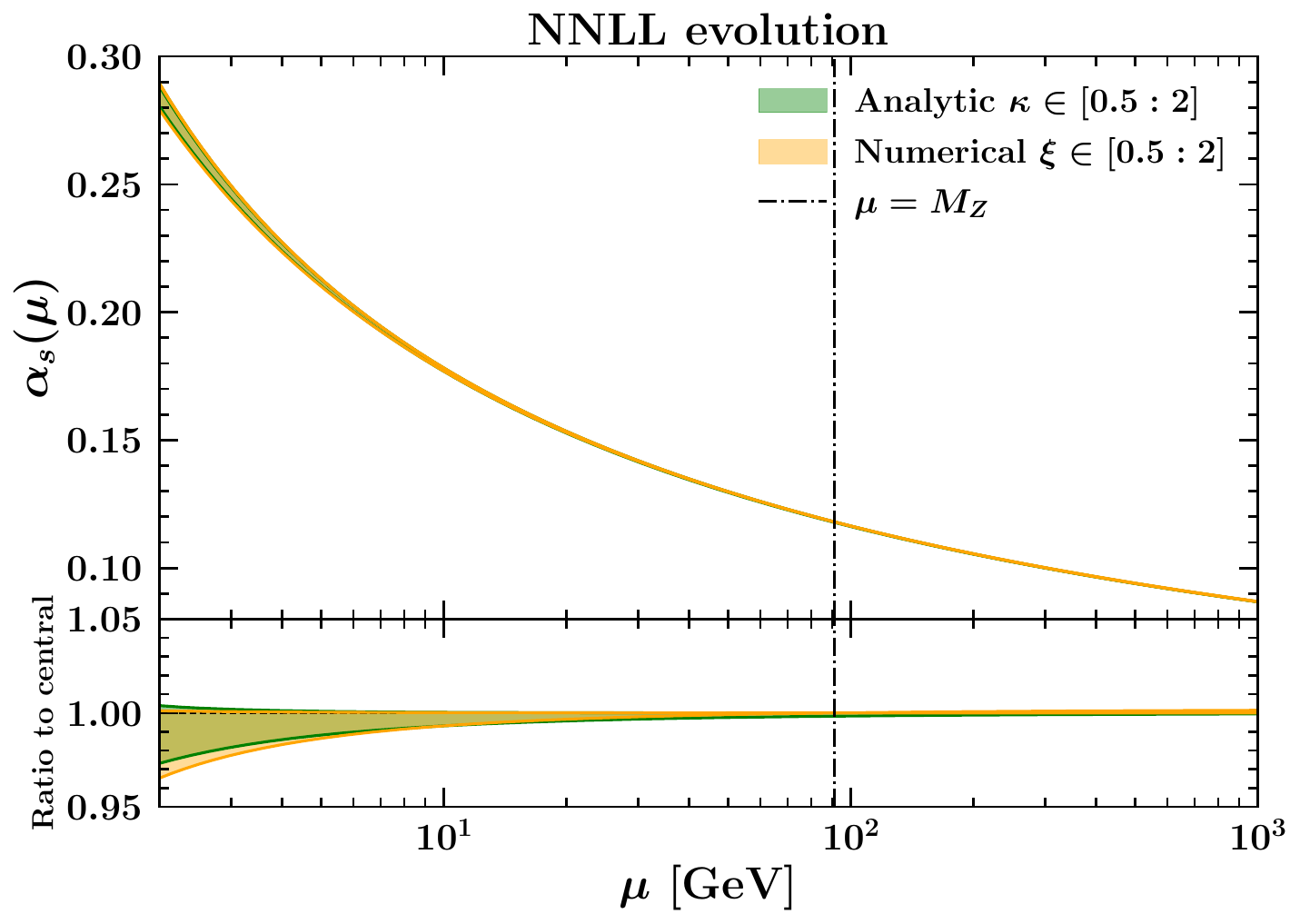}
    \includegraphics[width=0.49\textwidth]{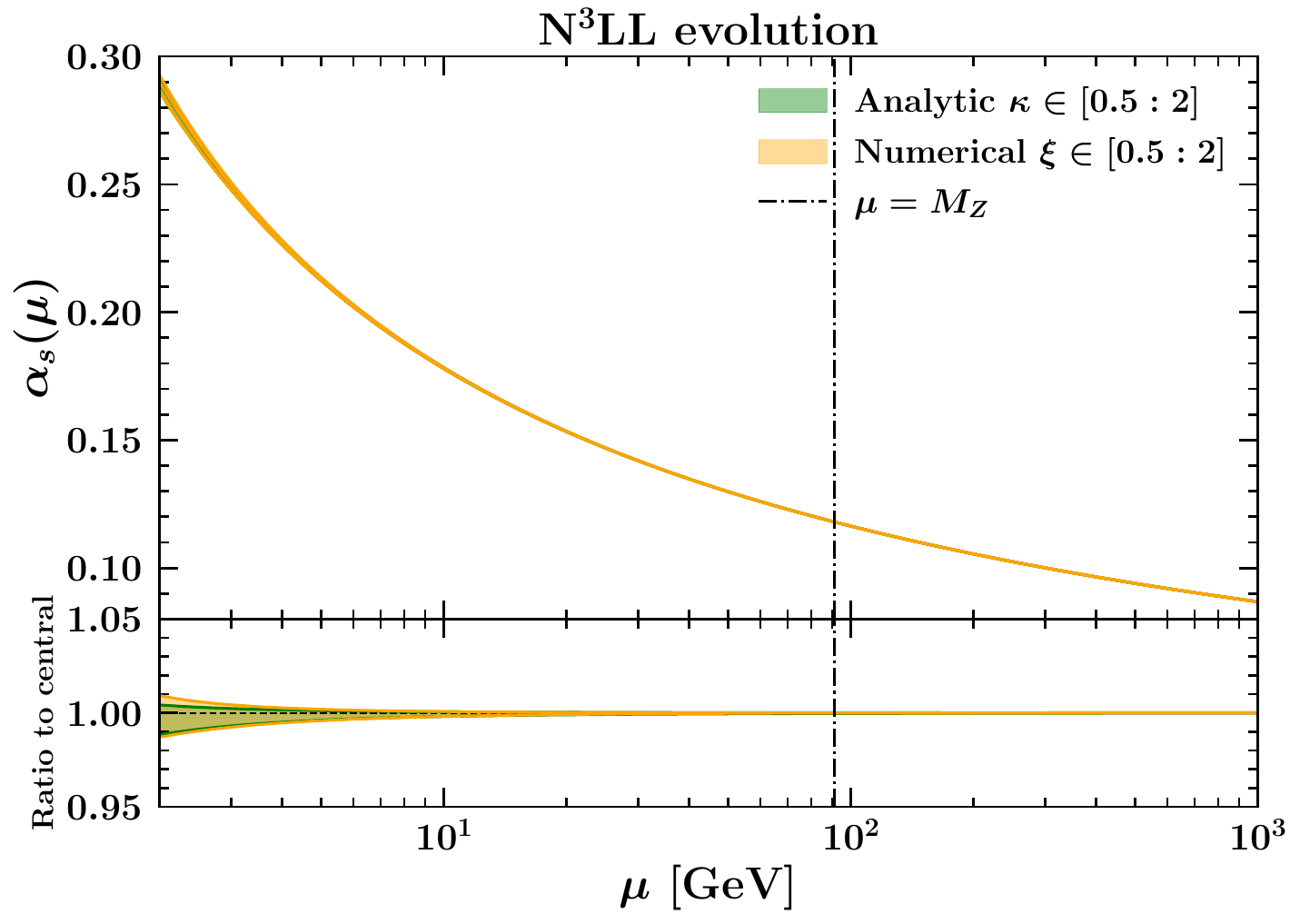}
    \caption{Analytic and numerical evolutions of the strong coupling
      $\alpha_s$ in the range $\mu\in[2:1000]$~GeV at NLL (top-left
      plot), NNLL (top-right plot), and N$^3$LL (bottom plot)
      accuracies. The boundary condition is set at
      $\alpha_s(M_Z)=0.118$. Uncertainty bands correspond to
      variations of the parameters $\kappa$ and $\xi$ for analytic and
      numerical solutions, respectively, in the ranges
      $\kappa,\xi\in[0.5:2]$. Lower panels display the bands
      normalised to the respective central-value curves obtained with
      $\kappa=\xi=1$. The vertical dot-dashed line indicates the scale
      $\mu=M_Z$ where the boundary condition is set. At this scale the
      uncertainty on numerical solutions reduces to
      zero.}\label{fig:AlphasScale}
  \end{centering}
\end{figure}
We finally move to presenting quantitative results for the analytic
and numerical evolutions of the strong
coupling. Fig.~\ref{fig:AlphasScale} shows the running of $\alpha_s$
computed at NLL (top-left), NNLL (top-right), and N$^3$LL (bottom)
accuracies, over a wide range in energy, obtained using
$\alpha_s(M_Z)=0.118$ as a boundary condition. The upper panel of each
plot displays the analytic (green) and the numerical (orange) runnings
of $\alpha_s$ along with bands that respectively correspond to
variations of $\kappa$ and $\xi$ in the range $[0.5:2]$. Lower panels
show the same curves normalised to the central-value predictions
obtained with $\kappa=\xi=1$. In accordance with the correspondence
between $\kappa$ and $\xi$ discussed above, we observe that the
uncertainties on analytic and numerical solutions at any given order
are generally comparable in size. We also see that the size of the
bands shrinks as one moves from NLL to N$^3$LL, as expected. In
particular, we observe that at NNLL and N$^3$LL uncertainties are well
below the 1\% level for most of the values of $\mu$
considered. However, for scales of the order of 2~GeV, the NNLL bands
reach the 3-4\% level, while the N$^3$LL ones remain at the 1\%
level. We note that the uncertainty on the numerical solution is zero
at $\mu=M_Z$, due to the fact that this corresponds to the
boundary-condition energy and thus no theoretical uncertainty related
to the running is present here. However, this feature is not shared by
the analytic solution.

\section{Parton distribution functions}
\label{sec:coll}

We next consider the case of Eq.~(\ref{eq:RGEproto}) in which the
quantity $R$ is identified with a Mellin moment of a flavour
non-singlet collinear PDF $f$, and $\gamma$ is its anomalous
dimension~\cite{Gribov:1972ri,Altarelli:1977zs,Dokshitzer:1977sg,Curci:1980uw,Furmanski:1980cm,Moch:2004pa,Vogt:2004mw}.\footnote{The
  discussion is generalisable to flavour-singlet distributions and $x$
  space, using anomalous dimension matrices and path-ordered
  exponentials where appropriate.}  The analytic solution takes the
form~\cite{Bertone:2022sso}
\begin{equation}
\label{eq:gensoldglap}
f^{{\rm N}^{k}{\rm LL}}(\mu) = g_0^{(\gamma), {\rm N}^k{\rm LL}}({\lambda},\kappa)\exp\left[\sum_{l=0}^{k}a_s^l(\mu_0)
  {g}_{l+1}^{(\gamma)}({\lambda},\kappa)\right]f(\mu_0)\,.
\end{equation}
The $g^{(\gamma)}$-functions depend on the variable $\lambda$ defined
in Eq.~(\ref{eq:lambdadefalphas}) and on the resummation scale
parameter $\kappa$ introduced in Eq.~(\ref{eq:SplitResScale}). In
order to achieve N$^k$LL accuracy, one needs to truncate
$g_0^{(\gamma), {\rm N}^k{\rm LL}}$ at $\mathcal{O}(\alpha_s^k)$ and
include functions ${g}_{i}^{(\gamma)}$ in the exponential up to
$i=k+1$. App.~\ref{App1} collects these functions up to NNLL accuracy.
The ${g}^{(\gamma)}$-functions that enter the exponential in
Eq.~(\ref{eq:gensoldglap}) are entirely defined in terms of the
${g}^{(\beta)}$-functions responsible for the analytic evolution of
$\alpha_s$ and of the LO coefficient of the PDF anomalous dimension
$\gamma_0$. Perturbative coefficients of the anomalous dimension
beyond LO contribute to $g_0^{(\gamma), {\rm N}^k{\rm LL}}$ in such a
way that coefficients up to $\gamma_k$ are necessary to achieve
N$^k$LL accuracy.

As in the case of the strong coupling, numerical solutions to the RGE
for PDFs are obtained by displacing the argument of $\alpha_s$ in the
expansion of the anomalous dimension by a factor of $\xi$, so that its
variations can be used to estimate the theoretical systematic
uncertainty. The resulting anomalous dimension valid up to NNLL
accuracy reads
\begin{equation}
  \label{eq:andimscale}
  \begin{array}{rcl}
  \displaystyle \gamma(\alpha_s(\mu)) &=&\displaystyle   a_s(\xi\mu)   \gamma_0   +a_s^2(\xi\mu)
                                          \left[\gamma_1 - \beta_0 \gamma_0 \ln\xi\right] \\
    \\
    &+&\displaystyle a_s^3(\xi\mu)  \left[\gamma_2 - \left(\beta_1 \gamma_0 + 2 \beta_0 \gamma_1 \right)\ln\xi+ \beta_0^2 \gamma_0 \ln^2\xi    \right]   +\mathcal{O}(\alpha_s^4)\,.
  \end{array}
\end{equation}
Analogously to the running of $\alpha_s$, one can show that variations
of $\xi$ in numerical solutions are in one-to-one correspondence with
variations of $\kappa$ in analytic solutions obtained through
Eq.~(\ref{eq:gensoldglap}).

In view of the phenomenological assessment of theoretical
uncertainties at the level of physical observables presented in
Sec.~\ref{sec:numapplications}, it is interesting to analytically
study the behaviour of resummation-scale uncertainties on PDFs. To
this purpose, let $f(\nu;\xi)$ be the PDF moment at the scale $\nu$
evolved from the scale $\nu_0$ using $\xi$ as resummation-scale
parameter. Defining
\begin{equation}
  \Delta f(\nu;\xi) = f(\nu;\xi) - f(\nu;1)\,,
\end{equation}
and using Eq.~(\ref{eq:andimscale}) truncated at
$\mathcal{O}(\alpha_s^2)$ so that the resulting evolution is NLL
accurate, we find
\begin{equation}
  \Delta f(\nu,\xi) =
  (a_s^2(\xi\nu)-a_s^2(\xi\nu_0))\ln \xi \left(\gamma_1+\frac12b_1\gamma_0 - \frac12\beta_0 \gamma_0\ln \xi\right)
  f(\nu;1) +\mathcal{O}(\alpha_s^3)\,.
  \label{eq:Deltafxi}
\end{equation}
A few comments are in order. First, we find
$\Delta f\sim\mathcal{O}(\alpha_s^2)$, that is, a subleading effect at
NLL accuracy, as expected.  Second, $\Delta f\sim\ln \xi$, so that it
approaches zero as $\xi\rightarrow 1$, again as expected. Finally,
$\Delta f\sim a_s^2(\xi\nu)-a_s^2(\xi\nu_0)$, which has important
consequences.\footnote{In general, one can show that, at N$^k$LL,
  $\Delta f(\nu;\xi)$ is proportional to
  $a_s^{k+1}(\xi\nu)-a_s^{k+1}(\xi\nu_0)$.} If $\nu\simeq \nu_0$, no
matter how small $\nu$ and thus how large $a_s(\xi\nu)$, $\Delta f$
remains small. If, instead, $\nu\gg \nu_0$ (or $\nu\ll \nu_0$),
$\Delta f$ will be dominated by the small scale, no matter how large
the large scale is. This behaviour of resummation-scale variations is
very specific and differs from that of usual renormalisation- and
factorisation-scale variations. Indeed, let us assume that PDFs are
measured (fitted) at the scale $\nu_0\simeq 1$~GeV, as it is often the
case. It follows that resummation-scale variations in the computation
of a physical observable at the scale $\nu$ will tend to produce a
small departure from the central scale when $\nu\simeq \nu_0$ and to
grow as $\nu$ becomes large. This is opposite to the typical trend of
renormalisation- and factorisation-scale variations, which are large
at small $\nu$ and small at large $\nu$. We will present quantitative
evidence of this behaviour in Sec.~\ref{sec:f2andfL} when studying DIS
structure functions (see also Ref.~\cite{Bertone:2022sso}). We
conclude that RGE solutions give rise to an additional source of
theoretical uncertainties, which is not accounted for by variations of
renormalisation and/or factorisation scales. Indeed, while the latter
estimate uncertainties specific to the region of the hard-scattering
scale $\nu$, the former accounts for the ``cumulative'' uncertainty
that stems from evolution between $\nu_0$ and $\nu$.

\begin{figure}[t]
  \begin{centering}
    \includegraphics[width=0.49\textwidth]{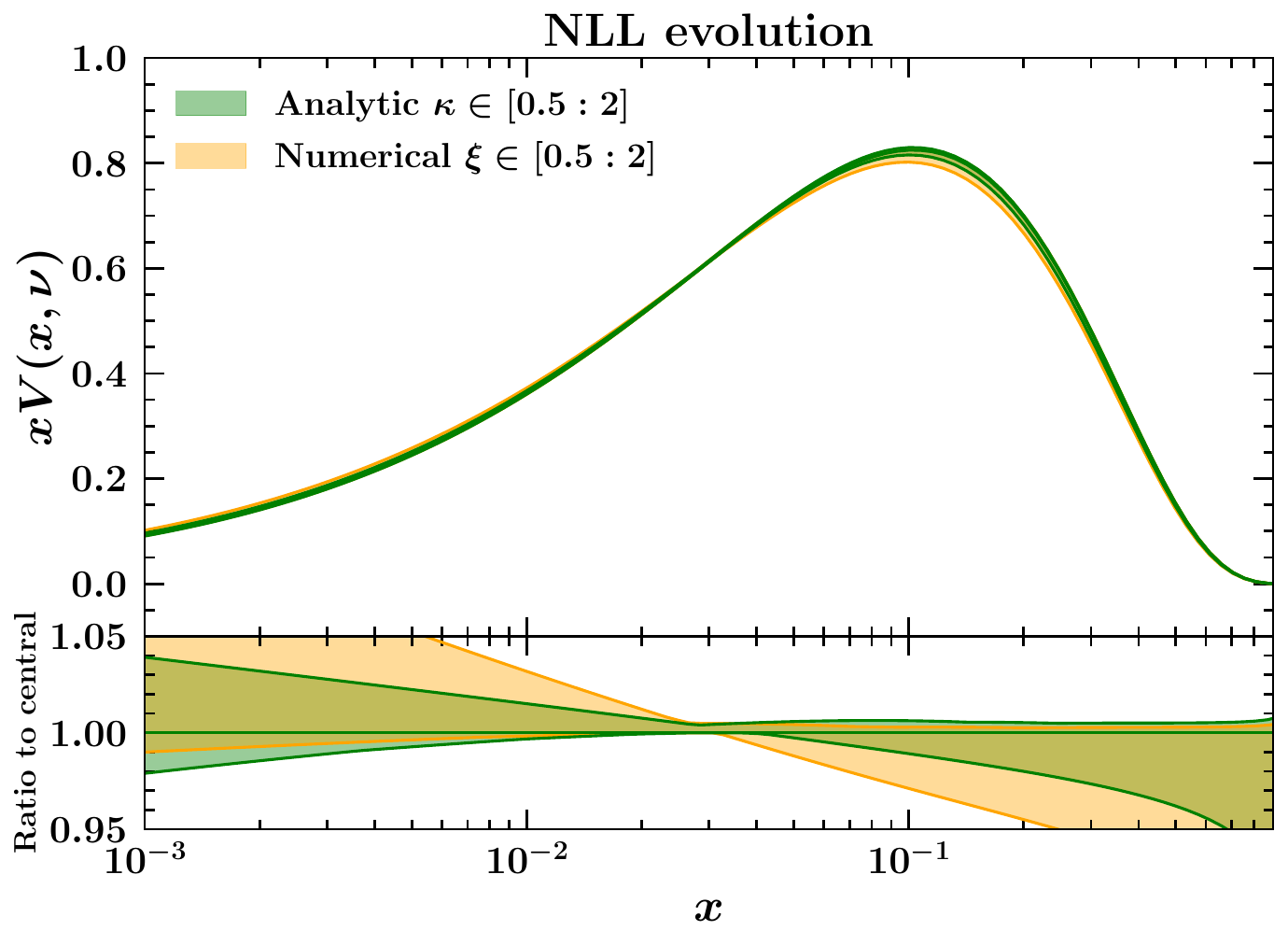}
    \includegraphics[width=0.49\textwidth]{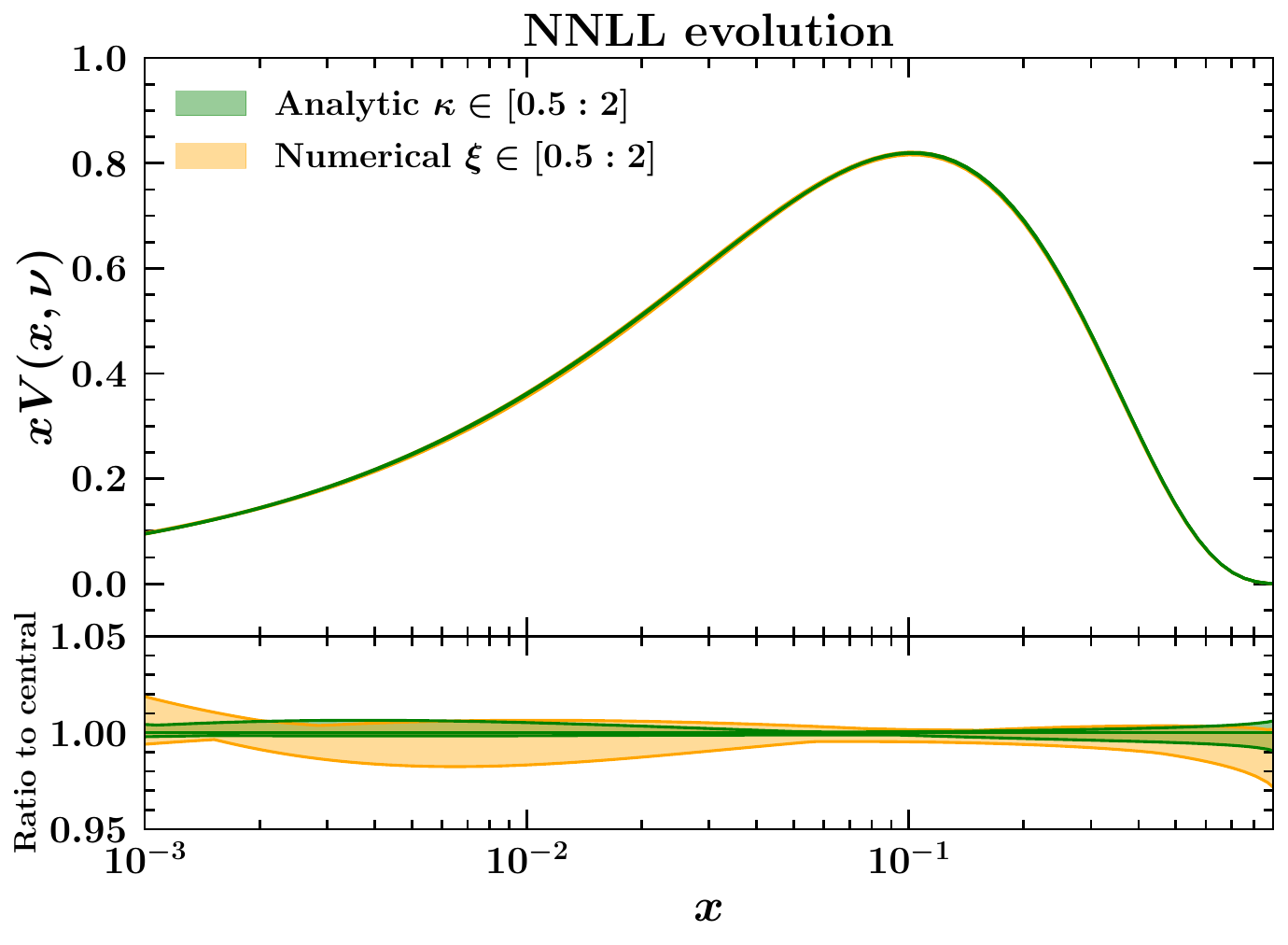}
    \caption{Total-valence PDF $xV$ at $\nu=100$~GeV evolved
      analytically and numerically, at NLL (left plot) and NNLL (right
      plot) accuracy, displayed over the range $x\in[10^{-3}:1]$. The
      boundary-condition distributions at $\nu_0=\sqrt{2}$~GeV and all
      other theory settings are taken from
      Ref.~\cite{Dittmar:2005ed}. Uncertainty bands correspond to
      variations of the parameters $\kappa$ and $\xi$ for analytic and
      numerical solutions, respectively, in the ranges
      $\kappa,\xi\in[0.5:2]$. Lower panels display the bands
      normalised to the respective central-value curves obtained with
      $\kappa=\xi=1$.}\label{fig:PDFsScale}
  \end{centering}
\end{figure}

Finally, we present a numerical comparison between analytic and
numerical PDF evolution using the settings of the benchmark exercise
of Ref.~\cite{Dittmar:2005ed}. To this purpose, we have used the
public code {\tt MELA}~\cite{Bertone:2015cwa}. In
Fig.~\ref{fig:PDFsScale}, we show the total valence distribution
$V=\sum_q(q-\overline{q})$ as a function of $x$ evolved from
$\nu_0=\sqrt{2}$~GeV to $\nu=100$~GeV at NLL (left) and NNLL
(right). Each curve has an uncertainty band deriving from variations
of the resummation-scale parameters $\kappa$ and $\xi$ in the analytic
and numerical solutions, respectively. As usual, the upper panel of
each plot shows the actual distributions along with their uncertainty
bands, while the lower panel shows the ratio to the $\kappa=\xi=1$
curves. As expected, we see that the size of the uncertainty bands is
reduced as one moves from NLL to NNLL. In addition, at each of the two
orders considered, analytic and numerical bands are comparable in
size, showing again the correspondence between variations of $\kappa$
in the analytic solutions and $\xi$ in the numerical ones. A similar
behaviour is observed also for the other PDF combinations.

To conclude this section, we point out that resummation-scale
uncertainties for the evolution of $\alpha_s$ and PDFs were treated as
distinct in our previous
studies~\cite{Bertone:2022sso,Bertone:2022ope}. However, the analysis
presented above suggests that this is not necessary. Indeed, the
variable $\lambda$, defined in Eq.~(\ref{eq:lambdadefalphas}) and
entering the analytic evolution of PDFs, arises directly from the
evolution of the coupling. As a consequence, we deduce that variations
of $\kappa$ in analytic solutions for PDFs (and by extension of $\xi$
in numerical solutions) are sufficient to estimate the total RGE
systematic uncertainties associated with the evolutions of both
$\alpha_s$ and PDFs.

\section{Transverse-momentum-dependent distributions}
\label{sec:sud}

In this section, we examine the evolution of TMDs from the standpoint
of RGE systematic uncertainties. We give a more detailed presentation
than in the cases of $\alpha_s$ and PDFs in the previous two
sections. Indeed, unlike $\alpha_s$ and PDFs, TMDs were not treated in
our previous work of Ref.~\cite{Bertone:2022sso}. Moreover, new
features arise in the case of TMDs, which are associated with the
double-logarithmic nature of the evolution. These will help us
illustrate further aspects of the resummation scales and the physical
meaning of the different scales that enter resummed theoretical
predictions.
 
In the following, we start by recalling the TMD evolution equations
(Sec.~\ref{sec:tmd}), then express the solution to these equations
through the Sudakov form factor (Sec.~\ref{sec:sud_ff}), and present
quantitative results for RGE systematics
(Secs.~\ref{sec:rge-effects-in-sud} and \ref{sudhyst}). In the light
of these results, we comment on the matching of TMDs onto collinear
PDFs (Sec.~\ref{sec:collmatch}).

\subsection{Evolution}
\label{sec:tmd}

TMD factorisation~\cite{Collins:2011zzd} is commonly carried out in
impact-parameter space $b$, that is, the Fourier-conjugated variable
of the partonic transverse momentum $k_T$. We thus consider the TMD
for the partonic species $i$, $F_i\equiv F_i(x, b,\mu,\zeta)$, which
depends on the longitudinal momentum fraction $x$, the impact
parameter $b$, the renormalisation scale $\mu$, and the rapidity scale
$\zeta$. Contrary to collinear distributions, TMDs do not mix upon
evolution.  Therefore, in order to simplify the notation, we drop the
flavour index $i$. Since in what follows we will mostly be interested
in the dependence of $F$ on the scales $\mu$ and $\zeta$, we also drop
its dependence on $x$ and $b$.

The TMD $F$ fulfils the following evolution equations,
\begin{equation}\label{eq:eveqs}
  \begin{array}{l}
    \displaystyle \frac{\partial \ln F}{\partial \ln \sqrt{\zeta}} =
    \overline{K}(\mu)\,,\\
    \\
    \displaystyle \frac{\partial \ln F}{\partial \ln \mu} = \gamma(\mu,\zeta)\,,
  \end{array}
\end{equation}
where $\gamma$ and $\overline{K}$ are the anomalous dimensions of the
evolution in $\mu$ and $\sqrt{\zeta}$, respectively.

Appropriate boundary conditions are needed to solve these evolution
equations. With the definition
\begin{equation}
  \label{eq:mubdef}
  \mu_b= \frac{2e^{-\gamma_E}}{b}\,,
\end{equation}
where $\gamma_E$ is the Euler-Mascheroni constant, it turns out that
convenient boundary-condition scales for the differential equations in
Eq.~(\ref{eq:eveqs}) are $\mu_0=\mu_b$ and $\zeta_0=\mu_b^2$, such
that $F(\mu_0,\zeta_0) = F(\mu_b,\mu_b^2)$ is assumed to be
known. Without loss of generality, we also set $\zeta = M^2$ as a
final rapidity scale,\footnote{The choice of $\zeta$ is actually
  immaterial. Any physical observable is written as the convolution of
  two TMDs, which may be computed at different rapidity scales
  $\zeta_1$ and $\zeta_2$. However, their convolution only depends on
  the product $\zeta_1\zeta_2=M^4$, which is fixed by kinematic
  constraints.}  where $M$ is the hard-scattering scale of the process
under consideration. Moreover, we also set $\mu=M$. As we will argue
below, variations of the resummation-scale parameter $\kappa$
introduced in Sec.~\ref{sec:emergent} automatically account for
variations of $\mu$ around $M$.

Under these conditions, the differential equations in
Eq.~(\ref{eq:eveqs}) are solved by
\begin{equation}\label{eq:solTMD1}
  F(M,M^2) = \exp\left[ \overline{K} (M)\ln\frac{M}{\mu_b}+\int_{\mu_b}^{M}\frac{d\mu'}{\mu'}\gamma(\mu',\mu_b^2)\right]F(\mu_b, \mu_b^2)\,.
\end{equation}
Besides, the equality of the crossed second derivatives of $F$
yields
\begin{equation}
  \label{eq:crossder}
  \displaystyle \frac{\partial \ln F}{\partial \ln\mu \ln \sqrt{\zeta}}
  = \frac{d \overline{K}}{d\ln \mu} = \frac{d\gamma}{d\ln\sqrt{\zeta}} = - \gamma_K(a_s(\mu))\, ,
\end{equation}
where $\gamma_K$ is the cusp anomalous
dimension. Eq.~(\ref{eq:crossder}) can be solved for $\overline K$ and
$\gamma$ using respectively $\mu=\mu_b$ and $\zeta=\mu^2$ as
boundary-condition scales, yielding
\begin{equation}\label{eq:anomdimevol}
  \begin{array}{l}
    \displaystyle \overline{K}(\mu) = K(a_s(\mu_b)) -
    \int_{\mu_b}^{\mu}\frac{d\mu'}{\mu'}\gamma_K(a_s(\mu'))\,,\\
    \\
    \displaystyle \gamma(\mu,\zeta) = \gamma_F(a_s(\mu)) - \gamma_K(a_s(\mu))\ln\frac{\sqrt{\zeta}}{\mu}\,,
  \end{array}
\end{equation}
with $K(a_s(\mu_b))=\overline{K}(\mu_b)$ and
$\gamma_F(a_s(\mu)) = \gamma(\mu,\mu^2)$. The relations in
Eq.~(\ref{eq:anomdimevol}) can be plugged into Eq.~(\ref{eq:solTMD1}),
finally giving
\begin{equation}
  \label{eq:solTMD2}
  F(M,M^2) = \exp\left\{ K(a_s(\mu_b))\ln\frac{M}{\mu_b}+\int_{\mu_b}^{M}\frac{d\mu'}{\mu'}\left[\gamma_F(a_s(\mu')) - \gamma_K(a_s(\mu'))\ln\frac{M}{\mu'}\right]\right\}F(\mu_b,\mu_b^2)\,.
\end{equation}
All of the anomalous dimensions in Eq.~(\ref{eq:solTMD2}) admit
perturbation expansions as power series in $\alpha_s$, as follows,
\begin{equation}
  \begin{array}{l}
    \displaystyle K(a_s(\mu)) = \sum_{n=0}^{\infty}a_s^{n+1}(\mu)K^{(n)}\,,\\
    \\
    \displaystyle  \gamma_F(a_s(\mu)) =
    \sum_{n=0}^{\infty}a_s^{n+1}(\mu)\gamma_F^{(n)}\,,\\
    \\
    \displaystyle  \gamma_K(a_s(\mu)) =
    \sum_{n=0}^{\infty}a_s^{n+1}(\mu)\gamma_K^{(n)}\,.
  \end{array}
  \label{eq:AnomDimExps}
\end{equation}

\subsection{Sudakov form factor}
\label{sec:sud_ff}

The solution in Eq.~(\ref{eq:solTMD2}) can be written as
\begin{equation}
  \label{eq:solTMD3}
  F(M,M^2) = \exp\left[\frac12S(M,\mu_b)\right]F(\mu_b,\mu_b^2)\,    ,
\end{equation}
where $S$ is referred to as {\it Sudakov form factor},
\begin{equation}
  \label{eq:SudFF}
  S(M,\mu_b) = 2K(a_s(\mu_b))\ln\frac{M}{\mu_b}+2\int_{\mu_b}^{M}\frac{d\mu'}{\mu'}\left[\gamma_F(a_s(\mu')) - \gamma_K(a_s(\mu'))\ln\frac{M}{\mu'}\right]\,   .
\end{equation}
Exploiting the identity
\begin{equation}
  \label{eq:CSexpr}
  K(a_s(\mu_b))\ln\frac{M}{\mu_b} = -\int_{\mu_b}^{M}\frac{d\mu'}{\mu'}\left[a_s(\mu') \beta(a_s(\mu'))\frac{dK(a_s(\mu'))}{da_s(\mu')} \ln\frac{M}{\mu'}  - K(a_s(\mu'))\right]\,,
\end{equation}
the Sudakov form factor in Eq.~(\ref{eq:SudFF}) can be written as
\begin{eqnarray}
  \label{sud-rewr} 
  S(M,\mu_b) &=& \int_{\mu_b}^{M}\frac{d\mu'^2}{\mu'^2}\left[\gamma_F(a_s(\mu')) +K(a_s(\mu')) \right.
                 \nonumber\\
             &-&  \left. \frac12\left(\gamma_K(a_s(\mu'))+a_s(\mu') \beta(a_s(\mu'))\frac{dK(a_s(\mu'))}{da_s(\mu')}\right)\ln\frac{M^2}{\mu'^2}\right]\,  .
\end{eqnarray}
Defining
\begin{equation}
  \label{eq:expansions}
  \begin{array}{rcl}
    \displaystyle  A(a_s(\mu))&=&\displaystyle
                                  \frac12\gamma_K(a_s(\mu))+\frac12
                                  a_s(\mu)
                                  \beta(a_s(\mu))\frac{dK(a_s(\mu))}{da_s(\mu)}\,,\\
    \\
    \displaystyle  B(a_s(\mu))&=&\displaystyle
                                  -\gamma_F(a_s(\mu))
                                  -K(a_s(\mu))\,,
  \end{array}
\end{equation}
Eq.~(\ref{sud-rewr}) takes the form
\begin{equation}\label{eq:SudFF2}
  S(M,\mu_b) = - \int_{\mu_b^2}^{M^2}\frac{d\mu'^2}{\mu'^2}\left[A(a_s(\mu'))\ln\frac{M^2}{\mu'^2}+B(a_s(\mu'))\right]\,, 
\end{equation}

As a consequence of Eqs.~(\ref{eq:AnomDimExps})
and~(\ref{eq:expansions}), the functions $A$ and $B$ admit the
perturbative expansions
\begin{equation}
  \begin{array}{l}
    \displaystyle  A(a_s(\mu)) =
    \sum_{n=1}^{\infty}a_s^{n}(\mu)A^{(n)}\,,\\
    \\
    \displaystyle  B(a_s(\mu)) =
    \sum_{n=1}^{\infty}a_s^{n}(\mu)B^{(n)}\,,
  \end{array}
  \label{eq:ABexps}
\end{equation}
and their perturbative coefficients can be related to those of the
anomalous dimensions $\gamma_K$, $\gamma_F$, and $K$. Since the
logarithmic accuracy N$^k$LL is achieved by computing $A$ at N$^k$LO
and $B$ at N$^{k-1}$LO, we have, up to N$^3$LL accuracy,
\begin{equation}
  \begin{array}{rcrclcl}
    \mbox{LL}:& &A^{(1)} &=& \displaystyle \frac12\gamma_K^{(0)}\,,&\qquad& \\
    \\
    \mbox{NLL}:& &A^{(2)} &=& \displaystyle
                              \frac12\gamma_K^{(1)}+\frac12\beta_{0}K^{(0)}\,,&\qquad
                                                                          &
                                                                            B^{(1)}
                                                                            =
                                                                            -\gamma_F^{(0)}-K^{(0)}\,,\\
    \\
    \mbox{NNLL}:& &A^{(3)} &=& \displaystyle
                               \frac12\gamma_K^{(2)}+\frac12\beta_{1}K^{(0)}+\beta_{0}K^{(1)}\,,&\qquad
                                                                          &
                                                                            B^{(2)}
                                                                            =
                                                                            -\gamma_F^{(1)}-K^{(1)}\,,\\
    \\
    \mbox{N}^3\mbox{LL}:& &A^{(4)} &=& \displaystyle
                                       \frac12\gamma_K^{(3)}+\frac12\beta_{2}K^{(0)}+\beta_{1}K^{(1)}+\frac{3}2\beta_{0}K^{(2)}\,,&\qquad
                                                                          &
                                                                            B^{(3)}
                                                                            =
                                                                            -\gamma_F^{(2)}-K^{(2)}\,.
  \end{array}
  \label{eq:AnDimToAB}
\end{equation}
Because $K^{(0)} = 0$ (see App.~\ref{App2}), at LL and NLL the $A$ and
$B$ coefficients are proportional, respectively, to the $\gamma_K$ and
$\gamma_F$ coefficients. Beyond NLL, this proportionality no longer
holds~\cite{Becher:2010tm} due to the contribution of the rapidity
anomalous dimension $K$.

Similarly to the cases of $\alpha_s$ and PDFs, the Sudakov form factor
in Eq.~(\ref{eq:SudFF2}) (or equivalently in Eq.~(\ref{eq:SudFF}))
encoding the evolution of TMDs can be computed analytically or
numerically. Once again, analytic solutions are obtained by resorting
to perturbative expansions, at the price that the evolution equations
in Eq.~(\ref{eq:eveqs}) are violated by subleading terms.

\subsection{RGE systematics}
\label{sec:rge-effects-in-sud}

We now exploit the analytic running of the strong coupling in
Sec.~\ref{sec:runningcoupling} to obtain the analytic solution to the
RGEs that govern the evolution of TMDs. By making the following change
of integration variable,
\begin{equation}
  t = a_s(\mu_0)\beta_{0}\ln\frac{\kappa \mu'}{\mu_0}\,,
\end{equation}
where $\mu_0$ is the reference scale at which the coupling 
$\alpha_s$ is measured, the 
Sudakov form factor $S$ in Eq.~(\ref{eq:SudFF2}) can be rewritten as
\begin{equation}
  \label{eq:SudFF31}
  S(M,\mu_b) = -\frac{2}{a_s(\mu_0)\beta_{0}}\int_{\lambda_{M}}^{\lambda}dt \left[\frac{2 t A(a_s(t))}{a_s(\mu_0)\beta_{0}}-\overline{B}(a_s(t))\right]\,, 
\end{equation}
where we have set
\begin{equation}
  \lambda_{M} = a_s(\mu_0)
  \beta_{0}\ln\frac{\kappa M}{\mu_0}\quad\mbox{and}\quad \lambda = a_s(\mu_0)
  \beta_{0}\ln\frac{\kappa \mu_b}{\mu_0}\,,
\end{equation}
and, with slight abuse of notation, denoted by $a_s(t)$ the coupling
evaluated in terms of the ${g}^{(\beta)}$-functions (see
Eq.~(\ref{eq:alphasgs})) with argument $t$. Moreover, we have defined
\begin{equation}\label{eq:Bdisplacement}
  \overline{B}(a_s) = 2A(a_s)\ln\frac{\kappa M}{\mu_0}+B(a_s) = \sum_{n=1}^{\infty}a_s^{n}\overline{B}^{(n)}\,.
\end{equation}

We now assume that $\mu_0\simeq M$ and $\kappa\simeq 1$, so that
$\lambda_{M}\ll 1$. Splitting the integral in Eq.~(\ref{eq:SudFF31})
as follows,
\begin{equation}\label{eq:Sfinal}
  \begin{array}{rcl}
  \displaystyle S(M,\mu_b) &=& \displaystyle
                                   -\frac{2}{a_s(\mu_0)\beta_{0}}\int_{0}^{\lambda}dt \left[\frac{2 t
    A(a_s(t))}{a_s(\mu_0)\beta_{0}} -\overline{B}(a_s(t))\right]\\
    \\
    &+& \displaystyle \frac{2}{a_s(\mu_0)\beta_{0}}\int_0^{\lambda_{M}}dt
    \left[\frac{2 t A(a_s(t))}{a_s(\mu_0)\beta_{0}}-\overline{B}(a_s(t))\right]\,,
  \end{array}
\end{equation}
we are allowed to compute the second integral perturbatively,
expanding the running coupling around $t=0$. In addition, the
exponential of this term (see Eq.~(\ref{eq:solTMD3})) can be further
expanded and truncated at the appropriate order.

The explicit analytic evaluation of the integrals in
Eq.~(\ref{eq:Sfinal}) finally leads to an expression for the Sudakov
form factor of the form
\begin{equation}\label{eq:gfunc}
  \exp(S) = g_0(a_s(\mu_0))\exp\left[L g_1(\lambda) + g_2(\lambda,\kappa) + a_s(\mu_0) g_3(\lambda,\kappa) + a_s^2(\mu_0) g_4(\lambda,\kappa) + \dots\right]\,,
\end{equation}
where $L=\ln(\kappa\mu_b/\mu_0)$, and the $g$-functions are expressed
in terms of the perturbative coefficients of the functions $A$ and
$B$. The function $g_n$, with $n>0$, in Eq.~(\ref{eq:gfunc}) resums
the N$^{n-1}$LL tower of logarithms and can be computed in terms of
the functions $g_{k}^{(\beta)}$, with $k\leq n$, responsible for the
analytic evolution of $\alpha_s$.  The function $g_0$, on the other
hand, admits a perturbative expansion that is to be truncated at
$\mathcal{O}(a_s^n)$ to achieve N$^{n+1}$LL
accuracy~\cite{Bacchetta:2019sam}. All ingredients necessary to obtain
a N$^3$LL accurate Sudakov form factor are given in App.~\ref{App2}.
 
The form of the Sudakov form factor in Eq.~(\ref{eq:gfunc})
corresponds to the customary formula obtained in the literature on
transverse-momentum resummation: see, \textit{e.g.},
Ref.~\cite{Bozzi:2005wk}. A comparison of the $g$-functions given in
App.~\ref{App2} with those of Ref.~\cite{Bozzi:2005wk} shows that the
results agree. In this respect, the calculation which we have
performed in this section can be viewed as a derivation of
Eq.~(\ref{eq:gfunc}) alternative to that given in transverse-momentum
resummation, based on a different, but equivalent, methodology which
starts from the TMD evolution equations in Eq.~(\ref{eq:eveqs}).
 
Interestingly, the agreement between our results and those of
Ref.~\cite{Bozzi:2005wk} is achieved by making the identifications
$\mu_{\rm R}=\mu_0$ and $Q=\mu_0/\kappa$, where $\mu_{\rm R}$ and $Q$
are, respectively, the renormalisation and resummation scales in the
terminology of Ref.~\cite{Bozzi:2005wk}.  This observation sheds light
on the meaning of the different scales which enter resummed
theoretical predictions.  As regards the renormalisation scale
$\mu_{\rm R}$ of Ref.~\cite{Bozzi:2005wk}, this is to be identified,
in our analysis, with the reference scale $\mu_0$ at which the strong
coupling is measured. In other words, $a_s(\mu_0)$ is the measured
value of the strong coupling used as an input to its running. To be
definite, a common choice is $\alpha_s(M_Z)=0.118$, so that
$\mu_0=M_Z$ and $4\pi a_s(\mu_0)=0.118$. In this respect, $\mu_0$ does
not have the same meaning as a ``standard'' renormalisation scale used
in fixed-order calculations.

As regards the resummation scale $Q$, introduced in
Ref.~\cite{Bozzi:2005wk} directly at the level of Sudakov form factor,
our analysis illustrates that this can be traced back to the
resummation-scale parameter $\kappa$ introduced in
Eq.~(\ref{eq:SplitResScale}) and used in
Eq.~(\ref{eq:lambdadefalphas}) to parameterise subleading corrections
to the evolution of the strong coupling. Indeed, as illustrated by the
analyses of $\alpha_s$ and PDFs in Secs.~\ref{sec:runningcoupling}
and~\ref{sec:coll}, the presence of a resummation scale is
\textit{not} a prerogative of the resummation of Sudakov (double)
logarithms. As a matter of fact, it emerges whenever a resummation
takes place as a consequence of the truncation of the anomalous
dimensions that govern the relevant RGEs. The resummation scale thus
defined originates from the running of the strong coupling, which
feeds through the resummation as an expansion parameter of the
anomalous dimensions.

We will next move to considering the numerical solution to
Eq.~(\ref{eq:SudFF2}). Analogously to the previous sections, this is
achieved by displacing the argument of $\alpha_s$ from $\mu$ to
$\xi\mu$ in the perturbative expansion of the functions $A$ and $B$ in
Eq.~(\ref{eq:ABexps}). We refrain from showing the corresponding
expansions. The integral in Eq.~(\ref{eq:SudFF2}) is then computed
numerically. As in the cases of $\alpha_s$ and PDFs, variations of
$\xi$ in numerical solutions correspond to variations of $\kappa$ in
analytic solutions and can thus be used to estimate RGE systematic
uncertainties.

\subsection{Perturbative uncertainty on the Sudakov form factor}
\label{sudhyst}

\begin{figure}[h]
  \begin{centering}
    \includegraphics[width=0.49\textwidth]{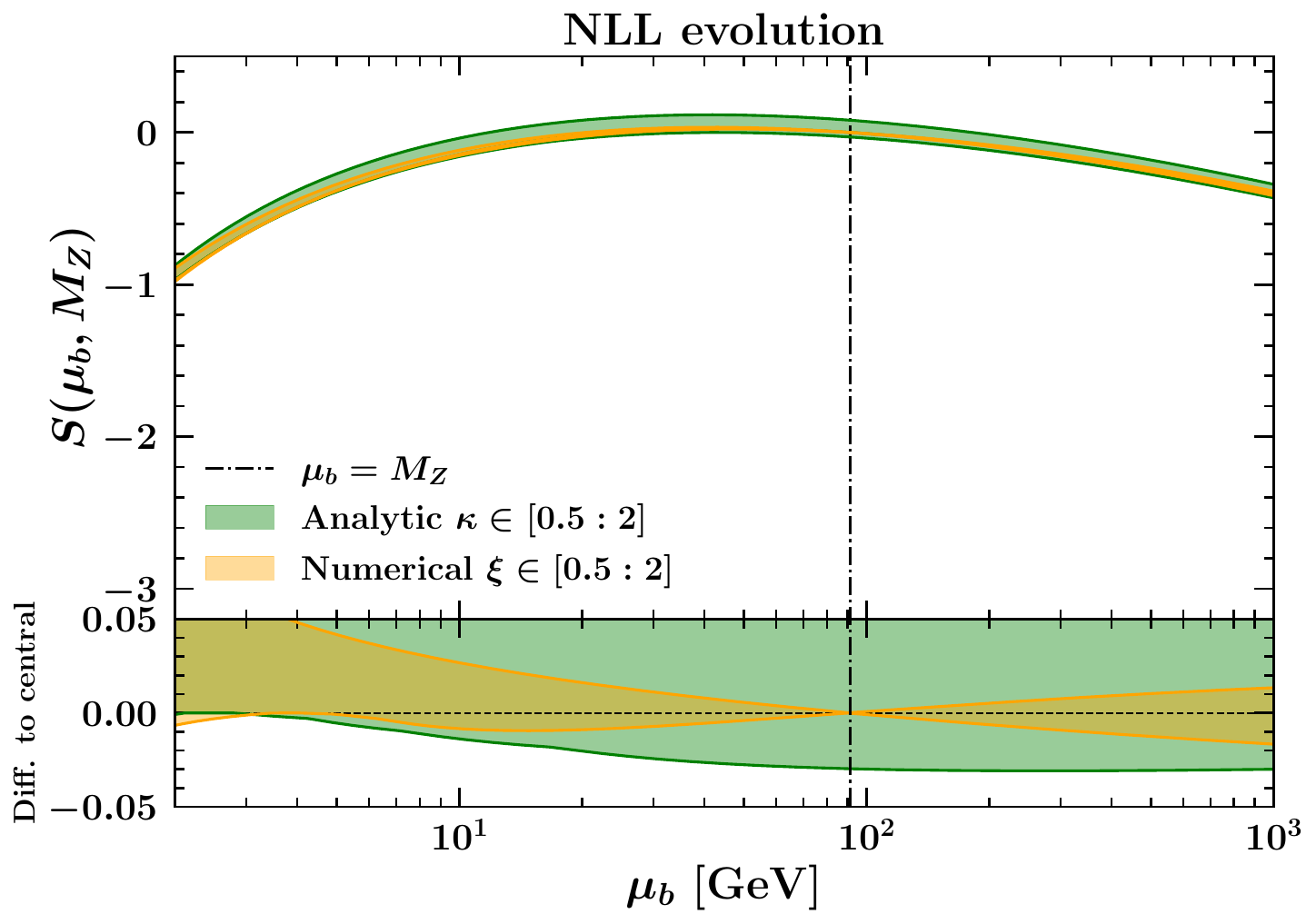}
    \includegraphics[width=0.49\textwidth]{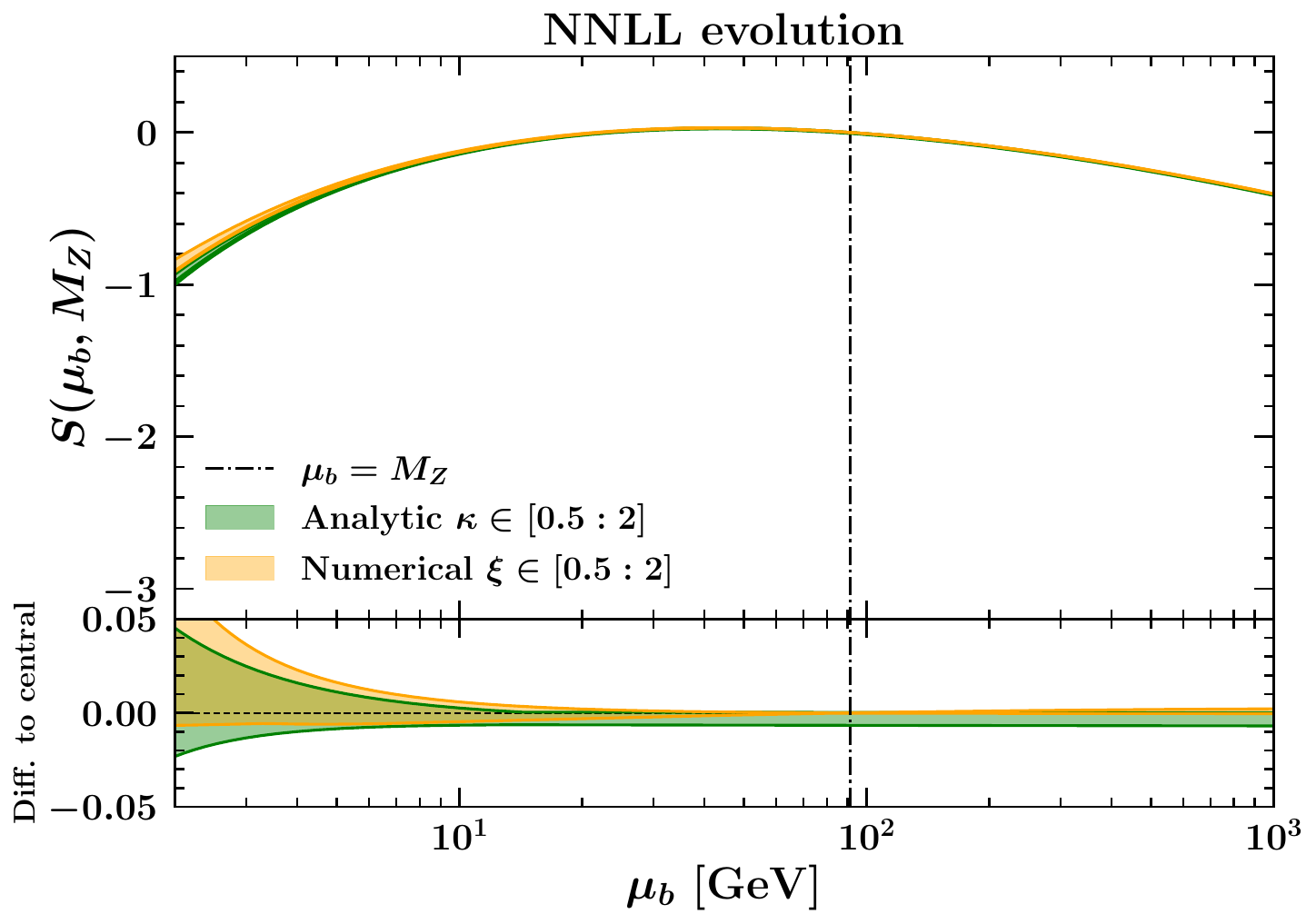}
    \includegraphics[width=0.49\textwidth]{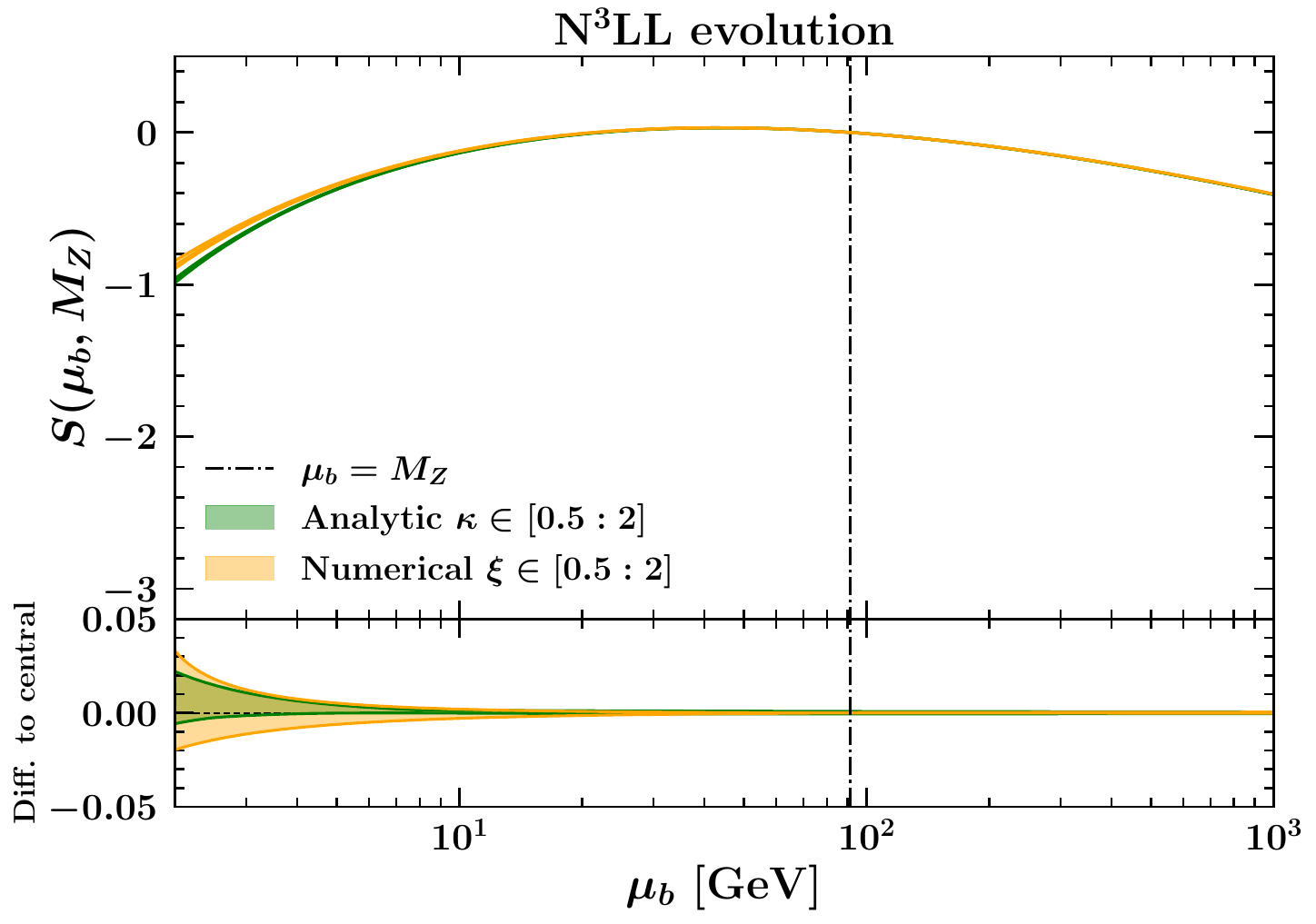}
    \caption{Analytic and numerical Sudakov form factors as functions
      of $\mu_b$ in the range $\mu_b\in[2:1000]$~GeV at NLL (top-left
      plot), NNLL (top-right plot), and N$^3$LL (bottom plot)
      accuracies. The boundary condition of the strong coupling is set
      at $\alpha_s(M_Z)=0.118$. Uncertainty bands correspond to
      variations of the parameters $\kappa$ and $\xi$ for analytic and
      numerical solutions, respectively, in the range
      $\in[0.5:2]$. Lower panels display the bands normalised to the
      respective central-value curves obtained with
      $\kappa=\xi=1$.}\label{fig:SudakovAnVsNum}
  \end{centering}
\end{figure}

As an application of the formalism discussed in the previous
subsection, we now assess the perturbative uncertainty on the Sudakov
form factor from variations of the resummation scales. In view of
applications to the transverse-momentum spectrum in DY production
presented in the next sections, we only consider the case of quark
TMDs, leaving gluons aside. Fig.~\ref{fig:SudakovAnVsNum} shows
results for $S$ as a function of $\mu_b$, obtained with
$\alpha_s(M_{Z})=0.118$ as a boundary condition to the strong
coupling, for NLL, NNLL, and N$^3$LL accuracies. The bands correspond
to variations of the resummation-scale parameters $\kappa$ (analytic
solution) and $\xi$ (numerical solution) in the range $[0.5: 2]$. The
upper panel of each plot shows the results for $S$ along with their
bands, while the lower panel displays their ratio to the central-scale
curves obtained with $\kappa=\xi=1$. We see a significant reduction of
the theoretical systematic uncertainties when moving from NLL to
N$^3$LL. While at NLL the uncertainties obtained with the numerical
solution are significantly smaller than those obtained with the
analytic solution, the size of the uncertainty bands at NNLL and
N$^3$LL tends to align. At N$^3$LL, in particular, one can see from
the lower inset of the bottom plot of Fig.~\ref{fig:SudakovAnVsNum}
that in the small-$\mu_b$ region, where $\alpha_s$ is largest, the
bands are as large as 3-4\%, while they tend to shrink to a
sub-percent level as $\mu_b$ increases. We also note that the
uncertainty band for the numerical calculation shrinks to zero at
$\mu_b=M_Z$, where the boundary condition on the running of $\alpha_s$
is imposed. This does not happen in the analytic case.

Further insights on the manifestation of the {\it perturbative hysteresis}~\cite{Bertone:2022sso} in the analytically computed Sudakov form factor can be found in Appendix~\ref{app:hyst}.

\subsection{Matching on collinear distributions}\label{sec:collmatch}

For small enough values of the impact parameter $b$, TMDs can be
matched onto collinear PDFs via an operator product expansion written
in terms of perturbatively computable coefficients.
Specifically, the expansion of the TMD distribution $F_i$ can be
written as
\begin{equation}
  F_i(x,b,\mu,\zeta) = \sum_{j}
  \int_x^1\frac{dy}{y}\overline{C}_{ij}(y,b,\mu,\zeta)
  f_{j}\left(\frac{x}{y},\mu\right) \equiv F(\mu,\zeta) \,,
  \label{eq:matchingformula}
\end{equation}
where $f_j$ is a collinear PDF, and the functions $\overline{C}$ admit
the perturbative expansion
\begin{equation}
  \label{eq:Cexpansion}
  \overline{C}_{ij}(y,b,\mu,\zeta) =\sum_{n=0}^{\infty}
  a_s^n(\mu)\sum_{m=0}^{2n}\sum_{l=0}^{m} \bar{C}_{ij}^{(n,m,l)}(y)\ln^{m-l}\left(\frac{\mu}{\mu_b}\right)\ln^{l}\left(\frac{\sqrt{\zeta}}{\mu_b}\right)\,.
\end{equation}
We can now use the matching formula Eq.~(\ref{eq:matchingformula}) in
the r.h.s.~of Eq.~(\ref{eq:solTMD3}) where, due to the scale choice,
the logarithmic terms in Eq.~(\ref{eq:Cexpansion}) vanish, giving
\begin{equation}
  \label{Ccentralexp}
  \overline{C}_{ij}(y,b,\mu_b,\mu_b^2) =\sum_{n=0}^{\infty}
  a_s^n(\mu_b)\bar{C}_{ij}^{(n,0,0)}(y)\equiv C_{ij}(a_s(\mu_b))\,,
\end{equation}
that is, a quantity free of any potentially large logarithms and whose
scale dependence is entirely driven by the strong coupling. Therefore,
dropping the flavour indices and denoting the Mellin convolution with
the symbol $\otimes$, we have
\begin{equation}
  \label{eq:Fmub}
  F(\mu_b,\mu_b^2)=C(a_s(\mu_b))\otimes f(\mu_b)\,.
\end{equation}

When computing the Fourier transform with respect to the impact
parameter $b$ to obtain physical observables, both the matching
coefficient function $C$ and the PDF $f$ are evaluated at the scale
$\mu_b$, defined in Eq.~(\ref{eq:mubdef}), which can get any positive
value. We then resort to the RGEs to write both $C(a_s(\mu_b))$ and
$f(\mu_b)$ as resulting from the evolution applied to some fixed
reference scale $M$.

The PDF $f$ obeys the RGE discussed in Sec.~\ref{sec:coll}, whose
solution from $M$ to $\mu_b$ allows us to formally write\footnote{As
  also noted in Sec.~\ref{sec:coll}, this solution implies a number of
  subtleties related to the flavour structure and the presence of the
  Mellin convolution.}
\begin{equation}
  \label{eq:PDFevol}
  f(\mu_b) =
  \exp\left[-\int_{\mu_b}^{M}\frac{d\mu'}{\mu'}\gamma(a_s(\mu'))\right]\otimes
  f(M)\,.
\end{equation}
The coefficient $C$, depending on the scale $\mu$ only through the
strong coupling, obeys the following evolution equation\footnote{The
  logarithm of $C$ is to be interpreted as the logarithm of a matrix
  in flavour space. In addition, the solution of this evolution
  equation is subject to the same caveats as the PDF evolution
  discussed above.}
\begin{equation}
  \frac{d\ln C(a_s(\mu))}{d\ln\mu} =a_s(\mu)\beta(a_s(\mu)) \frac{d\ln C(a_s(\mu))}{da_s(\mu)}\,.
\end{equation}
This equation can be used to evolve $C$ from $M$ to $\mu_b$ as
follows,
\begin{equation}
  \label{eq:Cevol}
  C(a_s(\mu_b)) =\exp\left[-\int_{\mu_b}^{M}\frac{d\mu'}{\mu'}
    a_s(\mu')\beta(a_s(\mu')) \frac{d\ln
      C(a_s(\mu'))}{da_s(\mu')}\right]\otimes C(a_s(M))\,.
\end{equation}

Using Eqs.~(\ref{eq:PDFevol}) and~(\ref{eq:Cevol}) in
Eq.~(\ref{eq:Fmub}), we find
\begin{equation}
  F(\mu_b,\mu_b^2)=\exp\left[-\int_{\mu_b}^{M}\frac{d\mu'}{\mu'} \left(a_s(\mu')\beta(a_s(\mu')) \frac{d\ln C(a_s(\mu'))}{da_s(\mu')}+\gamma(a_s(\mu'))\right)\right]\otimes C(a_s(M))\otimes f(M)\,.
\end{equation}
Finally, we can plug this expression into Eq.~(\ref{eq:solTMD3}) and
reabsorb the exponential due to the evolution of the PDF $f$ and the
coefficient $C$ into the Sudakov form factor. Since these evolutions
are single-logarithmic, their effect can be reabsorbed into the
function $B$ so that
\begin{equation}
  \label{eq:solTMD4}
  F(M,M^2) = \exp\left[\frac12\widetilde{S}(M,\mu_b)\right] \otimes C(a_s(M))\otimes f(M)\,,
\end{equation}
with
\begin{equation}
  \widetilde{S}(M,\mu_b) = - \int_{\mu_b^2}^{M^2}\frac{d\mu'^2}{\mu'^2}\left[A(a_s(\mu'))\ln\frac{M^2}{\mu'^2}+\widetilde{B} (a_s(\mu'))\right]\,,
\end{equation}
where
\begin{equation}
  \widetilde{B} (a_s(\mu))={B} (a_s(\mu))+2 a_s(\mu)\beta(a_s(\mu)) \frac{d\ln C(a_s(\mu))}{da_s(\mu)}+2 \gamma(a_s(\mu'))\,.
\end{equation}

Customarily, the coupling $a_s(M)$ used for the expansion of $C$ is
conveniently expressed in terms of the reference value $a_s(\mu_0)$,
assuming $\mu_0\simeq M$ and using the expanded evolution truncated to
the relevant order. Analogously, $f(M)$ is usually expressed in terms
of $f(\nu_0)$, where $\nu_0$ may be regarded as a scale at which PDFs
are fitted, by assuming $\nu_0\simeq M$ and using the corresponding
expanded evolution. Finally, this allows us to write
\begin{equation}
  C(a_s(M))\otimes f(M)=\widetilde{C}(a_s(\mu_0))\otimes f(\nu_0)\,,
\end{equation}
where the perturbative coefficients in the expansion of the function
$\widetilde{C} $
\begin{equation}
  \widetilde{C}(a_s(\mu_0))= \sum_{n=0}^{\infty}a_s^n(\mu_0) \widetilde{C}^{(n)}
\end{equation}
are related, up to N$^3$LL, to those of the function $C$ in
Eq.~(\ref{Ccentralexp}) as follows,
\begin{equation}
  \begin{array}{rcl}
    \widetilde{C}^{(0)}&=&\displaystyle \bar{C}^{(0,0,0)}\,,\\
    \\
    \widetilde{C}^{(1)}&=&\displaystyle \bar{C}^{(1,0,0)}+
                           \bar{C}^{(0,0,0)}\otimes\gamma_{0}\ln\frac{M}{\mu_0}\,,\\
    \\
    \widetilde{C}^{(2)}&=&\displaystyle
                           \bar{C}^{(2,0,0)}+\bar{C}^{(1,0,0)}\otimes\left(
                           \gamma_{0} + \beta_0\right)\ln \left(\frac{M}{\nu_0}\right)\\
    \\
                       &+&\displaystyle  \bar{C}^{(0,0,0)} \otimes\left[\frac{1}{2} \gamma ^{(0)} \otimes\left(\beta _0+\gamma ^{(0)}\right)\ln ^2\left(\frac{M}{\nu _0}\right)+\gamma ^{(1)} \ln \left(\frac{M}{\nu _0}\right)+\beta _0 \gamma ^{(0)} \ln \left(\frac{M}{\mu _0}\right) \ln \left(\frac{M}{\nu _0}\right)\right]\,.
  \end{array}
\end{equation}
Finally, the TMD distribution is evaluated as
\begin{equation}
  \label{eq:solTMD5}
  F(M,M^2) = \exp\left[\frac12\widetilde{S}(M,\mu_b)\right]\otimes
  \widetilde{C}(a_s(\mu_0))\otimes f(\nu_0)\,.
\end{equation}

This result, when used to compute physical cross sections, reproduces
the usual implementation of transverse-momentum resummation (see,
\textit{e.g.}, Refs.~\cite{Bozzi:2005wk, Camarda:2019zyx,
  Bizon:2018foh}). The advantage of this procedure is that it
formulates the resummation in a purely a\-na\-ly\-ti\-cal fashion,
thus limiting as much as possible the amount of possibly expensive
numerical computations. It also exposes all the relevant scales ($M$,
$\mu_0$, $\nu_0$, $\mu_b$) allowing for their variations as a proxy to
estimate theoretical uncertainties.

On the other hand, this procedure is also affected by limitations
which may obscure the meaning of some of the scales. In particular,
the scales $\mu_0$ and $\nu_0$ are taken to be of the order of the
physical hard scale $M$. Our analysis illustrates that $\mu_0$ and
$\nu_0$, which are usually denoted by $\mu_{\rm R}$ and $\mu_{\rm F}$
in transverse-momentum-resummation literature, correspond to the
reference scales at which strong coupling and PDFs are measured
(fitted). We thus observe that the traditional implementation of
transverse-momentum resummation implies that both the strong coupling
and the PDFs are measured at a scale of the order of the hard scale of
the process. To make a specific example, consider DY production with
an invariant mass of the lepton pair $M$ around the mass of the $Z$
boson $M_Z$. In this case, the formulation discussed above requires
$\mu_0\simeq M_Z$ and $a_s$ is indeed typically measured at this scale
with value $\alpha_s(M_Z)=0.118$. However, it is also necessary that
$\nu_0\simeq M_Z$, but this constraint is usually not fulfilled. As a
matter of fact, modern PDF fits are performed assuming
$\nu_0~\simeq 1$~GeV, which is far apart from $M_Z$. A possibility to
address this issue would be to extract PDFs by setting $\nu_0$ to a
scale of order $M_Z$.  Of course, for hard scales $M$ far away from
$M_Z$ the problem would arise again.

Conversely, a numerical formulation of transverse-momentum resummation
allows us to relax the constraints $\mu_0\simeq \nu_0 \simeq M$, while
still being able to provide a reliable estimate of perturbative
corrections by means of variations of the parameter $\xi$. Indeed, as
we discussed above, our analysis indicates that the scales $\mu_0$ and
$\nu_0$ are \textit{not} to be considered as arbitrary scales on the
same footing as the usual renormalisation and factorisation scales
that appear in fixed-order calculations. Specifically, in our
approach, there is little to be gained in varying them because, in
principle, this would require re-measuring $\alpha_s$ and PDFs at the
new scales. Rather, it is more sensible to keep $\mu_0$ and $\nu_0$
fixed and vary $\alpha_s$ and PDFs within their experimental
uncertainties.

\section{Physical observables}\label{sec:PhysicalObservables}

We devote this section to setting up the stage for our analysis of RGE
systematics in physical observables for precision physics at
high-energy colliders. We will consider the structure functions in
inclusive electron-proton ($ep$) deep-inelastic-scattering (DIS)
through a neutral vector boson with negative virtuality $Q$
(Sec.~\ref{subsec:DIS}), and the transverse-momentum ($q_T$) spectrum
of a lepton pair with invariant mass $M$ in inclusive Drell-Yan (DY)
production in proton-proton ($pp$) collisions
(Sec.~\ref{subsec:qTDY}). The former observable is characterised by
single-logarithmic enhancements, which are tamed by means of the
eveolution of $\alpha_s$ and PDFs discussed in
Secs.~\ref{sec:runningcoupling}
and~\ref{sec:coll}.\footnote{Additional logarithmic corrections,
  requiring further resummations beyond those in
  Secs.~\ref{sec:runningcoupling} and~\ref{sec:coll}, arise in DIS for
  Bjorken-$x_{\rm B}$ regions near $x_{\rm B} \to 0$ and
  $x_{\rm B} \to 1$. We do not address them here.} The latter
observable features two regimes: $q_T\simeq M$ and $q_T\ll M$. The
region $q_T\simeq M$, similarly to DIS, is characterised by
single-logarithmic enhancements related to $\alpha_s$ and PDFs that
are resummed accordingly. In contrast, the region $q_T\ll M$ is
affected by double logarithms that are resummed by means of the
Sudakov form factor treated in Sec.~\ref{sec:sud}, controlling the
evolution of TMD distributions. In the following, we will discuss how
to construct the respective observables and account for all sources of
perturbative uncertainties.

In our discussion, we assume that the strong coupling is known at a
reference scale $\mu_0$ ($\alpha_s(\mu_0)$) and the PDFs are known at
a reference scale $\nu_0$ ($f(\nu_0)$). These reference-scale
quantities come with associated uncertainties, which do not have a
perturbative origin. They can be incorporated in our calculations, as
we will do below, but they do not constitute the focal point of our
discussion. Rather, our focus is on perturbative RGE systematic
uncertainties, associated with the evolution of $\alpha_s$ and PDFs
from the reference scales, $\mu_0$ and $\nu_0$, to the scales $\mu$
and $\nu$ relevant to the physical observables. To this purpose, we
will use numerical solutions at the appropriate theoretical accuracy
with a value of the resummation-scale parameter $\xi$ of order
${\cal O} (1)$.\footnote{Having established in the previous sections
  the equivalence between analytic and numerical solutions for all
  relevant RGEs, in the applications that follow we will employ the
  numerical solutions throughout.}  Therefore, the evolved $\alpha_s$
and PDFs will also depend parametrically on $\xi$ (\textit{i.e.},
$\alpha_s\equiv \alpha_s(\mu;\xi)$ and $f\equiv f(x,\mu;\xi)$), whose
value can be varied to estimate the ensuing uncertainty.

Next, we address the question of defining the perturbative accuracy of
a given calculation. In this respect, we will use a fixed-order
counting (LO, NLO, NNLO, etc.) to label the relative accuracy in terms
of powers of $\alpha_s$ of partonic cross sections and anomalous
dimensions. We will instead use the logarithmic counting (LL, NLL,
NNLL, etc.) to label the towers of logarithms that are resummed by
means of RGEs. When combining fixed-order and resummed calculations to
compute a physical observable, the two accuracies are not
independent. Indeed, resummation usually comes with a fixed-order
(\textit{i.e.}, not logarithmically enhanced) component which also
contributes towards the logarithmic counting. Examples are the $g_0$
functions in the evolution of both PDFs (Eq.~(\ref{eq:g0PDF})) and
TMDs (Eq.~(\ref{eq:g0TMD})), as well as the matching functions in the
case of TMDs (Eq.~(\ref{eq:Cexpansion})). As a consequence, when
combined with a fixed-order calculation, the accuracy of resummation
is required to be such that it does not deteriorate the fixed-order
accuracy. We will substantiate this discussion below when discussing
the specific cases of DIS structure functions and DY $q_T$ spectrum.

\subsection{Inclusive deep-inelastic electron-proton
  scattering}\label{subsec:DIS}

One of the simplest examples of observable at high-energy colliders,
and yet very important for PDF determinations, is that of the
structure functions $\mathcal{F}=F_2,F_L,xF_3$ in (unpolarised)
inclusive deep-inelastic $ep$ scattering. These observables obey
\textit{collinear factorisation}\footnote{Appropriate generalisations
  of the factorisation are needed in DIS for regions near
  $x_{\rm B} \to 0$ and $x_{\rm B} \to 1$.}  and therefore they are
computed as Mellin convolutions, denoted by the symbol $\otimes$, over
the partonic longitudinal-momentum fraction $x$ between a set of
partonic cross sections $\hat{\mathcal{F}}$ and the PDFs
$f$:\footnote{In Eq.~(\ref{eq:DISconvolution}), we are implicitly
  neglecting the mass $m_h$ of heavy quarks (mostly charm and
  bottom). Corrections due to the finiteness of $m_h$ can be accounted
  for and they take the form of powers of $m_h/Q$ in the partonic
  cross sections. Large logarithms of the same ratio are resummed into
  the heavy-quark PDFs.}
\begin{equation}\label{eq:DISconvolution}
  \mathcal{F}(Q)= \hat{\mathcal{F}}\left(a_s(\mu_{\rm R};\xi),\frac{\mu_{\rm R}}{Q},\frac{\mu_{\rm F}}{Q}\right)\otimes
  f(\mu_{\rm F};\xi)\,,
\end{equation}
where $Q$ is the (negative) virtuality of the neutral vector boson
($Z/\gamma^*$) that mediates the process. To simplify the notation, we
have dropped from Eq.~(\ref{eq:DISconvolution}) all non-scale
dependences; these include the Bjorken variable $x_{\rm B}$, the
partonic longitudinal-momentum fraction $x$, and the flavour
indices. The partonic cross section $\hat{\mathcal{F}}$ at N$^p$LO
accuracy is computed in perturbation theory truncating its expansion
at order $p$ in $\alpha_s$:\footnote{When $\mathcal{F}=F_L$, the
  $\mathcal{O}(\alpha_s^0)$ contribution to $\hat{\mathcal{F}}$ is
  zero. Therefore, its perturbative expansion effectively starts from
  $n=1$ so that, strictly speaking, a truncation of the series in
  Eq.~(\ref{eq:CDISexpansion}) at order $p$ for $F_L$ should be
  labelled as N$^{p-1}$LO. However, in DIS measurements $F_L$ is
  usually combined with $F_2$ and $xF_3$ whose perturbative expansions
  instead do start from $n=0$. For this reason, we truncate the
  perturbative series of $F_L$ at the same absolute order $p$ as $F_2$
  and $xF_3$ and yet label its accuracy as N$^p$LO.}
\begin{equation}\label{eq:CDISexpansion}
  \hat{\mathcal{F}}\left(a_s(\mu_{\rm R};\xi),\frac{\mu_{\rm R}}{Q},\frac{\mu_{\rm F}}{Q}\right) =
  \sum_{n=0}^p a_s^n(\mu_{\rm
    R};\xi)\sum_{i=0}^{n}\sum_{j=0}^i \hat{\mathcal{F}}^{(n,i,j)}\ln^{i-j}\left(\frac{\mu_{\rm R}}{Q}\right) \ln^{j}\left(\frac{\mu_{\rm F}}{Q}\right)\,.
\end{equation}
Strong coupling $a_s$ and PDFs $f$ are evolved using the solution to
their respective RGEs. As noted above, the logarithmic accuracy of the
evolutions is bound by the fixed-order accuracy of the partonic cross
section $\hat{\mathcal{F}}$. Using Eq.~(\ref{eq:gensoldglap}), one has
that N$^k$LL accuracy is achieved by computing $g_0^{(\gamma)}$ at
$\mathcal{O}(\alpha_s^k)$. As a consequence, the minimal accuracy of
PDF (and $\alpha_s$) evolution that preserves the N$^k$LO accuracy of
the partonic cross section is N$^k$LL. Conventionally, computations of
$\mathcal{F}$ that combine N$^k$LO partonic cross sections with
N$^k$LL evolutions are simply labelled as N$^k$LO.

Finally, we note that the expansion in Eq.~(\ref{eq:CDISexpansion})
shows that renormalisation and factorisation scales, $\mu_{\rm R}$ and
$\mu_{\rm F}$, have to be of order ${\cal O} (Q)$ to avoid spoiling
the convergence of the perturbative series. Alongside, we already
argued that the resummation-scale parameter $\xi$ involved in the
evolution of $\alpha_s$ and PDFs is to be of order ${\cal O}
(1)$. Thus, an estimate of the perturbative uncertainties on
$\mathcal{F}$ can be achieved by varying $\mu_{\rm R}$ and
$\mu_{\rm F}$ around $Q$, and $\xi$ around 1, by a conventional factor
of two. We will present quantitative results for $F_2$, $F_L$, and
$xF_3$ at NNLO accuracy in Sec.~\ref{sec:f2andfL}.

\subsection{Transverse-momentum spectrum in Drell-Yan
  production}\label{subsec:qTDY}

The inclusive production of a lepton pair with large invariant mass
$M$ in $pp$ collisions with centre-of-mass energy $\sqrt{s}$ is
referred to as DY production. The transverse-momentum ($q_T$) spectrum
of the lepton pair is a phenomenologically interesting observable, in
which $q_T$ introduces an additional scale on top of $M$ that allows
one to define two possible regimes: $q_T \simeq M$ and $q_T\ll M$. In
the following sections, we will discuss theoretical predictions in
both regimes, finally showing how to merge them.

\subsubsection{Large-$q_T$ regime}

For $q_T \simeq M$, we are in a single-scale regime in which the
$q_T$-differential cross section obeys collinear factorisation, and
therefore shares many features with the DIS structure functions
discussed in Sec.~\ref{subsec:DIS}.  Schematically, it factorises as
\begin{equation}\label{eq:DYconvolution}
  \frac{d\sigma}{dq_T}(q_T\simeq M,M,s)=
  \frac{d\hat{\sigma}}{dq_T}\left(\frac{q_T}{M},\frac{M}{s},a_s(\mu_{\rm R};\xi),\frac{\mu_{\rm
        R}}{M},\frac{\mu_{\rm F}}{M}\right)\otimes f_1(\mu_{\rm
    F};\xi)\otimes f_2(\mu_{\rm F};\xi)\,,
\end{equation}
where we have omitted non-scale variables and flavour indices, and, as
usual, used the shorthand symbol $\otimes$ to indicate Mellin
convolutions over the partonic longitudinal-momentum fractions. For
$q_T\neq 0$, the partonic cross section at N$^p$LO accuracy reads
\begin{equation}\label{eq:DYqTexpansion}
  \begin{array}{c}
    \displaystyle \frac{d\hat{\sigma}}{dq_T}\left(\frac{q_T}{M},\frac{M}{s},a_s(\mu_{\rm R};\xi),\frac{\mu_{\rm
    R}}{M},\frac{\mu_{\rm F}}{M}\right) =\\
    \\
    \displaystyle  \sum_{n=0}^p a_s^{n+1}(\mu_{\rm
    R};\xi)\sum_{i=0}^{n}\sum_{j=0}^iD^{(n,i,j)}\left(\frac{q_T}{M},\frac{M}{s}\right)\ln^{i-j}\left(\frac{\mu_{\rm R}}{M}\right) \ln^{j}\left(\frac{\mu_{\rm F}}{M}\right)\,.
  \end{array}
\end{equation}
As in the case of DIS, N$^k$LO accuracy at the level of the observable
in Eq.~(\ref{eq:DYconvolution}) is obtained by combining N$^k$LO
partonic cross sections with N$^k$LL evolution for $\alpha_s$ and
PDFs. However, we are also allowed to use N$^{k+1}$LL evolutions with
N$^k$LO partonic cross sections, which we will do below. The reason
for this stems from the way large- and small-$q_T$ regions are
combined, see Sec.~\ref{eq:matchinsetup}. Similarly to the case of DIS
structure functions, perturbative uncertainties are estimated by
varying $\mu_{\rm R}$ and $\mu_{\rm F}$ around the hard scale $M$, and
$\xi$ around one by a conventional factor of two. Numerical results
for the large-$q_T$ spectrum in DY production accurate at LO and NLO
will be presented in Sec.~\ref{sec:drellyan}.

\subsubsection{Small-$q_T$ regime}

For $q_T\ll M$, the $q_T$-differential DY cross section obeys
\textit{TMD factorisation}, \textit{i.e.}, it can be written as a
convolution over the partonic transverse momentum $k_T$ of two quark
TMDs. This convolution is disentangled under Fourier transformation
with respect to $k_T$, which introduces the conjugate variable $b$ and
brings the cross section to the form
\begin{equation}
  \frac{d\sigma}{dq_T}(q_T\ll M,M,s)=\sigma_0H\left(a_s(\mu;\xi),\frac{\mu}{M}\right)\int_0^{\infty} db\,\frac{b}{2}J_0(bq_T)F_1(b,\mu,\zeta_1) F_2(b,\mu,\zeta_2)\,.
\end{equation}
As above, in order to simplify the notation, non-scale variables and
flavour indices are omitted. In the expression above, $\sigma_0$ is
the Born cross section and $J_0$ is the Bessel function of the first
kind. The function $H$ is related to the form factor for the
transition $Z/\gamma^*\rightarrow q\overline{q}$ and its perturbative
expansion reads
\begin{equation}
  H\left(a_s(\mu;\xi) ,\frac{\mu}{M}\right)= 1+\sum_{n=1}^pa_s^n(\mu;\xi) \sum_{i=0}^{2n}H^{(n,i)}\ln^i\left(\frac{\mu}{M}\right)\,,
\end{equation}
where the truncation power $p$ is determined by the target logarithmic
accuracy (see below). As discussed in Sec.~\ref{sec:tmd}, we can set
$\zeta_1=\zeta_2=M^2$ without loss of generality. Moreover, based on
the correspondence between variations in $\kappa$ and $\xi$
illustrated in Sec.~\ref{sudhyst}, we can set $\mu=M$, as long as
$\xi$ is allowed to vary.  We thus rewrite the cross section as
\begin{equation}
  \frac{d\sigma}{dq_T}(q_T\ll M,M,s)=\sigma_0H\left(a_s(M;\xi),1\right)\int_0^{\infty}
  db\,\frac{b}{2}J_0(bq_T)F_1(b,M,M^2) F_2(b,M,M^2)\,.
  \label{eq:DYconvolutionsmallqT}
\end{equation}
RGE systematic uncertainties on the cross section
(\ref{eq:DYconvolutionsmallqT}) due to the evolution of $\alpha_s$,
PDFs, and TMDs can be simultaneously estimated through variations of
the resummation-scale parameter $\xi$.

The theoretical accuracy of Eq.~(\ref{eq:DYconvolutionsmallqT}) is
determined by the accuracy of TMDs and hard factor $H$. In turn, the
accuracy of TMDs is determined by Sudakov form factor $S$ and the
matching functions $C$, as well as by the evolution of $\alpha_s$ and
PDFs.  Numerical results for the small-$q_T$ spectrum in DY production
at NNLL and N$^3$LL accuracy will be presented in
Sec.~\ref{sec:drellyan}. For definiteness, predictions at NNLL
(N$^3$LL) are obtained combining Sudakov form factor at NNLL
(N$^3$LL), hard factor $H$ and matching functions $C$ at
$\mathcal{O}(\alpha_s)$ ($\mathcal{O}(\alpha_s^2)$), PDFs at NLL
(NNLL), and $\alpha_s$ at NNLL (N$^3$LL).

\subsubsection{Matching small-$q_T$ and large-$q_T$
  regions}\label{eq:matchinsetup}

The formulas in Eqs.~(\ref{eq:DYconvolution})
and~(\ref{eq:DYconvolutionsmallqT}), which are valid, respectively, in
the regions $q_T\simeq M$ and $q_T\ll M$, can be consistently combined
to produce predictions accurate over a broader range in $q_T$. This
procedure is referred to as \textit{matching}. Schematically, the
\textit{additive} form of matching reads as follows,
\begin{equation}
  \frac{d\sigma}{dq_T}(q_T,M,s) = 
  \frac{d\sigma}{dq_T}(q_T\simeq
  M,M,s)+\frac{d\sigma}{dq_T}(q_T\ll M,M,s)-\left.\frac{d\sigma}{dq_T}\right|_{\rm d.c.}(q_T,M,s)\,,
  \label{eq:matchingqT}
\end{equation}
where the last term in the r.h.s.~is the counter-term (labelled by the
subscript ``d.c.'', which stands for double-counting) needed to
subtract the contribution present in both small- and large-$q_T$
terms. Expressions for the counter-term up to
$\mathcal{O}(\alpha_s^2)$ can be found, \textit{e.g.}, in
Refs.~\cite{Bozzi:2005wk, Camarda:2021ict, Bizon:2018foh}.  We stress
that also the counter-term is affected by theoretical systematic
uncertainties, which can be estimated in a manner analogous to the
fixed-order cross section in Eq.~(\ref{eq:DYconvolution}),
\textit{i.e.}, by varying $\mu_{\rm R}$, $\mu_{\rm F}$, and $\xi$.

The implementation of Eq.~(\ref{eq:matchingqT}) may be accompanied by
a damping function $f$ suppressing the difference between the
small-$q_T$ term and the counter-term when $q_T\simeq M$. The reason
for this is that this difference, although subleading in $\alpha_s$,
may have a numerically non-negligible effect.
Eq.~(\ref{eq:matchingqT}) is then modified to
\begin{equation}
  \begin{array}{rcl}
    \displaystyle \frac{d\sigma}{dq_T}(q_T,M,s) &=& \displaystyle  
                                                    \frac{d\sigma}{dq_T}(q_T\simeq
                                                    M,M,s)\\
    \\
                                                &+&\displaystyle  f(q_T,M)\left[\frac{d\sigma}{dq_T}(q_T\ll
                                                    M,M,s)-\left.\frac{d\sigma}{dq_T}\right|_{\rm
                                                    d.c.}(q_T,M,s)\right]\,,
  \end{array}
  \label{eq:matchingqTdamp}
\end{equation}
where $f$ has to be such that
\begin{equation}
  f(q_T,M)\mathop{\longrightarrow}_{q_T\ll M} 1\quad\mbox{and}\quad f(q_T,M)\mathop{\longrightarrow}_{q_T\simeq M} 0\,. 
\end{equation}
For the numerical applications presented in Sec.~\ref{sec:drellyan},
we take the functional form for $f$ given in
Ref.~\cite{Catani:2015vma}, which reads
\begin{equation}
  f(q_T,M)=  \left\{
    \begin{array}{ll}
      1 & \quad q_T<k M , \\
      \\
      \displaystyle \exp\left[- \frac{(k M
      - q_T)^2}{\delta^2 M^2}\right]&\quad q_T>k M , 
    \end{array}
  \right.
\end{equation}
with $k=0.5$ and $\delta=0.25$. This damping function ensures that,
when $q_T\simeq M$, the second line of Eq.~(\ref{eq:matchingqTdamp})
is exponentially suppressed leaving only the first line, as
appropriate. The matching formula (\ref{eq:matchingqTdamp}) also
implies that a cancellation between large-$q_T$ term and counter-term
is to take place at $q_T\ll M$, so as to leave only the small-$q_T$
piece. From a numerical point of view, this is challenging because
both the large-$q_T$ term and the counter-term are logarithmically
divergent for $q_T\rightarrow 0$. Therefore, it is of crucial
importance that their cancellation is numerically very precise in
order not to spoil the small-$q_T$ result. We will study this aspect
quantitatively in Sec.~\ref{sec:drellyan}.

When matching the small-$q_T$ and large-$q_T$ regimes, the accuracies
of the single components in the r.h.s. of Eq.~(\ref{eq:matchingqT})
(or Eq.~(\ref{eq:matchingqTdamp})) are not independent. The
large-$q_T$ and double-counting components have to be computed at the
same fixed-order accuracy. This imposes a constraint on the minimal
logarithmic accuracy of the small-$q_T$ calculation. Specifically, at
N$^k$LO accuracy for the large-$q_T$ and double-counting components,
the small-$q_T$ needs to include fixed-order components, encoded in
the hard function $H$ and matching functions $C$, to order no lower
than $\mathcal{O}(\alpha_s^{k+1})$. In a standard counting, this
corresponds to a resummation accuracy of N$^{k+2}$LL. However, one can
also use the so-called ``primed'' convention by which the N$^{k+1}$LL
accurate Sudakov form factor $\exp(S)$ is associated with
$\mathcal{O}(\alpha_s^{k+1})$ $H$ and $C$, and $\alpha_s$ and PDF
evolution at N$^{k+1}$LL. This produces N$^{k+1}$LL$^\prime$
predictions at small $q_T$ that, like N$^{k+2}$LL, can be matched with
N$^{k}$LO large-$q_T$ predictions. When considering matched
predictions in Sec.~\ref{sec:drellyan}, we will consider the following
combinations of small- and large-$q_T$ calculations: NNLL+LO,
NNLL$^\prime$+NLO, and N$^3$LL+NLO.

\section{Numerical applications}
\label{sec:numapplications}

In Sec.~\ref{sec:PhysicalObservables}, we have thoroughly described
the observables that we will consider and the theoretical setup for
their computation, including the definition of the accuracy and the
strategy to estimate theoretical uncertainties. In this section, we
will present the ensuing numerical results. However, before moving to
discussing them, we first specify the settings that are common to all
computations.

Throughout this section, we will use
$\alpha_s(M_Z)=0.118$~\cite{ParticleDataGroup:2020ssz}, with
$M_Z=91.1876$~GeV, as an input to the evolution of the strong
coupling, and the MSHT20 PDF set~\cite{Bailey:2020ooq} at
$\nu_0 = 2$~GeV as an input to the evolution of PDFs. In order to
focus the discussion on the estimate of theoretical uncertainties, we
will always use the NNLO set from the MSHT20 family as an
initial-scale set of PDFs, regardless of the accuracy of the
calculation.

All evolutions are performed in the variable-flavour-number scheme
(VFNS), in which the number of active flavours $n_f$ that enters the
perturbative coefficients of the anomalous dimensions changes by one
unit when crossing a heavy-quark threshold. In the case of $\alpha_s$
and PDFs, also appropriate matching conditions at threshold are
applied. The values of charm and bottom thresholds are taken from the
MSHT20 PDF set and are $m_c=1.4$~GeV and $m_b=4.75$~GeV. No top-quark
threshold is considered, so that the maximum value attainable for
$n_f$ is 5.

All numerical calculations presented below use the code {\tt
  APFEL++}~\cite{Bertone:2013vaa, Bertone:2017gds}. A private
implementation of the calculation of Ref.~\cite{Gonsalves:1989ar}
extracted from the {\tt DYTURBO} code~\cite{Camarda:2019zyx} is used
for the computation of the large-$q_T$ DY cross section. The {\tt
  LHAPDF} interface~\cite{Buckley:2014ana} is used to access the
MSTH20 PDF set at $\nu_0=2$~GeV.

We next discuss the single observables.

\subsection{Deep-inelastic structure functions}
\label{sec:f2andfL}

In this subsection, we present quantitative results for the DIS
neutral-current structure functions $F_2$, $F_L$ and $xF_3$ at NNLO,
and estimate the size of theoretical uncertainties by means of
variations of the resummation-scale parameter $\xi$, the factorisation
scale $\mu_{\rm F}$, and the renormalisation scale
$\mu_{\rm R}$.\footnote{In this subsection, as well as in the
  following one, each band associated with $\xi$, $\mu_{\rm F}$, or
  $\mu_{\rm R}$ variations is obtained by keeping the other two scales
  fixed at their central values. Moreover, the bands are obtained as
  envelopes of a scan between the quoted bounds, and \textit{not} as a
  differences between predictions with central scales and the bounds
  of the variation ranges. This prevents possible non-monotonicities
  of scale variations from underestimating the corresponding
  uncertainties.} A thorough description of the theory settings can be
found in Sec.~\ref{subsec:DIS}. We also note that a benchmark of the
numerical implementation of the DIS structure functions used here
against the code {\tt HOPPET}~\cite{Salam:2008qg} has been recently
presented in Ref.~\cite{Bertone:2024dpm}

\begin{figure*}[h]
  \begin{center}
    \includegraphics[width=0.49\textwidth]{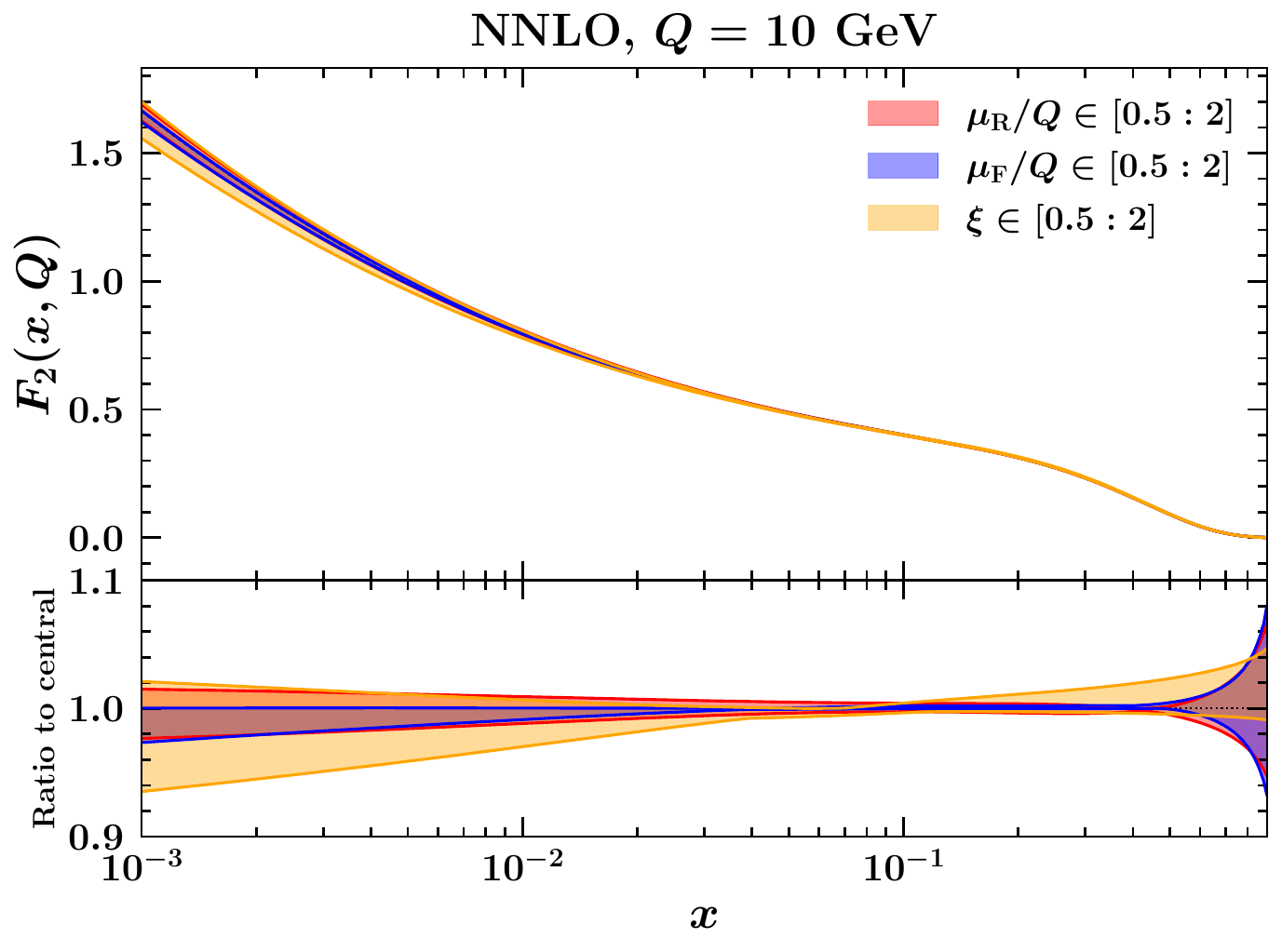}
    \includegraphics[width=0.49\textwidth]{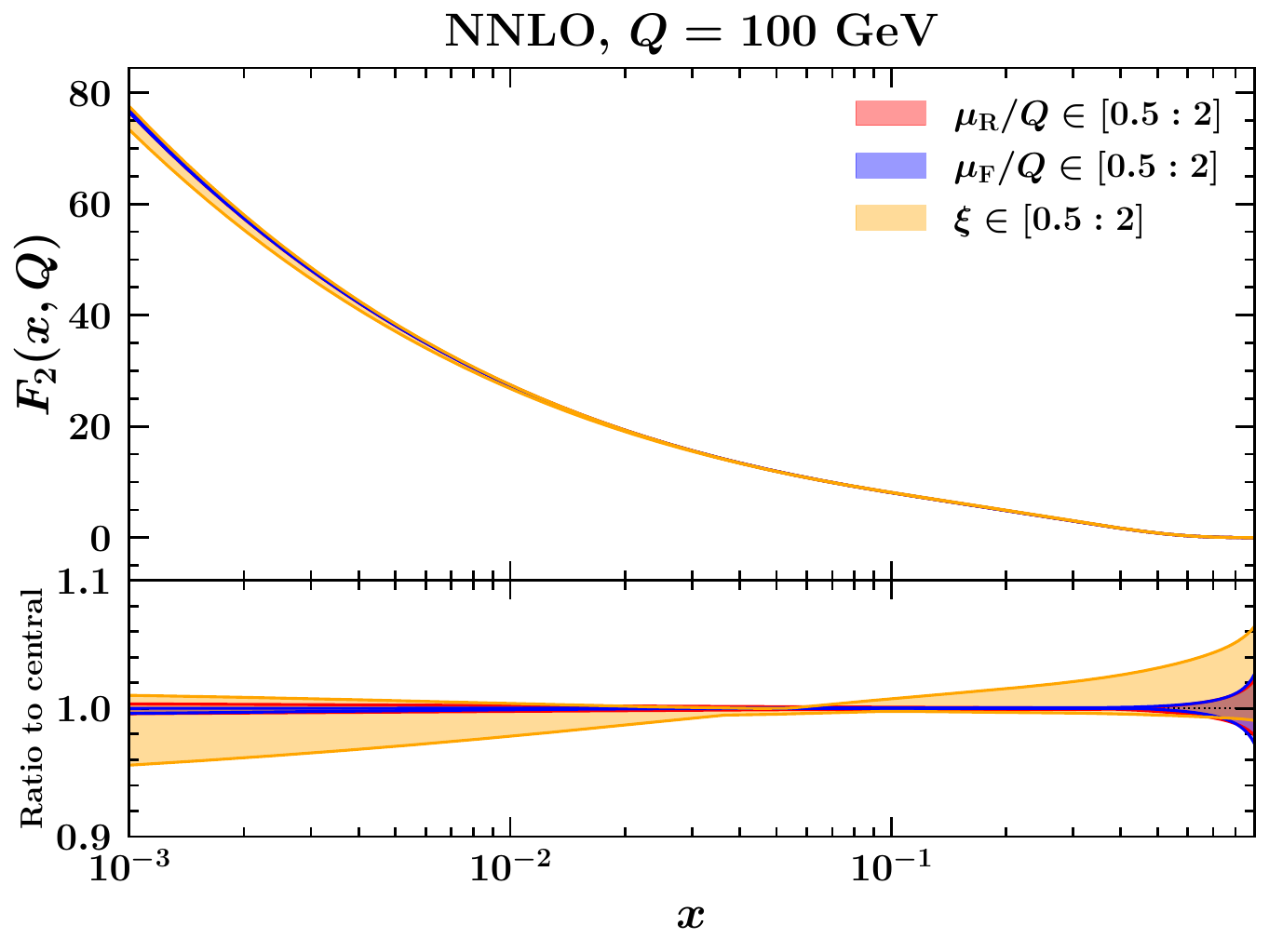}
    \caption{The DIS $F_2$ structure function at NNLO plotted versus
      $x$ for two different values of $Q$. Uncertainty bands
      associated with variations of renormalisation and factorisation
      scales, $\mu_{\rm R}$ and $ \mu_{\rm F}$, and resummation-scale
      parameter, $\xi$, are also displayed. The lower insets show the
      predictions normalised to the central-scale curves.}
    \label{fig:f2_muFmuRresscale}
  \end{center}
\end{figure*}
\begin{figure*}[h]
  \begin{center}
    \includegraphics[width=0.49\textwidth]{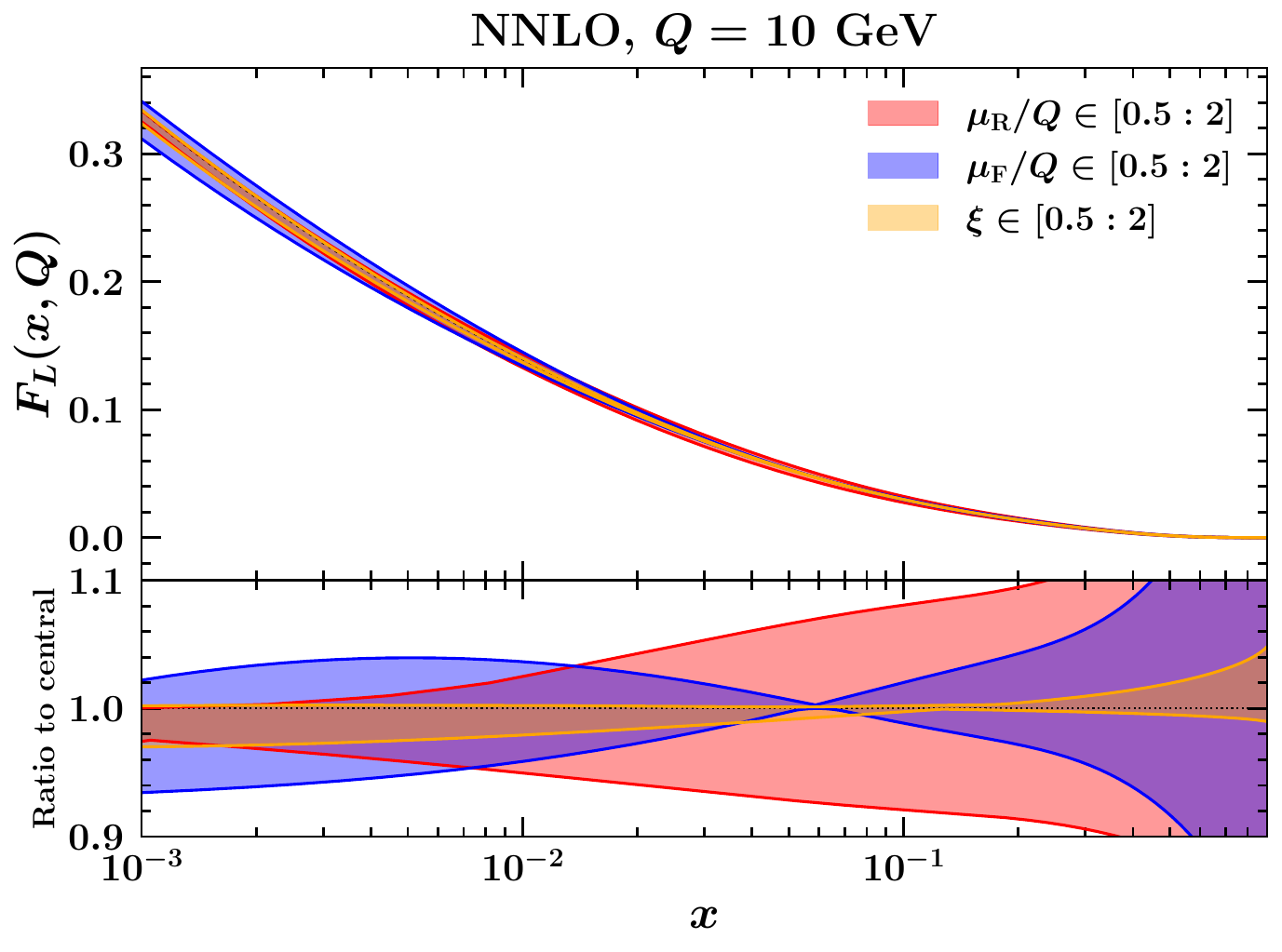}
    \includegraphics[width=0.49\textwidth]{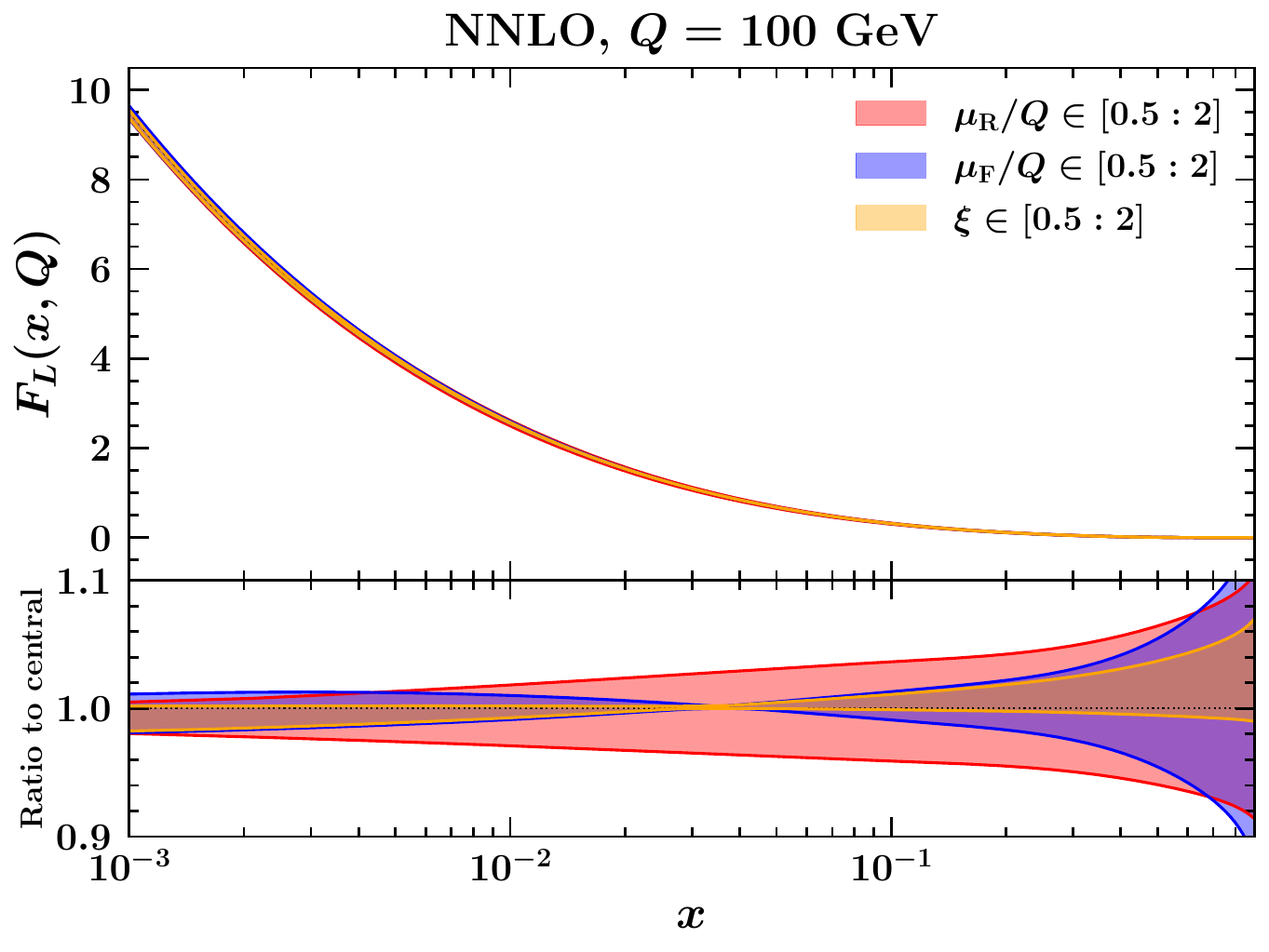}
    \caption{Same as Fig.~\ref{fig:f2_muFmuRresscale} for $F_L$.}
    \label{fig:fL_muFmuRresscale}
  \end{center}
\end{figure*}
\begin{figure*}[h]
  \begin{center}
    \includegraphics[width=0.49\textwidth]{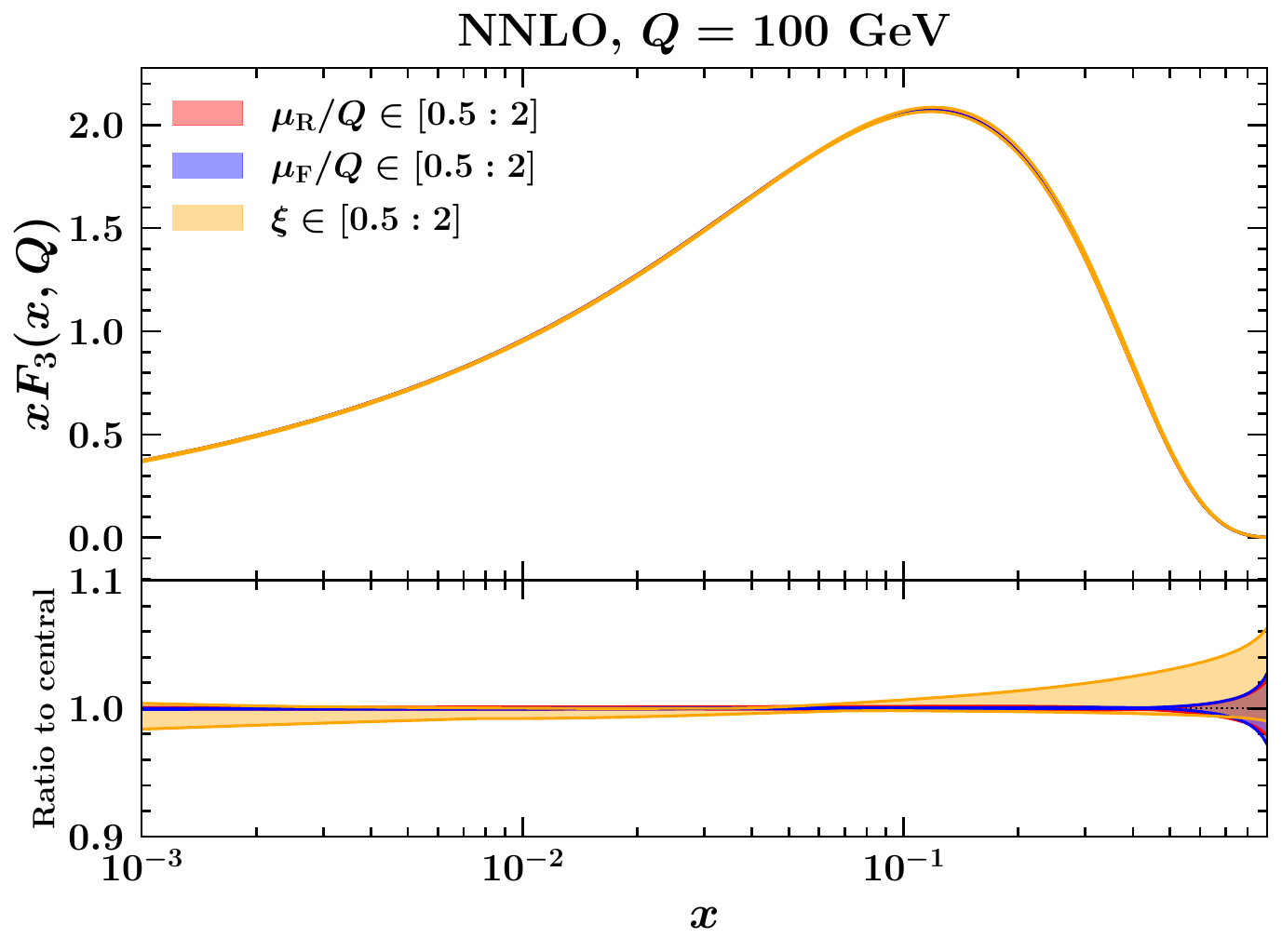}
    \includegraphics[width=0.49\textwidth]{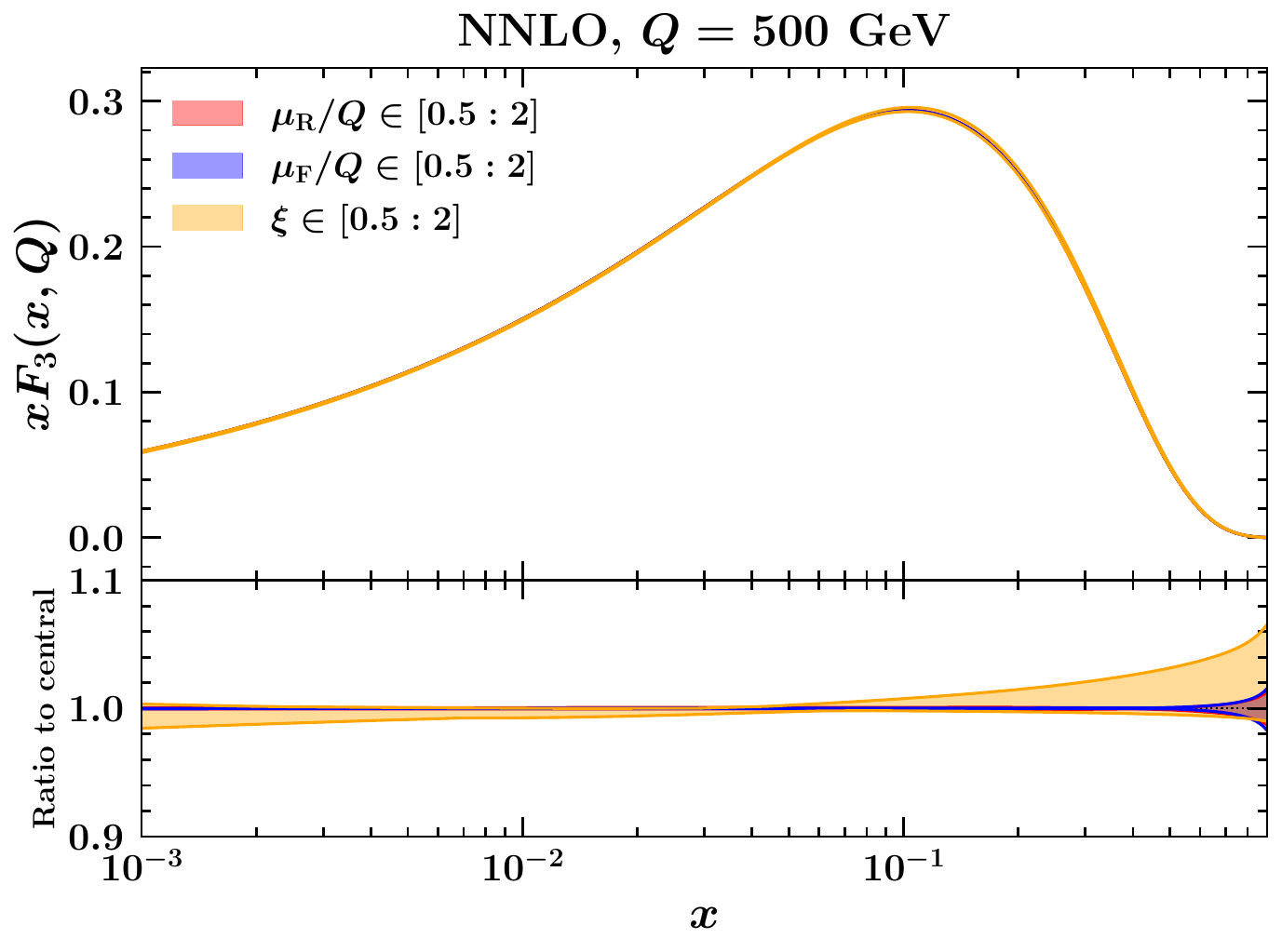}
    \caption{Same as Fig.~\ref{fig:f2_muFmuRresscale} for $xF_3$.}
    \label{fig:f3_muFmuRxi1}
  \end{center}
\end{figure*}

In Figs.~\ref{fig:f2_muFmuRresscale}, \ref{fig:fL_muFmuRresscale},
and~\ref{fig:f3_muFmuRxi1}, we show the $x$-dependence of $F_2$,
$F_L$, and $xF_3$, respectively, over a wide interval ranging from
$x=10^{-3}$ up to $x=0.9$, and for two different values of the
virtuality $Q$ of the exchanged vector boson. For $F_2$ and $F_L$ in
Figs.~\ref{fig:f2_muFmuRresscale} and~\ref{fig:fL_muFmuRresscale}, we
chose $Q=10,100$~GeV, \textit{i.e.}, one scale significantly below
$M_Z$ characterised by mostly photon exchange, and one scale above
$M_Z$ where the $Z$ contribution is dominant. For $xF_3$ in
Fig.~\ref{fig:f3_muFmuRxi1}, instead, we chose the scales
$Q=100,500$~GeV because, at scales below $M_Z$, this structure
function is suppressed by the $Z$ propagator. In these figures, the
upper panel of each plot shows the value of the structure function
along with its theoretical uncertainty bands, while the lower panel
shows its ratio to the central-scale curve obtained with
$\mu_{\rm R}=\mu_{\rm F}=Q$ and $\xi=1$.

We observe that the $\xi$ bands are sizeable across the board and are
generally comparable to $\mu_{\rm R}$ and $\mu_{\rm F}$ bands,
confirming that RGE-related uncertainties are non-negligible and
should therefore be taken into account. As a consequence of the fact
that the $\mathcal{O}(\alpha_s^0)$ contribution to $F_L$ is
identically zero, $\xi$ variations for this structure function are
relatively smaller than $\mu_{\rm R}$ and $\mu_{\rm F}$
variations. The picture is different for $F_2$ and $xF_3$, which
instead have a non-vanishing $\mathcal{O}(\alpha_s^0)$ contribution
and where the $\xi$ bands are generally larger than those associated
with $\mu_{\rm R}$ and $\mu_{\rm F}$. We also observe that $\xi$
variations remain approximately of the same relative size for both
values of $Q$ considered, while $\mu_{\rm R}$ and $\mu_{\rm F}$
variations tend to shrink as the scale $Q$ increases. This is an
expected behaviour. Indeed, for an N$^k$LO calculation, the size of
$\mu_{\rm R}$ and $\mu_{\rm F}$ variations is roughly proportional to
$\alpha_s^{k+1}(Q)\ln^{k+1}(\mu_{\rm R,F}/Q)$. Therefore, their size
decreases proportionally to a power of $\alpha_s(Q)$, becoming smaller
and smaller as $Q$ grows. Variations of the parameter $\xi$ have a
completely different behaviour that we discussed in
Sec.~\ref{sec:coll} (see Eq.~(\ref{eq:Deltafxi})). The main point here
is that these variations generate terms proportional to
$(\alpha_s^{k+1}(\xi Q)-\alpha_s^{k+1}(\xi \nu_0))\ln^{k+1}\xi$, with
$\nu_0=2$~GeV. The consequence is that, for $Q\gg \nu_0$ and
$\xi \sim 1$, the uncertainty is dominated by
$\alpha_s^{k+1}(\xi \nu_0)$ no matter how large $Q$ is.

\begin{figure*}[h]
  \begin{center}
    \includegraphics[width=0.6\textwidth]{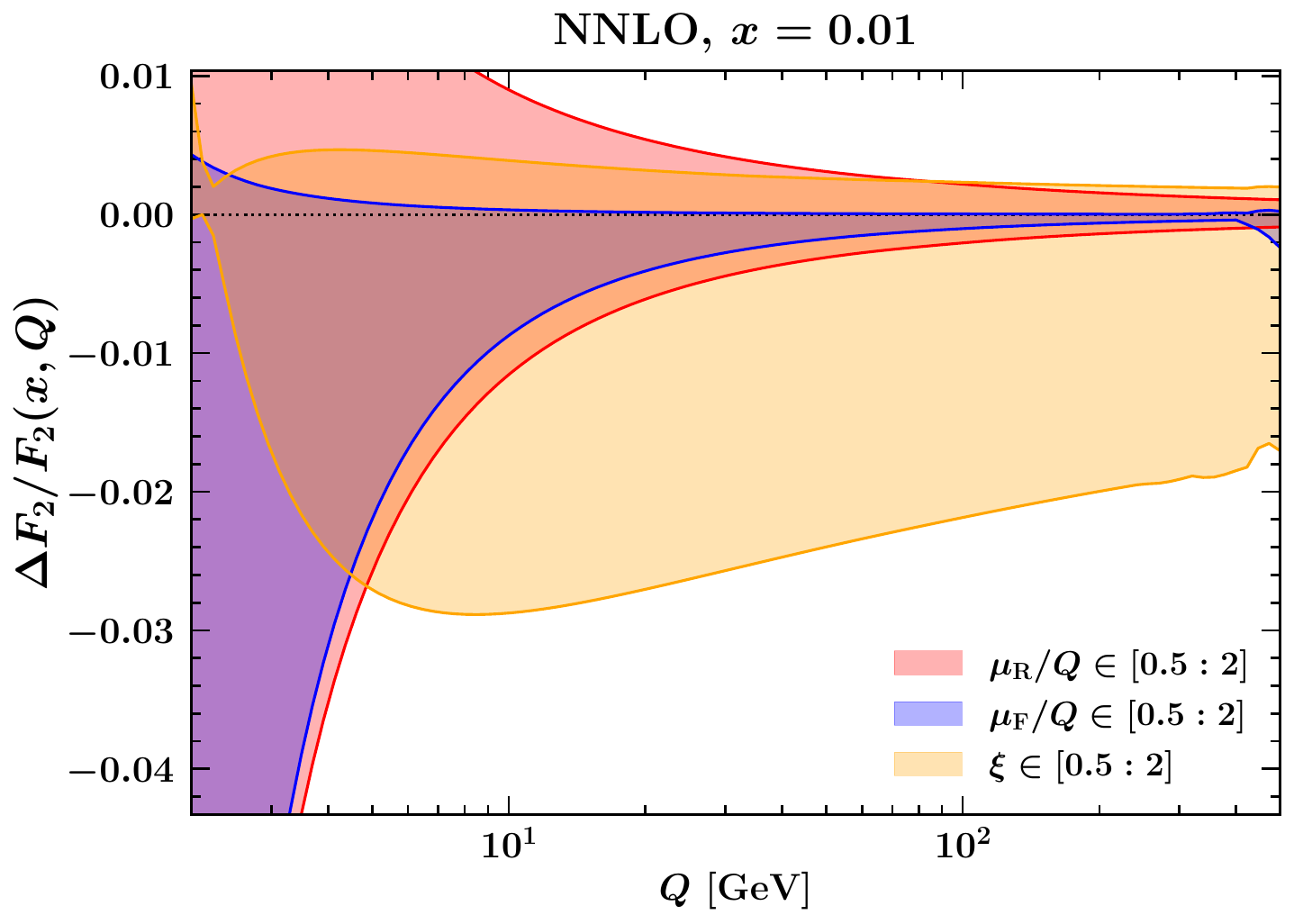}
    \caption{$Q$-dependence of the ratio $\Delta F_2 / F_2$ at NNLO at
      $x=10^{-2}$ associated with variations of renormalisation and
      factorisation scales, $\mu_{\rm R}$ and $ \mu_{\rm F}$, and
      resummation-scale parameter, $\xi$.}
    \label{fig:f2ofQ_muFmuRresscale}
  \end{center}
\end{figure*}

In order to study the $Q$ behaviour of theoretical uncertainties more
in detail, in Fig.~\ref{fig:f2ofQ_muFmuRresscale} we show
$\Delta F_2 / F_2$ as a functions $Q$ in a range between $Q=2$~GeV and
$Q=500$~GeV at $x=10^{-2}$, where $F_2$ is the structure function
computed with central scales and $\Delta F_2$ is the deviation caused
by each variation. As expected, uncertainties generated by variations
of $\mu_{\rm R}$ and $\mu_{\rm R}$ are largest at small scales and
decrease as $Q$ increases. On the contrary, the uncertainty related to
$\xi$ variations is small at low values of $Q$, \textit{i.e.}, close
to the PDF initial scale, and tends to grow at large $Q$, remaining
roughly of the same relative size for $Q\gtrsim 10$~GeV. A similar
behaviour is observed for other kinematic configurations and structure
functions.

\subsection{Drell-Yan transverse momentum distribution}
\label{sec:drellyan}

We now turn to the numerical implementation of our framework in the
case of the DY $q_{T}$ spectrum. We will first address the large-$q_T$
region, then consider the small-$q_T$ region, and finally present
results for the matched calculation. As above, we will put particular
emphasis on the estimate of theoretical uncertainties.

\begin{figure*}[t!]
  \begin{center}
    \includegraphics[width=0.49\textwidth]{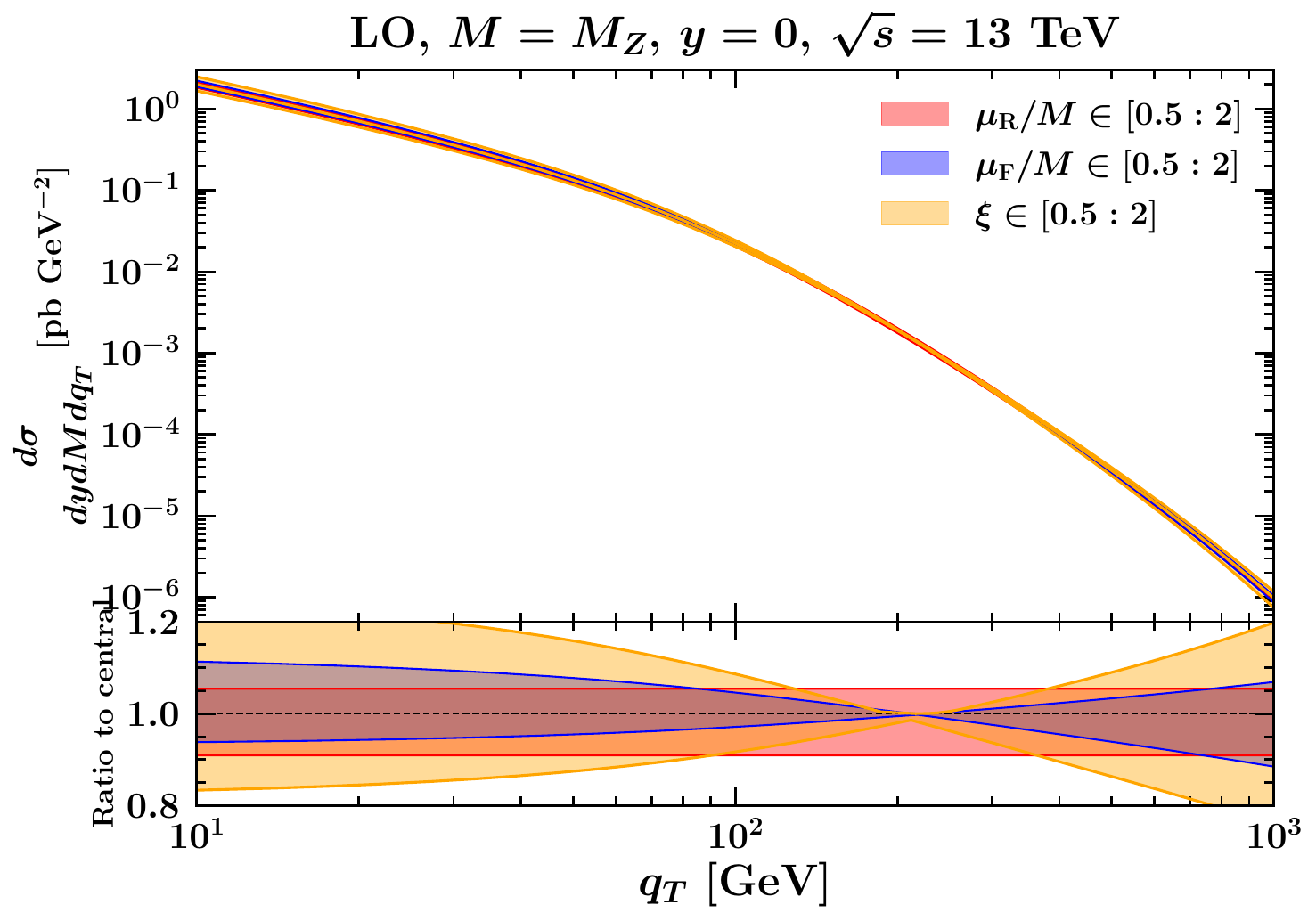}
    \includegraphics[width=0.49\textwidth]{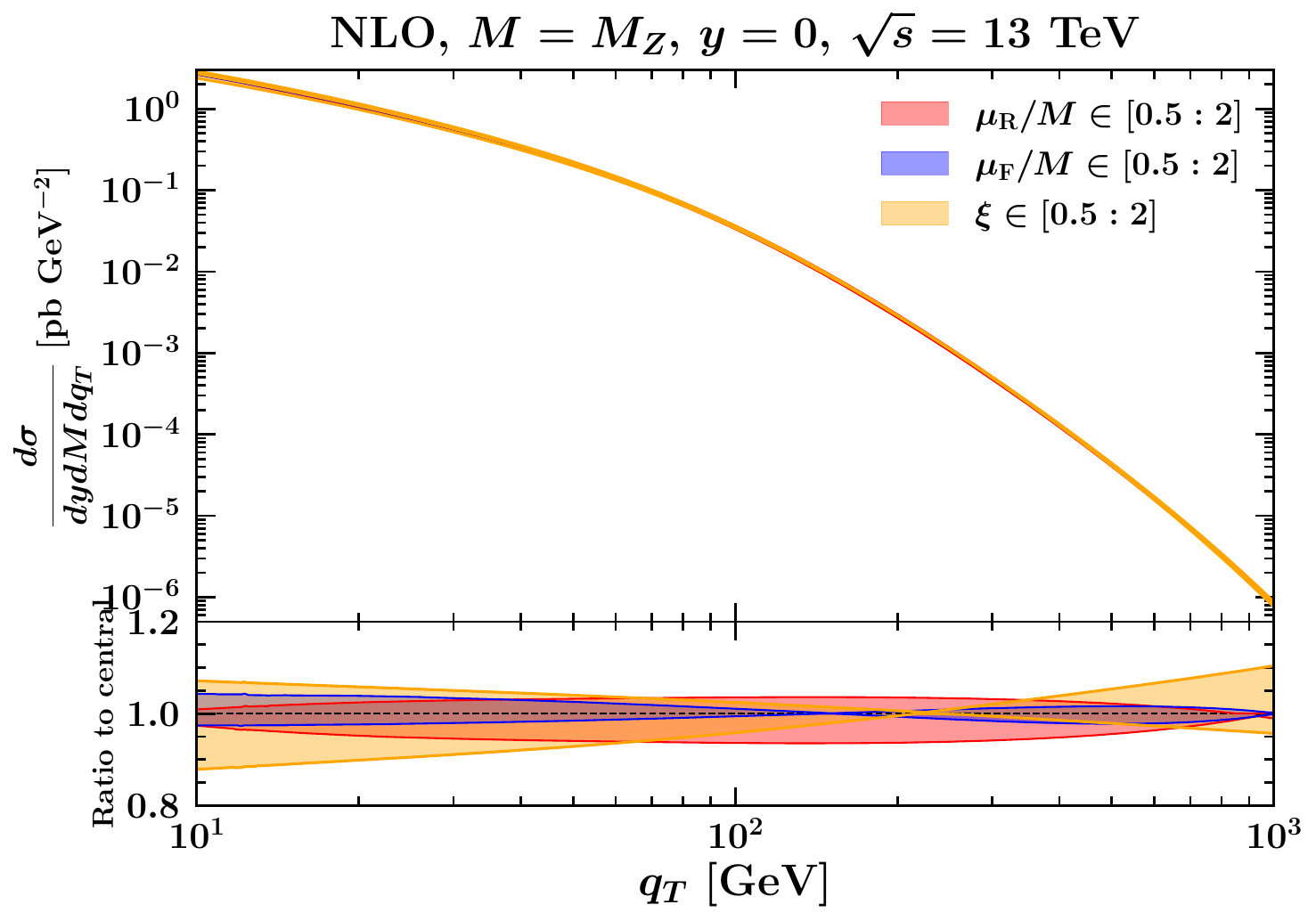}
    \caption{DY large-$q_T$ spectrum of a lepton pair with invariant
      mass $M$ equal to $M_Z$ and central rapidity ($y=0$) at LO
      (left) and NLO (right) accuracies at the 13~TeV LHC. The curves
      include uncertainty bands associated with variations of
      resummation scale-parameter $\xi$, renormalisation scale
      $\mu_{\rm R}$, and factorisation scale $\mu_{\rm F}$. The upper
      insets display the cross sections, while the lower insets show
      their ratios to central-scale predictions.}
    \label{fig:DYpT_muFmuRxi_fixed}
  \end{center}
\end{figure*}

Fig.~\ref{fig:DYpT_muFmuRxi_fixed} shows predictions for the $q_T$
spectrum of a lepton pair with invariant mass $M=M_Z$ and central
rapidity, $y=0$, in DY production in $pp$ collisions at
$\sqrt{s}=13$~TeV in a region ranging between $q_T=10$~GeV and
$q_T=1000$~GeV. Predictions are computed at LO (left panel) and NLO
(right panel) and include uncertainty bands relative to variations of
renormalisation scale $\mu_{\rm R}$, factorisation scale
$\mu_{\rm F}$, and resummation-scale parameter $\xi$. The upper insets
display the cross sections, while the lower insets show their ratios
to central-scale predictions. The expected reduction of all of the
band widths when moving from LO to NLO is observed. Analogously to DIS
structure functions, we also note that the $\xi$ band is generally of
the same size of, or even larger than, the $\mu_{\rm R}$ and $\mu_{\rm F}$
bands. This provides a further example of the fact that RGE-related
uncertainties can be sizeable and thus need to be taken into proper
account.

\begin{figure*}[t!]
  \begin{center}
    \includegraphics[width=0.49\textwidth]{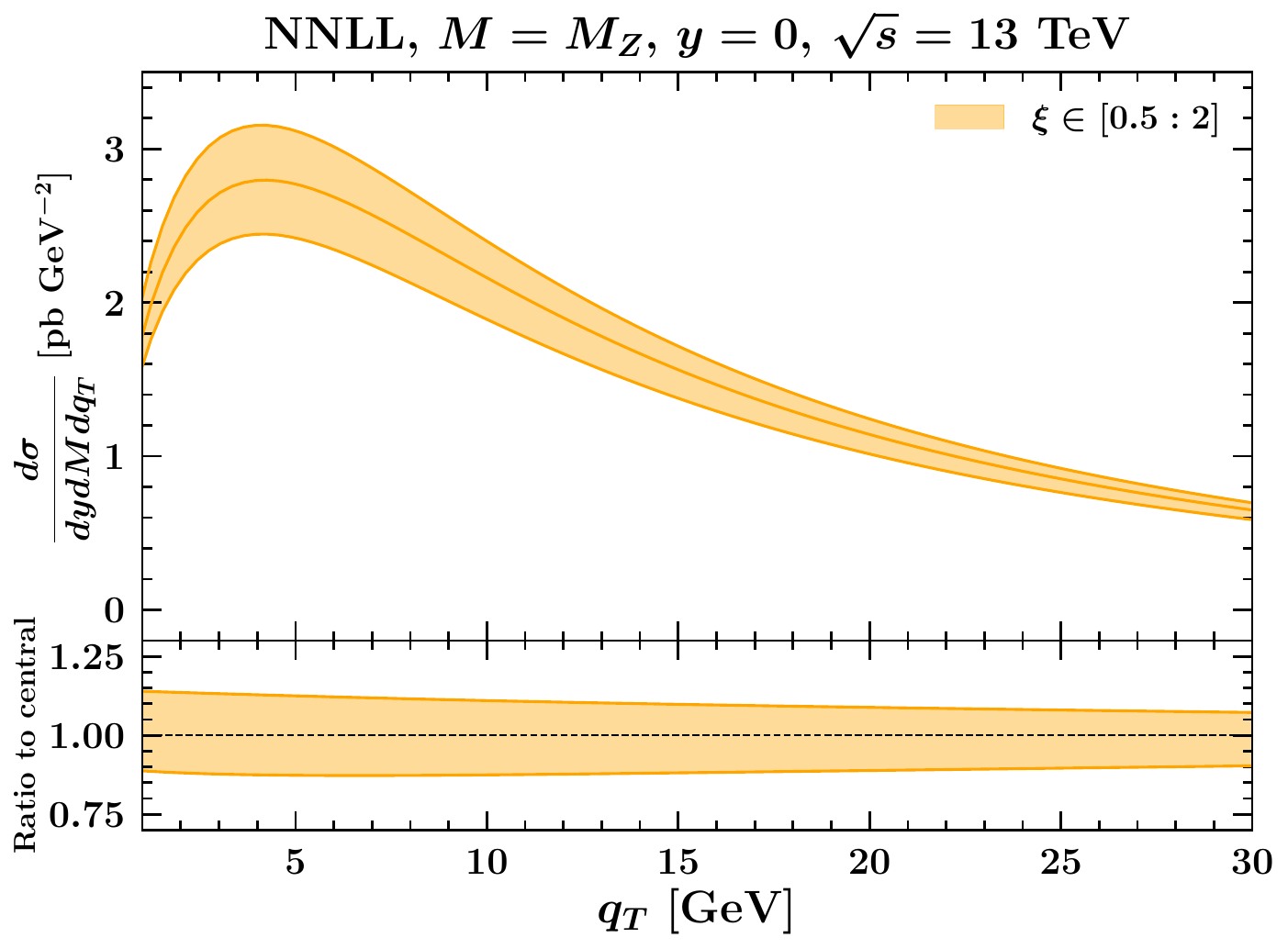}
    \includegraphics[width=0.49\textwidth]{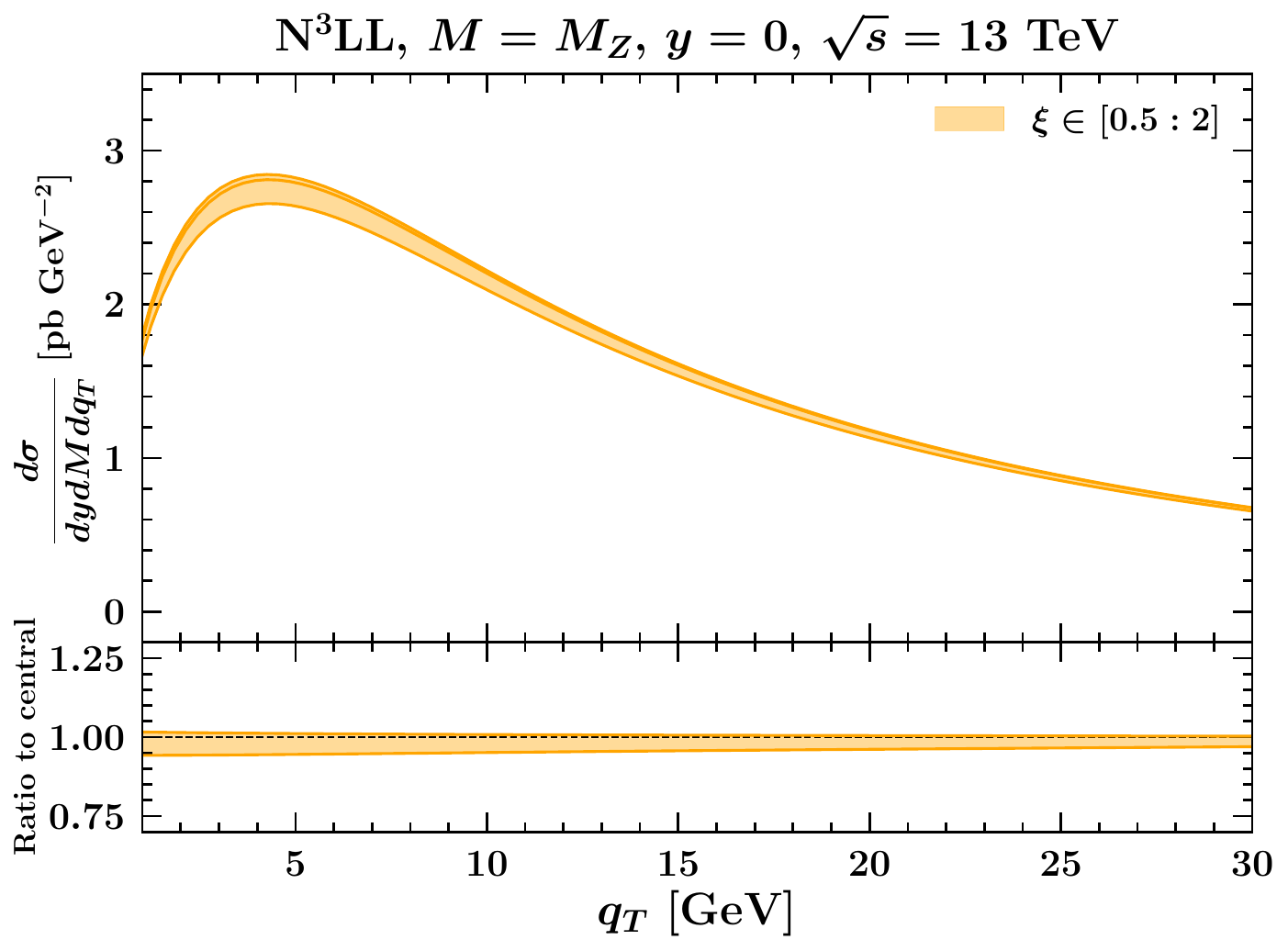}
    \caption{DY small-$q_T$ spectrum of a lepton pair with invariant
      mass $M$ equal to $M_Z$ and at central rapidity ($y=0$) at NNLL
      (left) and N$^3$LL (right) accuracies at the 13~TeV LHC. The
      curves include an uncertainty band associated with variations of
      resummation scale-parameter $\xi$. The upper insets display the
      cross sections, while the lower insets show their ratios to
      central-scale predictions.}
    \label{fig:DYpT_muFmuRxi_res}
  \end{center}
\end{figure*}

We now move to the small-$q_T$ region. We consider again the $q_T$
distribution of a lepton pair with invariant mass $M=M_Z$ and rapidity
$y=0$ in DY production in $pp$ collisions at $\sqrt{s}=13$~TeV. As
argued above, this regime is characterised by the resummation of large
logarithms of $q_T/M$. Therefore, its theoretical uncertainty can be
estimated by variations of the resummation-scale parameter $\xi$,
which accounts at once for the uncertainty on the evolution of
$\alpha_s$, PDFs, and TMDs. Fig.~\ref{fig:DYpT_muFmuRxi_res} shows
this cross section in the range $q_T\in[0:30]$~GeV at NNLL (left
panel) and N$^3$LL (right panel) along with its $\xi$ uncertainty. As
above, upper insets display the cross sections, while the lower insets
show their ratios to central-scale predictions. We observe that,
moving from NNLL to N$^3$LL, the uncertainty shrinks significantly, as
expected. Moreover, at both orders, bands are only mildly dependent on
$q_T$, remaining approximately of the same relative size across the
whole range considered.

We now address the question of matching small- and large-$q_T$
regimes. In Sec.~\ref{eq:matchinsetup}, we discussed the technical
details of this procedure. Before presenting results for the matched
cross sections, we study the numerical cancellation between
large-$q_T$ and double-counting contributions at $q_T\ll M$. Indeed, a
precise cancellation of these terms at small $q_T$ is a requirement in
order for the additive matching formula in
Eq.~(\ref{eq:matchingqTdamp}) to work accurately.  The cancellation
needs to be accurate for all possible choices of the scales.
\begin{figure*}[h!]
  \begin{center}
    \includegraphics[width=0.49\textwidth]{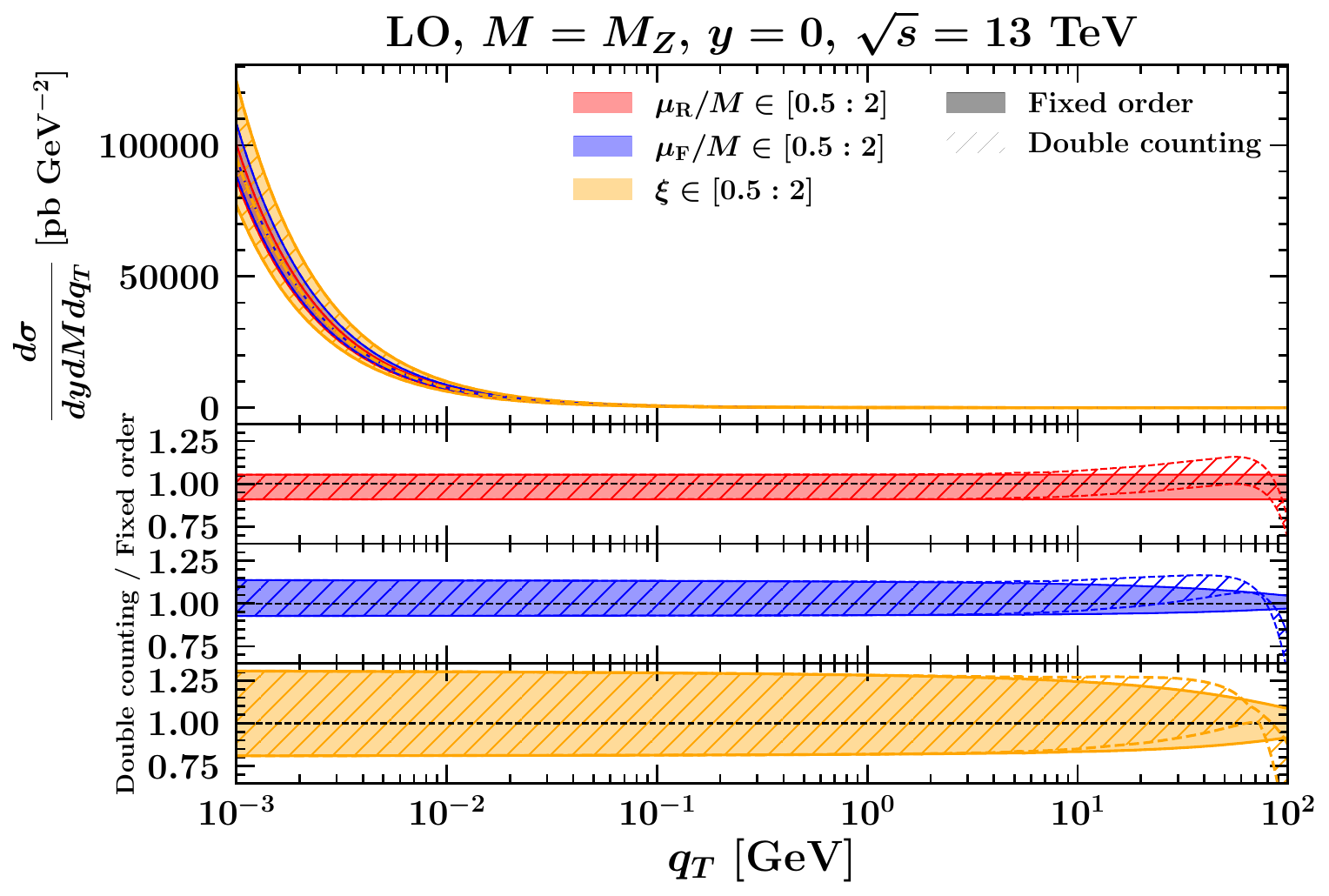}
    \includegraphics[width=0.49\textwidth]{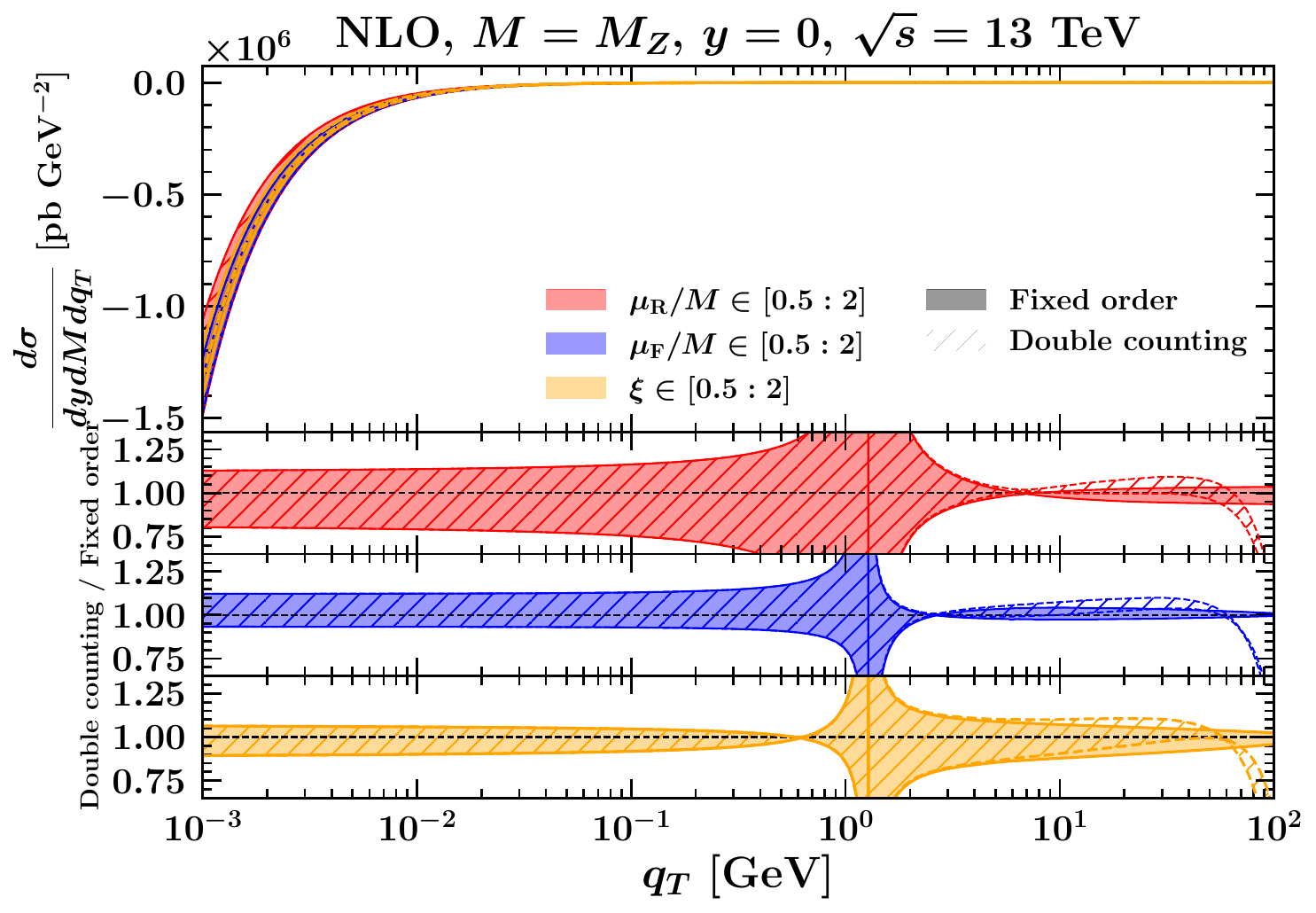}
    \caption{Large-$q_T$ (fixed order) vs. double-counting terms of
      Eq.~(\ref{eq:matchingqTdamp}) in the small-$q_T$ region for the
      DY spectrum with invariant mass $M=M_Z$ and rapidity $y=0$ at LO
      (left) and NLO (right) accuracies at the 13~TeV LHC . The curves
      include uncertainty bands associated with variations of
      resummation scale-parameter $\xi$, renormalisation scale
      $\mu_{\rm R}$, and factorisation scale $\mu_{\rm F}$. The upper
      insets display the cross sections, while the lower insets show
      their ratios to central-scale predictions.}
    \label{fig:DYpT_muFmuRxi_nonsing}
  \end{center}
\end{figure*}
To this purpose, in Fig.~\ref{fig:DYpT_muFmuRxi_nonsing} we show the
behaviour of both large-$q_T$ and double-counting cross sections in
the small-$q_T$ region at LO (left plot) and NLO (right plot). As
above, the kinematics is $M=M_Z$, $y=0$, and $\sqrt{s}=13$~TeV. Each
curve comes with three different bands associated with
renormalisation-scale (red), factorisation-scale (blue), and
resummation-scale-parameter (orange) variations. Plain (hatched) bands
correspond to the large-$q_T$ (double-counting) calculations. The
upper panel of each plot shows the cross sections, while the
underlying panels display ratios of each scale variation to the
respective central-scale curve. We first note that both LO and NLO
calculations are divergent as $q_T\rightarrow 0$, but while the LO is
positively divergent, the NLO tends to negative infinity. In spite of
this steep behaviour, as seen from the ratio panels, large-$q_T$ and
double-counting calculations converge nicely in this region at the
level of both central values and single scale variations. This
guarantees an accurate implementation of the matching formula,
Eq.~(\ref{eq:matchingqTdamp}), in the $q_T\ll M$ region.

\begin{figure*}[t!]
  \begin{center}
    \includegraphics[width=0.49\textwidth]{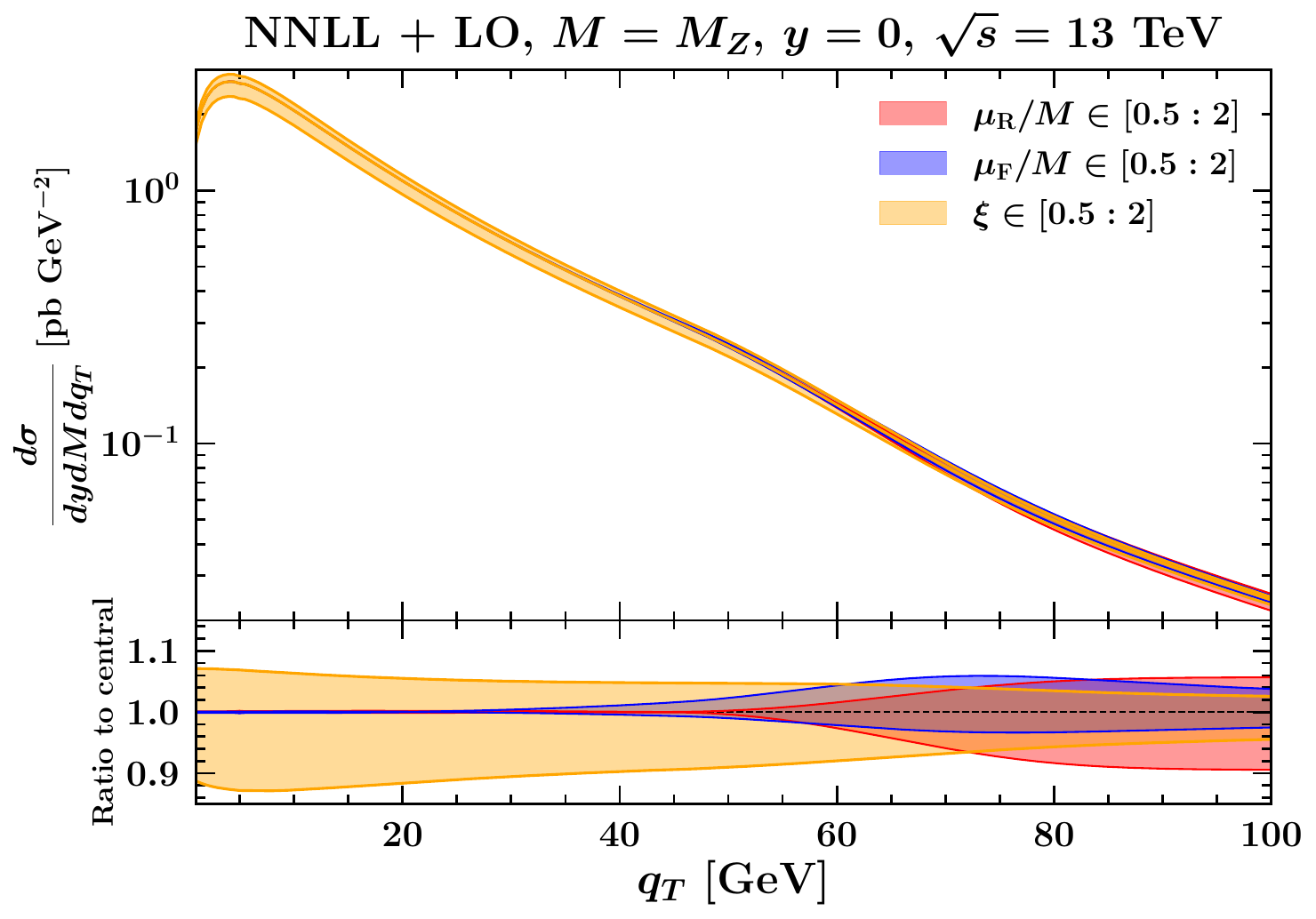}
    \includegraphics[width=0.49\textwidth]{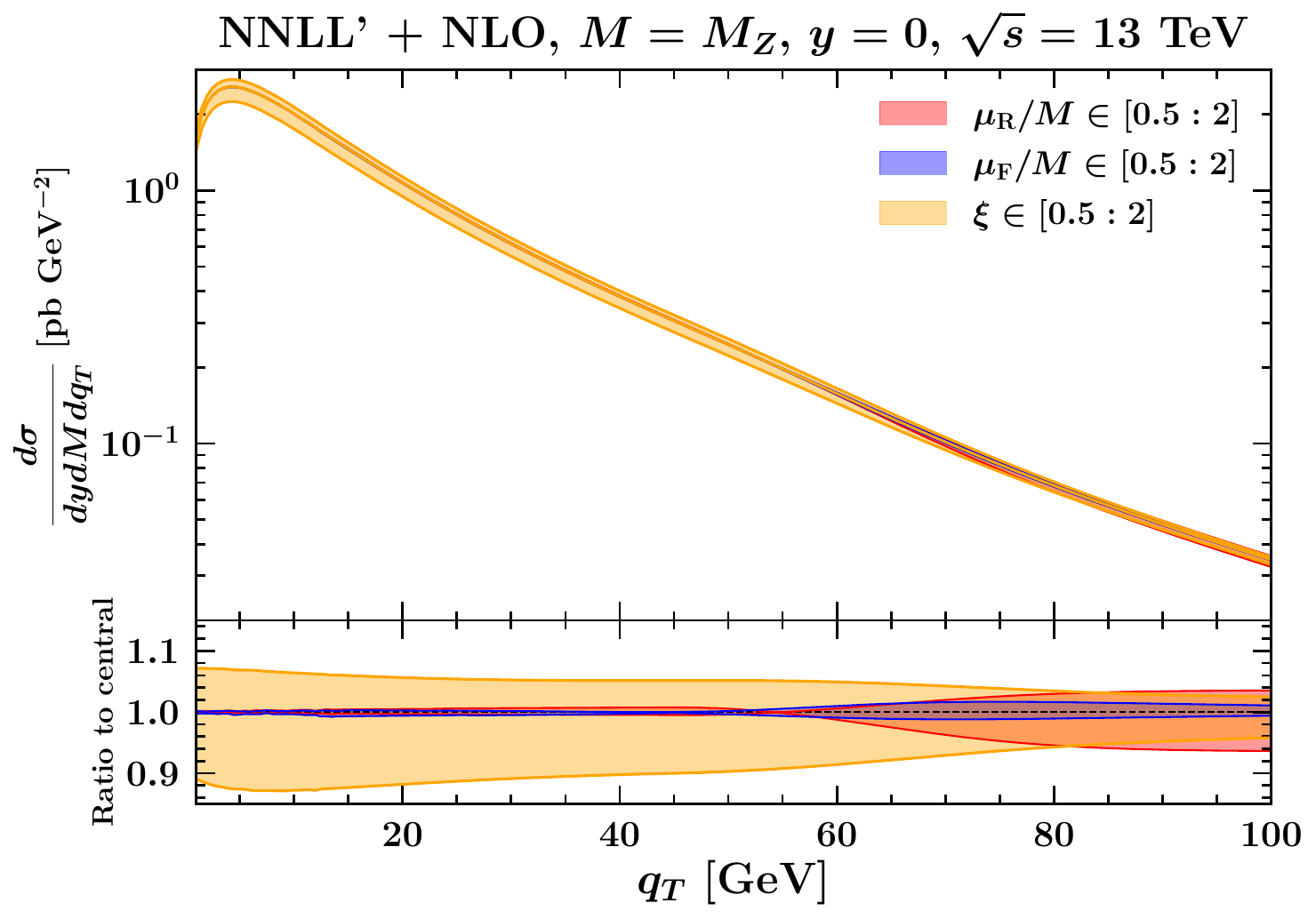}
    \includegraphics[width=0.49\textwidth]{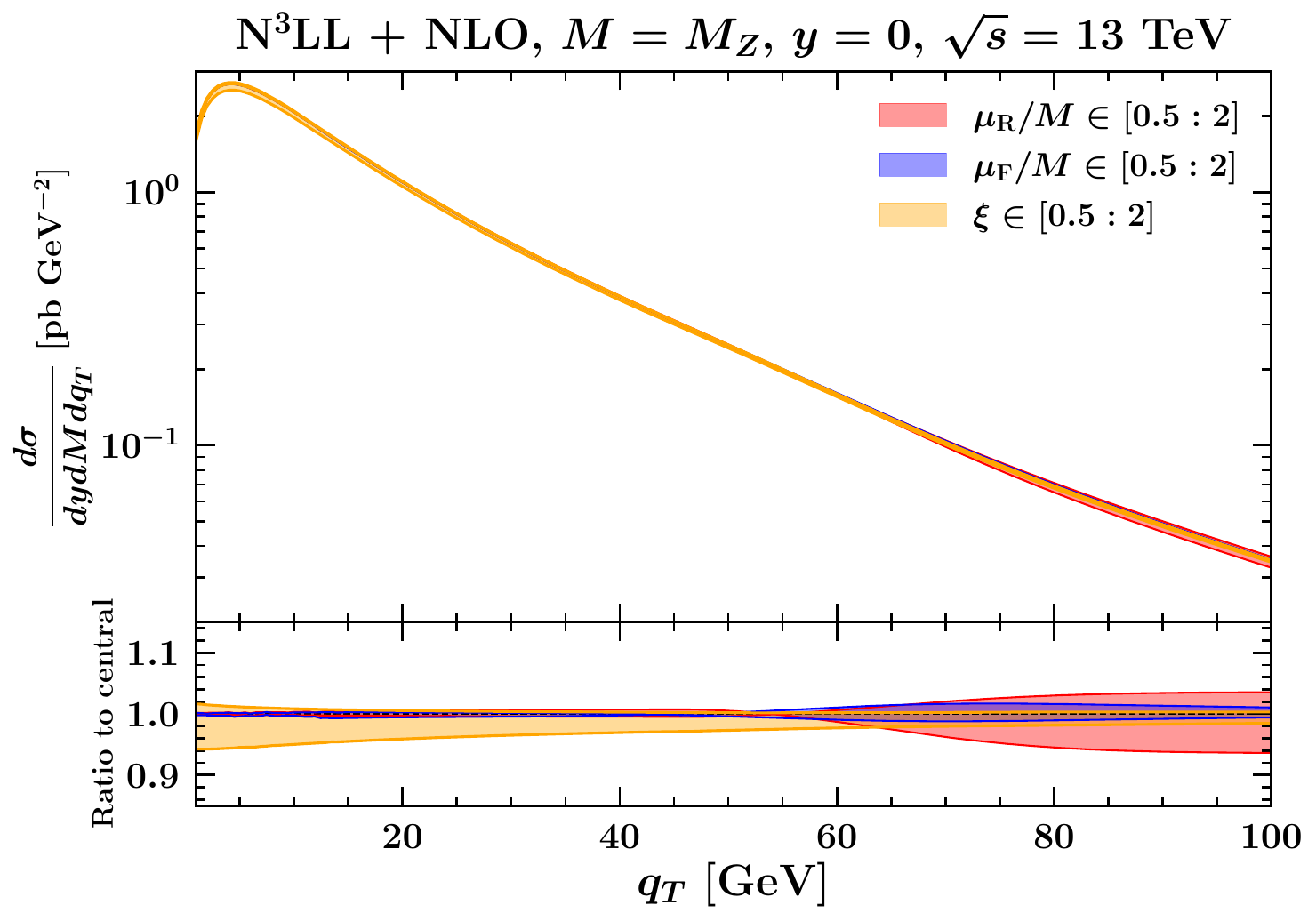}
    \caption{Matched result for the DY $q_T$ spectrum of a lepton pair with
      invariant mass $M$ equal to $M_Z$ and at central rapidity ($y=0$) at NNLL
      + LO (top left), NNLL' + NLO (top right), and N$^3$LL + NLO (bottom)
      accuracies at the 13~TeV LHC. The curves include an uncertainty band
      associated with variations of resummation scale-parameter $\xi$. The upper
      insets display the cross sections, while the lower insets show the ratio
      to central-scale predictions.}
    \label{fig:DYpT_muFmuRxi_match}
  \end{center}
\end{figure*}
Finally, in Fig.~\ref{fig:DYpT_muFmuRxi_match} we show matched
predictions for the $q_T$ distribution with the same kinematics as
above at NNLL + LO (top left), NNLL$^\prime$ + NLO (top right), and
N$^{3}$LL + NLO (bottom) accuracy. The logarithmic accuracy
corresponds to the small-$q_T$ calculation while the fixed-order one
to the large-$q_T$ calculation, with the counter-term chosen
accordingly. The matching allows us to consider at once predictions
that range between $q_T=0$~GeV and $q_T=100$~GeV. As usual, the upper
panels show the cross sections along with scale-variation curves,
while the lower panels show their ratios to central-scale
predictions. The uncertainty reduction in moving to higher
perturbative accuracies is manifest. In particular, while at NNLL + LO
uncertainties are roughly at the 10\% level all across the range in
$q_T$ considered, they reduce to about 5\% at N$^3$LL + NLO. We also
observe a clear predominance of $\xi$ uncertainties at small and
intermediate values of $q_{T}$ region (up to $q_{T}\sim 60$ GeV), with
renormalisation- and factorisation-scale uncertainties taking over at
larger $q_T$. This behaviour reflects the regions of importance of
resummed and fixed-order calculations.

\begin{figure*}[t!]
  \begin{center}
    \includegraphics[width=0.65\textwidth]{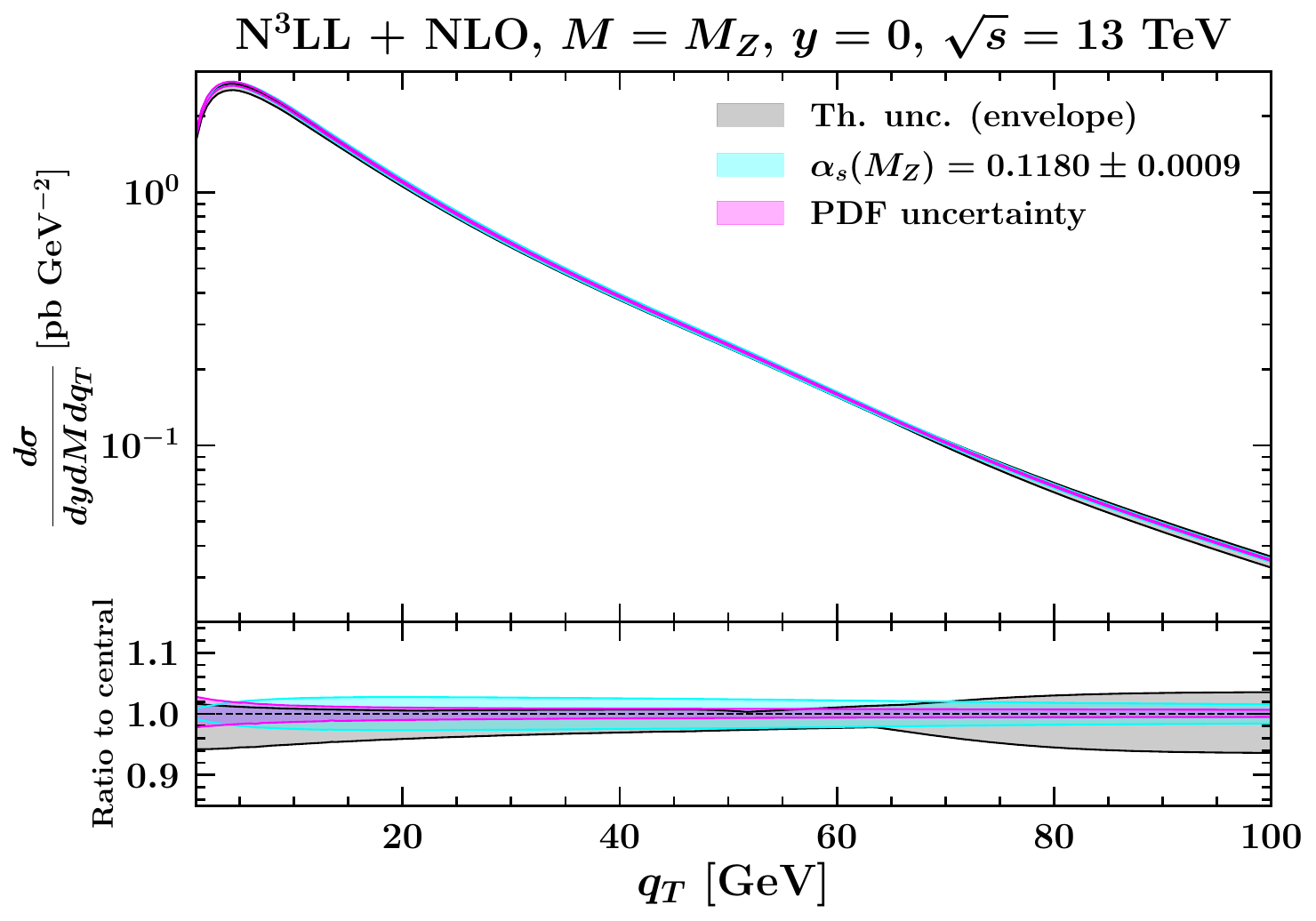}
    \caption{Same as the bottom plot of
      Fig.~\ref{fig:DYpT_muFmuRxi_match}, \textit{i.e.}, the N$^3$LL +
      NLO $q_T$ spectrum, where the global theoretical uncertainty
      (gray), obtained as the envelope of the variations of
      $\mu_{\rm R}$, $\mu_{\rm F}$, and $\xi$, is compared with the
      uncertainties associated with the boundary conditions of strong
      coupling (cyan) and PDFs (magenta).}
    \label{fig:DYpT_muFmuRxiAlphasPDFs_match}
  \end{center}
\end{figure*}
As a last study, in Fig.~\ref{fig:DYpT_muFmuRxiAlphasPDFs_match} we
consider the N$^3$LL + NLO spectrum and, on top of the
theoretical-uncertainty band (gray) obtained as an envelope of all
scale-variations bands ($\mu_{\rm R}$, $\mu_{\rm F}$, $\xi$), we also
show:
\begin{itemize}
\item in cyan the band associated to variations of the strong-coupling
  boundary condition within its uncertainty,
  $\alpha_s(M_Z)=0.1180\pm 0.0009$~\cite{ParticleDataGroup:2020ssz},
\item in magenta the band produced by the error members of the
  MSHT2020 PDF set~\cite{Bailey:2020ooq} at $\nu_0=2$~GeV.
\end{itemize}
The main observation is that all bands are roughly of the same size,
but with the theory uncertainty being generally larger. This plot
shows that an accurate estimate of theoretical uncertainties,
including those of RGE origin, is of crucial importance to achieve a
reliable comparison with the accurate data delivered by the LHC.

\section{Conclusions}
\label{sec:concl}

In this work, we have investigated the role of theoretical RGE
uncertainties in QCD in the context of high-energy collider
phenomenology. We have developed a theoretical framework, inspired by
soft-gluon and transverse-momentum resummations, able to take this RGE
systematics into account.

The discussion moves from the observation that \textit{analytic}
solutions to RGEs explicitly expose the potentially large logarithms
which are resummed, thus allowing resummation scales to be naturally
introduced (see, \textit{e.g.}, Eq.~(\ref{eq:SplitResScale})).  These
scales have the purpose of parameterising subleading corrections to
analytic solutions and enable us to estimate the corresponding
theoretical uncertainty. An alternative approach to solving RGEs is
the \textit{numerical} one.  While in this case it is impossible to
introduce resummation scales at the level of the solutions, we have
shown that this can still be done at the level of anomalous
dimensions.  To this end, we displace the argument of the coupling
$\alpha_s$ in their perturbative expansions so as to generate
different anomalous dimensions that differ by subleading terms (see,
\textit{e.g.}, Eq.~(\ref{eq:NLOanomalousdimensionShifted})). Varying
the amount of displacement allows us to compute numerical solutions
that are perturbatively equivalent but numerically different. The
difference provides an estimate of the associated theoretical
uncertainty. Importantly, we have argued that the two seemingly
different ways of estimating RGE-related uncertainties in analytic and
numerical solutions are in fact in one-to-one correspondence.

In this work, we have treated the single-logarithmic cases of
$\alpha_s$ and PDFs, already addressed in Ref.~\cite{Bertone:2022sso},
and extended the discussion to the double-logarithmic case of TMDs. In
all three cases, we have computed and compared analytic and numerical
solutions to the respective RGEs, focusing on the estimate of
uncertainties deriving from variations of the resummation scales. In
this respect, we have shown quantitatively the general agreement
between the two approaches up to high perturbative accuracies.

We then moved on to assessing the impact of RGE-related uncertainties
on the perturbative computation of physical observables at high-energy
colliders. In this context, we put particular emphasis on \textit{comparing
resummation-scale uncertainties to renormalisation- and
factorisation-scale uncertainties}(Sections~\ref{sec:coll}
and \ref{sec:numapplications}). We considered the cases of
structure functions in DIS and of the $q_T$ spectrum in DY production
in $pp$ collisions. The former observable is relevant to extractions
of PDFs from data, and so is the latter when $q_T$ is large enough. DY
at small values of $q_T$ is instead relevant to extractions of TMDs
from data, as well as to the precision determination of the
electroweak parameters of the Standard Model.

The computation of DIS structure functions obeys collinear
factorisation and is characterised by single-logarithmic resummations,
encoded in the running of $\alpha_s$ and PDFs. We considered the
neutral-current unpolarised structure functions $F_2$, $F_L$, and
$xF_3$ at NNLO accuracy. We observed that resummation-scale
uncertainties are generally sizeable and of the same order as (when
not larger than) renormalisation- and factorisation-scale
variations. A similar picture emerges when studying DY production at
large values of $q_T$ at LO and NLO accuracy. Indeed, also this
observable undergoes collinear factorisation and is thus affected by
single logarithms only.

DY production at small values of $q_T$ obeys TMD factorisation and is
characterised by the resummation of double logarithms. Also in this
case, we have carried out a detailed study of the impact of
resummation-scale uncertainties up to N$^3$LL accuracy. We have
examined the potential advantages of the numerical approach over the
analytic one, often employed in $q_T$-resummation literature. In
particular, we have discussed the meaning of the different scales
entering resummed predictions in the two cases. In our calculation,
based on numerical solutions, we employ appropriate reference scales
for each of the evolving quantities, and evaluate RGE-related
uncertainties only by means of resummation-scale variations.

In order to show that our framework can be consistently applied to
single- and double-logarithmic resummations at once, we have also
given a quantitative assessment of theoretical uncertainties when
large-$q_T$ and small-$q_T$ DY calculations are matched. Here we have
compared resummation-scale and renormalisation/factorisation-scale
uncertainties (with the latter only stemming from the large-$q_T$
calculation) over a wide range in $q_T$ relevant to DY production at
the LHC at 13~TeV. This allowed us to provide a comprehensive estimate
of theoretical uncertainties that affect this observable and to
compare them to the experimental uncertainties associated with the
measured values of $\alpha_s$ and PDFs.

Our results indicate that RGE effects are sizeable and can influence
ongoing collider phenomenology programmes. In particular, they are
relevant for the assessment of theoretical uncertainties in the
extraction of PDFs from data, and for precision physics in DY
production at the LHC.  In the longer term, they can affect the
planning of experiments at the next energy frontier, for which RGEs
are an essential component to relate lower-energy results to physics
studies in the kinematic region accessible at future machines.

\vskip 0.3 cm
\section*{Acknowledgments}

The authors would like to thank S. Camarda, T. Cridge, F. Giuli,
S. Glazov, P. Monni, and A. Vicini for stimulating
discussions. G.B. would like to express his deepest gratitude to the
late Stefano Catani: a caring mentor and colleague, an outstanding
scientist and, above all, a wonderful person.  V.B. was supported by
the European Union’s Horizon 2020 research and innovation programme
under grant agreement STRONG 2020 - No 824093.

\appendix
\section{Collection of formulas}\label{app:formulas}

In this appendix, we collect the $g$-functions relevant to the
implementation of the analytic evolution of $\alpha_s$
(App.~\ref{App}), PDFs (App.~\ref{App1}), and TMDs
(App.~\ref{App2}). Expressions are given in terms of known
perturbative coefficients, such as those of the $\beta$-function, of
the DGLAP splitting functions (in Mellin space), and the $A$- and
$B$-functions. A numerical implementation of these functions can be
found in the codes {\tt APFEL++}~\cite{Bertone:2013vaa,
  Bertone:2017gds} and {\tt MELA}~\cite{Bertone:2015cwa}.

\subsection{$g$-functions for the strong coupling}\label{App}

With reference to Eq.~(\ref{eq:alphasgs}), the $g^{(\beta)}$-functions
necessary to achieve N$^3$LL accurate evolution of the strong coupling
$\alpha_s$ are the following:
\begin{eqnarray}
{g}_1^{(\beta)}(\lambda) &=& \frac{1}{1-{\lambda}}\,,
                                              \nonumber\\
  {g}_2^{(\beta)}(\lambda , \kappa )&=&
                                        -\frac{b_1\ln(1-{\lambda})+\beta_{0}\ln\kappa}{(1-{\lambda})^2}\,,
                                        \nonumber\\
  {g}_3^{(\beta)}(\lambda , \kappa)&=& \left[ b_2
                                       {\lambda}-b_1^2\left(
                                       {\lambda}+\ln\left(1-{\lambda}\right)-\ln^2\left(1-{\lambda}\right)\right)
                                       \right.
                                       \nonumber\\
                                          &+& \left.
                                              \beta_{0}b_1\left(2\ln(1-{\lambda})-1\right)
                                              \ln\kappa+(\beta_{0})^2
                                              \ln^2\kappa \right] /
                                              (1-{\lambda})^3 \, ,
  \\
  {g}_4^{(\beta)}(\lambda , \kappa)&=& \displaystyle\big[(b_3 -b_2
                                       b_1) \lambda -\frac12 (b_1^3
                                       -2b_2 b_1+b_3) \lambda ^2+(2
                                       b_1(b_1^2
                                       -b_2 ) \lambda  -b_2 b_1) \ln \left(1-\lambda\right) \nonumber\\
                                          &+&\displaystyle b_1^3
                                              (\frac52 -\ln
                                              \left(1-\lambda\right))
                                              \ln
                                              ^2\left(1-\lambda\right)+ (2 (b_1^2-b_2) \lambda  \nonumber\\
                                          &+&b_1^2 (5 -3 \ln
                                              \left(1-\lambda\right))\ln
                                              \left(1-\lambda\right) -
                                              b_2 ) \beta
                                              _0\ln\kappa\nonumber\\
                                          &+&b_1 (-3 \ln
                                              \left(1-\lambda\right)+\frac52)
                                              \beta _0^2\ln^2\kappa-
                                              \beta _0^3 \ln^3\kappa
                                              \big]/(1-\lambda)^4\nonumber\,,
\end{eqnarray}
with $b_i=\beta_i/\beta_0$. Given our definition of RGE for $\alpha_s$
in Eq.~(\ref{eq:RGEalphas}), the relevant coefficients $\beta_i$
read~\cite{Herzog:2017ohr}
\begin{equation}
  \label{betacoefs}
  \small
  \begin{array}{rcl}
    \beta_{0} &= &  \displaystyle -2\bigg[\frac{11}{3} C_A - \frac{4}{3} T_R n_f\bigg] \,, \\
    \beta_{1} &= & \displaystyle -2\bigg[\frac{34}{3} C_A^2 - \frac{20}{3} C_A T_R n_f - 4 C_F T_R n_f \bigg]\,, \\
    \beta_{2} &= &  \displaystyle -2\bigg[\frac{2857}{54} C_A^3 
                   - \frac{1415}{27} C_As T_R n_f
                   - \frac{205}{9} C_F C_A T_R n_f + 2 C_F^2 T_R n_f 
    \\ & + & \displaystyle 
             \frac{44}{9} C_F T_R^2 n_f^2 
             + \frac{158}{27} C_A T_R^2 n_f^2 \bigg] \,, \\
    \beta_{3} &=  &\displaystyle 
                    -2\bigg[ C_A^4 \left(  \frac{150653}{486} - \frac{44}{9} \zeta_3 \right)   
                    + d_A^{abcd}d_A^{abcd} \left(  - \frac{80}{9} + \frac{704}{3}\zeta_3 \right)
    \\ & + & \displaystyle 
             C_A^3 T_R n_f  
             \left(  - \frac{39143}{81} + \frac{136}{3} \zeta_3 \right)
             + C_A^2 C_F T_R n_f 
             \left(  \frac{7073}{243} - \frac{656}{9}\zeta_3 \right)
    \\ & + & \displaystyle 
             C_A C_F^2 T_R n_f 
             \left(  - \frac{4204}{27} + \frac{352}{9}\zeta_3 \right)
             + d_F^{abcd}d_A^{abcd} n_f \left(  \frac{512}{9} - \frac{1664}{3}\zeta_3 \right)
    \\ & + & \displaystyle 
             46 C_F^3 T_R n_f 
             +  C_A^2 T_R^2 n_f^2 
             \left(  \frac{7930}{81} + \frac{224}{9}\zeta_3 \right)
             +  C_F^2 T_R^2 n_f^2 
             \left(  \frac{1352}{27} - \frac{704}{9}\zeta_3 \right)
    \\ & + & \displaystyle 
             C_A C_F T_R^2 n_f^2 
             \left(  \frac{17152}{243} + \frac{448}{9}\zeta_3 \right)
             + d_F^{abcd}d_F^{abcd} n_f^2 \left( - \frac{704}{9} + \frac{512}{3}\zeta_3 \right)
    \\ & + & \displaystyle 
             \frac{424}{243} C_A T_R^3 n_f^3 
             + \frac{1232}{243} C_F T_R^3 n_f^3 \bigg]\, ,
  \end{array}
\end{equation} 
where $n_f$ stands for the number of active quark flavours and
$\zeta_i\equiv\zeta(i)$ are the values of Riemann's
$\zeta$-function. The colour factors are given by
\begin{equation}
  \begin{array}{c}
    \displaystyle T_R = \frac{1}{2}\,, \quad C_A = N_c\,, \quad   C_F =
    \frac{N_c^2-1}{2 N_c} \,, \quad d_A^{abcd}d_A^{abcd} =
    \frac{N_c^2(N_c^2+36)}{24}\,,  
    \\
    \displaystyle d_F^{abcd}d_A^{abcd} = \frac{ N_c(N_c^2+6)}{48}\,, \quad d_F^{abcd}d_F^{abcd} = \frac{N_c^4-6N_c^2+18}{96\, N_c^2} \,,
  \end{array}
\end{equation}
with $N_c=3$.

\subsection{$g$-functions for PDFs}\label{App1}

The analytic solution for the evolution of PDFs is given by
Eq.~(\ref{eq:gensoldglap}). As discussed in Sec.~\ref{sec:coll}, that
formulation strictly applies to Mellin moments of a non-singlet PDF
combination. However, the singlet case can be accommodated by
promoting the $g^{(\gamma)}$-functions to $2\times2$ matrices that
couple the evolution of total-singlet and gluon PDFs. In turn, this
implies computing exponentials of $2\times2$ matrices, which can be
done in closed form~\cite{weisstein}. The $x$-space solution is then
obtained by inverse Mellin transform of the evolved PDF moments. This
technology is implemented in the code {\tt
  MELA}~\cite{Bertone:2015cwa}. An analytic computation directly in
$x$-space is harder to obtain due to the integro-differential nature
of the PDF evolution equations in this space. A discretised
path-ordered solution could achieve this goal but, to the best of our
knowledge, this approach has never been pursued so far.

The functions $g_i^{(\gamma)}$ with $i>0$ entering the exponential in
in Eq.~(\ref{eq:gensoldglap}) and necessary to achieve NNLL accuracy
are the following:
\begin{eqnarray}
  \small
  \label{eq:gfunctions-gamma}
  {g}_1^{(\gamma)}(\lambda) &=&   \frac{\gamma^{(0)}}{\beta_{0}}\ln\left({g}_{1}^{(\beta)}(\lambda)\right)=-\frac{\gamma^{(0)}}{\beta_{0}}\ln\left(1-{\lambda}\right)   \, ,
                                \nonumber\\
  {g}_2^{(\gamma)}(\lambda , \kappa )&=&   \frac{\gamma^{(0)}}{\beta_{0}}\frac{{g}_{2}^{(\beta)}({\lambda, \kappa})}{{g}_{1}^{(\beta)}(\lambda)}=-\frac{\gamma^{(0)}}{\beta_{0}}\frac{b_1\ln(1-{\lambda})+\beta_{0}\ln\kappa}{1-{\lambda}}
                                         \,   ,
                                         \nonumber\\
  {g}_3^{(\gamma)}(\lambda , \kappa)&=&
                                        \frac{\gamma^{(0)}}{\beta_{0}}\left[\frac{{g}_{3}^{(\beta)}(\lambda,\kappa)}{{g}_{1}^{(\beta)}(\lambda)}-\frac{1}{2}\left(\frac{{g}_{2}^{(\beta)}(\lambda,\kappa)}{{g}_{1}^{(\beta)}(\lambda)}\right)^2\right]
                                        \nonumber\\
                            &=&   \frac{\gamma^{(0)}}{\beta_{0}}   \left[  b_2{\lambda}-b_1^2\left({\lambda}+\ln\left(1-{\lambda}\right)-\frac{1}{2}\ln^2\left(1-{\lambda}\right)\right)
                                \right.
                                \nonumber\\
                            &+&  \left.  \beta_{0}b_1\left(\ln(1-{\lambda})-1\right)
                                \ln\kappa+\frac{1}{2}\beta_{0}^2
                                \ln^2\kappa \right]  /
                                (1-{\lambda})^2 \,.
\end{eqnarray}
The first equality in each of the equations above shows how the
$g^{(\gamma)}$-functions are related to the $g^{(\beta)}$-functions
presented in App.~\ref{App}.

The function $g_0^{(\gamma)}$ truncated at $\mathcal{O}(\alpha_s^2)$,
necessary to achieve NNLL accuracy, reads
\begin{eqnarray}
\label{eq:g0PDF}
  \small
  \displaystyle g_0^{(\gamma), {\rm NNLL}} ({\lambda} , \kappa) &=&
                                                                    \displaystyle 1 +
                                                                    \frac{a_s(\mu_0){\lambda}}{\beta_{0}(1-{\lambda})}\left(\gamma^{(1)}-b_1\gamma^{(0)}\right) 
  \\
                                                                &+&\displaystyle\frac{a_s^2(\mu_0)}{2\beta_0^2(1-\lambda)^2}\Big[b_1\left(\gamma^{(1)}
                                                                    \left(\beta_0\left((\lambda-2) \lambda-2\ln(1-\lambda)\right)-2\lambda^2\gamma^{(0)}\right)+2
                                                                    \beta_0^2\ln \kappa\gamma^{(0)}\right)\nonumber\\
                                                                &+&\displaystyle b_1^2\gamma^{(0)}\left(\lambda^2 \gamma^{(0)}-\beta_0\left((\lambda-2) \lambda-2 \ln(1-\lambda)\right)\right)+b_2\beta_0(\lambda -2)\lambda\gamma^{(0)}\nonumber\\
                                                                &+&\displaystyle\gamma^{(1)}\left(\lambda^2 \gamma^{(1)}-2\beta_0^2\ln \kappa\right)-\beta_0 (\lambda-2) \lambda\gamma^{(2)}\Big]
                                                                    \nonumber\,.
\end{eqnarray}

The numerical values of the coefficients $\beta_0$ and $b_i$ can be
found in App.~\ref{App}. The anomalous dimensions $\gamma_i$ for the
evolution of (unpolarised) PDFs are the Mellin transforms of the DGLAP
splitting functions. They are given in Mellin space as functions of
the complex moment variable $N$, conjugate to the partonic
longitudinal-momentum fraction $x$. The splitting functions up to
three loops, relevant to NNLL evolution, are computed in
Refs.~\cite{Moch:2004pa, Vogt:2004mw}.

We remark that the method developed in this paper, based on the
$g^{(\gamma)}$-functions presented above at NNLL accuracy, can be
readily extended to N$^3$LL accuracy, once the four-loop splitting
functions become available: for ongoing work see
Refs.~\cite{Moch:2023tdj,Falcioni:2023vqq,Falcioni:2023luc}, and
Refs.~\cite{McGowan:2022nag,Cridge:2023ryv,Cridge:2023ozx,NNPDF:2024nan}
for studies of the associated PDF phenomenology.

\subsection{$g$ functions for TMDs}\label{App2}

TMD evolution is encoded in the so-called Sudakov form factor $S$ in
Eq.~(\ref{eq:SudFF2}).  The computation of the factor $S$, leading to
Eq.~(\ref{eq:gfunc}), can be carried out by exploiting the analytic
running of the strong coupling discussed in
Sec.~\ref{sec:runningcoupling}, and whose perturbative ingredients are
collected in App.~\ref{App}. Below, we collect all $g$-functions
necessary to evolve TMDs up to N$^3$LL accuracy.  We also show how
they are related to the $g^{(\beta)}$-functions responsible for the
evolution of $\alpha_s$.

The functions $g_i$ with $i>0$ in Eq.~(\ref{eq:gfunc}) are given by
\begin{equation}
  g_1(\lambda)=-\frac{4A^{(1)}}{a_s(\mu_0)\beta_{0}^2L}\int_{0}^{\lambda}dt \, t\,{g}_1^{(\beta)}(t)=\frac{4A^{(1)} }{\beta_0}\frac{\lambda +\ln(1-\lambda)}{\lambda}\,,
\end{equation}
\begin{equation}
  \begin{array}{rcl}
    g_2(\lambda,\kappa) &=& \displaystyle \int_{0}^{\lambda}dt\left[ - \frac{4A^{(1)}}{\beta_{0}^2}\,t\,{g}_2^{(\beta)}(t,\kappa)-\frac{
                            4A^{(2)}}{\beta_{0}^2}\,
                            t({g}_1^{(\beta)}(t))^2 +\frac{2
                            \overline{B}^{(1)}}{\beta_{0}}{g}_1^{(\beta)}(t)\right]\\
                        &=&\displaystyle \frac{4\beta_{1}A^{(1)}}{\beta_{0}^3}\left(\frac12\ln^2(1-\lambda)+\frac{\ln(1-\lambda)}{1-\lambda}+\frac{\lambda}{1-\lambda}\right)\\
                        &-&\displaystyle \left(\frac{
                            4A^{(2)}}{\beta_{0}^2}-\frac{4A^{(1)}\ln\kappa}{\beta_{0}}\right)\left(\frac{\lambda}{1-\lambda}+\ln(1-\lambda)\right) -\frac{2
                            \overline{B}^{(1)}}{\beta_{0}}\ln(1-\lambda)\,,
  \end{array}
\end{equation}
\begin{equation}
  \begin{array}{rcl}
    g_3(\lambda,\kappa) &=&\displaystyle -\frac{2}{\beta_{0}} \int_{0}^{\lambda}dt
                            \bigg[\frac{2 A^{(1)} }{\beta_{0}}t{g}_3^{(\beta)}(t,\kappa)
                            +\frac{4 A^{(2)}}{\beta_{0}}t{g}_1^{(\beta)}(t){g}_2^{(\beta)}(t,\kappa)\\
                        &+&\displaystyle \frac{2 A^{(3)} }{\beta_{0}}t({g}_1^{(\beta)}(t))^3 -\overline{B}^{(1)}{g}_2^{(\beta)}(t,\kappa)-\overline{B}^{(2)}({g}_1^{(\beta)}(t))^2\bigg]\\
                        &=&\displaystyle -\frac{4 (\beta_{0}\beta_{2}-\beta_{1}^2)
                            A^{(1)} }{\beta_{0}^4} \left(\frac{\lambda (3 \lambda - 2)}{2
                            (1 - \lambda)^2} - \ln(1 - \lambda)\right)\\
                        &+&\displaystyle \frac{2 \beta_{1}^2 A^{(1)} }{\beta_{0}^4}
                            \left(\frac{2\lambda}{1 - \lambda} + \frac{2\ln(1 -
                            \lambda)}{1 - \lambda}
                            +\frac{ (1 - 2 \lambda) \ln^2(1 - \lambda)}{(1 - \lambda)^2}\right)\\
                        &+&\displaystyle\left(\frac{4\beta_{1}A^{(2)}}{\beta_{0}^3}-\frac{4 \beta_{1}A^{(1)}\ln\kappa }{\beta_{0}^2}\right)
                            \left(\frac{\lambda (3 \lambda-2)}{2(1 - \lambda)^2}-\frac{(1-2
                            \lambda) \ln(1 - \lambda)}{(1 - \lambda)^2}\right)\\
                        &-&\displaystyle \left(\frac{2 A^{(3)} }{\beta_{0}^2} -\frac{4 A^{(2)}\ln\kappa}{\beta_{0}}+2 A^{(1)} \ln^2\kappa-\frac{2 \beta_{1}A^{(1)}\ln\kappa }{\beta_{0}^2}\right)\frac{\lambda^2}{(1-\lambda)^2}\\
                        &-&\displaystyle \frac{2\beta_{1} \overline{B}^{(1)}}{\beta_{0}^2}\left(\frac{\lambda}{1-\lambda}+\frac{\ln(1-\lambda)}{1-\lambda}\right)+\left(\frac{2 \overline{B}^{(2)}}{\beta_{0}}-2 \overline{B}^{(1)}\ln\kappa\right)\frac{\lambda}{1-\lambda}\,,
  \end{array}
\end{equation}
\begin{equation}
  \small
  \begin{array}{rcl}
    g_4(\lambda,\kappa) &=&\displaystyle -\frac{2}{\beta_{0}} \int_{0}^{\lambda}dt
                            \bigg[\frac{2 A^{(1)} }{\beta_{0}}t{g}_4^{(\beta)}(t,\kappa)
                            +\frac{2 A^{(2)}}{\beta_{0}}t\left(2{g}_1^{(\beta)}(t){g}_3^{(\beta)}(t,\kappa)+({g}_2^{(\beta)}(t,\kappa))^{2}\right)\\
                        &+&\displaystyle \frac{6 A^{(3)}
                            }{\beta_{0}}t({g}_1^{(\beta)}(t))^2{g}_2^{(\beta)}(t,\kappa) +
                            \frac{2 A^{(4)} }{\beta_{0}}t({g}_1^{(\beta)}(t))^4 -
                            \overline{B}^{(1)}{g}_3^{(\beta)}(t,\kappa)\\
                        &-&\displaystyle 2\overline{B}^{(2)}{g}_1^{(\beta)}(t){g}_2^{(\beta)}(t,\kappa)-\overline{B}^{(3)}({g}_1^{(\beta)}(t))^3\bigg]\\
                        &=&\displaystyle -\frac{A^{(4)}}{3 (\lambda -1)^3 \beta_0^2}2 (\lambda -3)\lambda ^2\\
                        &+&\displaystyle \frac{A^{(3)}}{3 (\lambda -1)^3 \beta _0^2} \Bigg[6 (\lambda -3) \lambda^2\ln \kappa 
                            +\left(\lambda \left(5 \lambda ^2-15 \lambda +6\right)+(6-18 \lambda ) \ln (1-\lambda)\right) b_1\Bigg]\\
                        &+&\displaystyle \frac{A^{(2)}}{3(\lambda -1)^3 \beta _0^2} \Bigg[\left(18 \lambda ^2 \beta _0^2-6\lambda ^3 \beta _0^2\right) \ln ^2\kappa\\
                        &+&\displaystyle \left(-6 b_1 \beta _0 \lambda ^3+18 b_1 \beta_0 \lambda ^2+36 \ln (1-\lambda ) b_1 \beta _0 \lambda -12 b_1
                            \beta _0 \lambda -12 \ln (1-\lambda ) b_1 \beta _0\right) \ln\kappa\\
                        &-&\displaystyle 11 b_1^2 \lambda ^3+8 b_2 \lambda ^3+9 b_1^2 \lambda ^2+18 \ln ^2(1-\lambda ) b_1^2 \lambda +6 \ln (1-\lambda
                            ) b_1^2 \lambda -6 b_1^2 \lambda\\
                        &-&\displaystyle 6 \ln ^2(1-\lambda ) b_1^2-6 \ln (1-\lambda ) b_1^2\Bigg]\\
                        &+&\displaystyle \frac{A^{(1)}}{3(\lambda -1)^3 \beta _0^2} \Bigg[\left(2 \lambda ^3 \beta _0^3-6
                            \lambda ^2 \beta _0^3\right) \ln ^3\kappa+\left(6 \lambda  b_1 \beta _0^2-18 \lambda
                            \ln (1-\lambda ) b_1 \beta _0^2+6 \ln (1-\lambda ) b_1 \beta_0^2\right) \ln ^2\kappa\\
                        &+& \displaystyle \Big( 6 b_1^2 \beta _0 \lambda ^3-6 b_2 \beta_0 \lambda ^3+6 b_1^2 \beta _0 \lambda ^2-6 b_2 \beta _0 \lambda
                            ^2-18 \ln ^2(1-\lambda ) b_1^2 \beta _0 \lambda \\
                        &+&\displaystyle 12 \ln (1-\lambda) b_1^2 \beta _0 \lambda +6 \ln^2(1-\lambda) b_1^2 \beta_0 \Big) \ln \kappa\\
                        &+&\displaystyle 6 \ln (1-\lambda ) b_1^3 \lambda ^3+2 b_1^3\lambda ^3-12 \ln (1-\lambda ) b_1 b_2 \lambda ^3+5 b_1 b_2
                            \lambda ^3+6 \ln (1-\lambda ) b_3 \lambda ^3 \\
                        &-& \displaystyle 7 b_3 \lambda ^3+6 \ln (1-\lambda ) b_1^3 \lambda ^2+12 \ln (1-\lambda ) b_1 b_2
                            \lambda ^2-15 b_1 b_2 \lambda ^2-18 \ln (1-\lambda ) b_3 \lambda^2\\
                        &+&\displaystyle 15 b_3 \lambda ^2-6 \ln ^3(1-\lambda ) b_1^3 \lambda +6 \ln^2(1-\lambda ) b_1^3 \lambda -18 \ln (1-\lambda ) b_1 b_2\lambda +6 b_1 b_2 \lambda \\
                        &+& \displaystyle 18 \ln (1-\lambda ) b_3 \lambda -6b_3 \lambda +2 \ln ^3(1-\lambda ) b_1^3+6 \ln (1-\lambda ) b_1
                            b_2-6 \ln (1-\lambda ) b_3\Bigg]\\
                        &+&\displaystyle \frac{\overline{B}^{(3)}}{3 (\lambda -1)^3 \beta _0^2}\Bigg[-3 \beta _0 \lambda^3+9 \beta _0 \lambda ^2-6 \beta _0 \lambda \Bigg]\\
                        &+&\displaystyle \frac{\overline{B}^{(2)}}{3(\lambda -1)^3 \beta _0^2} \Bigg[\left(6 \beta _0^2 \lambda ^3-18\beta _0^2 \lambda ^2+12 \beta _0^2 \lambda \right)\ln \kappa \\
                        &+&\displaystyle 3 b_1 \beta _0 \lambda ^3-9 b_1
                            \beta _0 \lambda ^2-6 \ln (1-\lambda ) b_1 \beta _0 \lambda +6b_1 \beta _0 \lambda +6 \ln (1-\lambda ) b_1 \beta _0\Bigg]\\
                        &+&\displaystyle \frac{\overline{B}^{(1)}}{3(\lambda -1)^3 \beta _0^2} \Bigg[\left(-3 \lambda ^3 \beta _0^3+9
                            \lambda ^2 \beta _0^3-6 \lambda  \beta _0^3\right) \ln^2\kappa+\left(6 \lambda
                            \ln (1-\lambda ) b_1 \beta _0^2-6 \ln (1-\lambda ) b_1 \beta_0^2\right) \ln \kappa\\
                        &-&\displaystyle 3 b_1^2 \beta _0 \lambda ^3+3 b_2 \beta _0 \lambda ^3+3 b_1^2 \beta _0 \lambda ^2-3 b_2 \beta _0 \lambda ^2
                            +3 \ln ^2(1-\lambda ) b_1^2 \beta _0 \lambda -3 \ln ^2(1-\lambda ) b_1^2 \beta _0\Bigg] \,,
\end{array}
\end{equation}
with
\begin{equation}
  \overline{B}^{(n)}=2 A^{(n)}L+B^{(n)}\quad\mbox{and}\quad L=\ln\frac{\kappa M}{\mu_0}\,,
\end{equation}
(see Eq.~(\ref{eq:Bdisplacement})).

The function $g_0$ in Eq.~(\ref{eq:gfunc}) truncated at
$\mathcal{O}(\alpha_s^2)$ and relevant to TMD evolution at N$^3$LL is
given by
\begin{equation}\label{eq:g0TMD}
  \small
  \begin{array}{rcl}
    g_0(a_s(\mu_0)) &=&\displaystyle 1 -2 a_s(\mu_0)
                        \left(A^{(1)} L^2+B^{(1)} L\right)+\frac{1}{2} a_s^2(\mu_0)
                        \bigg[4 \left(A^{(1)} L^2+B^{(1)}
                        L\right)^2\\
                    &+&\displaystyle \frac{2}{3} \left(-2 A^{(1)} \beta _0
                        L^3+6 A^{(1)} \beta _0 L^2 \ln \kappa-6 A^{(2)}
                        L^2-3 \beta _0 B^{(1)} L^2+6 \beta _0 B^{(1)} L
                        \ln \kappa-6 B^{(2)} L\right)\bigg]\,.
  \end{array}
\end{equation}

As discussed in Sec.~\ref{sec:sud_ff}, the coefficients $A^{(n)}$ and
$B^{(n)}$ are related to the anomalous dimensions $\gamma_K$,
$\gamma_F$, and $K$ that govern the evolution of TMDs (see
Eqs.~(\ref{eq:eveqs}) and~(\ref{eq:anomdimevol})). In
Eq.~(\ref{eq:AnDimToAB}), we have derived these relationships up to
N$^3$LL accuracy. Below, we report the perturbative coefficients of
the TMD anomalous dimensions necessary to reconstruct the coefficients
$A^{(n)}$ and $B^{(n)}$ in the quark case up to N$^3$LL
accuracy~\cite{Collins:2017oxh, Henn:2019swt,
  vonManteuffel:2020vjv}. For the cusp anomalous dimension $\gamma_K$
we have
\begin{equation}
  \label{appen3-cusp}
  \small
  \begin{array}{rcl}
    \gamma_K^{(0)} &=&\displaystyle 8 C_F\,,\\
    \gamma_K^{(1)} &=&\displaystyle 8 C_F  \bigg[ \left( \frac{67}{9} - \frac{\pi^2}{3} \right) C_A - \frac{20}{9}  T_R  n_f \bigg]\,,\\
    \gamma_K^{(2)} &=&\displaystyle 8
                       C_F\bigg[\left(\frac{245}{6}-\frac{134 \pi
                       ^2}{27}+\frac{11 \pi^4}{45}+\frac{22
                       \zeta _3}{3}\right) C_A^2\\
                   &+&\displaystyle \left(-\frac{418}{27}+\frac{40 \pi ^2}{27}-\frac{56 \zeta _3}{3}\right) C_A T_R n_f+\left(-\frac{55}{3}+16 \zeta _3\right) C_F T_R n_f-\frac{16}{27} n_f^2 T_R^2\bigg]\,,\\
    \gamma_K^{(3)} &=&\displaystyle 2 C_F\bigg[n_f^3\left(\frac{64
                       \zeta _3}{27}-\frac{32}{81}\right)+n_f^2
                       C_A\left(-\frac{224 \zeta_2^2}{15} +\frac{2240 \zeta _3}{27}-\frac{608 \zeta
                       _2}{81}+\frac{923}{81}\right)\\
                   &+&\displaystyle n_f^2 C_F\left(\frac{64 \zeta _2^2}{5}-\frac{640
                       \zeta _3}{9}+\frac{2392}{81}\right)\\
                   &+&\displaystyle n_f C_A^2\left(\frac{2096 \zeta _5}{9}+\frac{448
                       \zeta _3 \zeta _2}{3}-\frac{352 \zeta _2^2}{15}-\frac{23104 \zeta
                       _3}{27}+\frac{20320 \zeta _2}{81}-\frac{24137}{81}\right)\\
                   &+&\displaystyle n_f C_A C_F\left(160 \zeta _5-128 \zeta _3 \zeta
                       _2-\frac{352 \zeta _2^2}{5}+\frac{3712 \zeta _3}{9}+\frac{440
                       \zeta _2}{3}-\frac{34066}{81}\right)\\
                   &+&\displaystyle n_f C_F^2\left(-320 \zeta _5+\frac{592 \zeta
                       _3}{3}+\frac{572}{9}\right)+n_f d_F^{abcd}
                       d_F^{abcd}\left(-\frac{1}{3} \left(1280 \zeta _5\right)-\frac{256
                       \zeta _3}{3}+256 \zeta _2\right)\\
                   &+&\displaystyle d_A^{abcd} d_F^{abcd}\left(-384 \zeta _3^2-\frac{7936
                       \zeta _2^3}{35}+\frac{3520 \zeta _5}{3}+\frac{128 \zeta _3}{3}-128
                       \zeta _2\right)\\
                   &+&\displaystyle C_A^3\left(-16 \zeta _3^2-\frac{20032 \zeta _2^3}{105}-\frac{3608 \zeta _5}{9}-\frac{352 \zeta _3 \zeta _2}{3}+\frac{3608 \zeta _2^2}{5}+\frac{20944 \zeta _3}{27}-\frac{88400 \zeta _2}{81}+\frac{84278}{81}\right)\bigg]\,.
  \end{array}
\end{equation}
For the anomalous dimension $\gamma_F$ we have
\begin{equation}
  \label{appen3-gaF}
  \small
  \begin{array}{rcl}
    \gamma_F^{(0)} &=&\displaystyle  6  C_F\,,\\
    \gamma_F^{(1)} &=&\displaystyle -\left[C_F^2\left(-3+4 \pi ^2-48
                       \zeta _3\right)+C_F
                       C_A\left(-\frac{961}{27}-\frac{11 \pi^2}{3}+52 \zeta _3\right)+ C_F T_R n_f\left(\frac{260}{27}+\frac{4 \pi ^2}{3}\right)\right]\,,\\
    \gamma_F^{(2)} &=&\displaystyle  -\bigg[C_F^3\left(-29-6 \pi
                       ^2-\frac{16 \pi^4}{5}-136 \zeta
                       _3+\frac{32 \pi ^2 \zeta _3}{3}+480 \zeta
                       _5\right)\\
                   &+&\displaystyle C_F^2 C_A\left(-\frac{151}{2}+\frac{410 \pi
                       ^2}{9}+\frac{494 \pi^4}{135}-\frac{1688 \zeta
                       _3}{3}-\frac{16 \pi ^2 \zeta _3}{3}-240 \zeta _5\right)\\
                   &+&\displaystyle C_F C_A^2\left(-\frac{139345}{1458}-\frac{7163 \pi
                       ^2}{243}-\frac{83 \pi^4}{45}+\frac{7052 \zeta
                       _3}{9}-\frac{88 \pi ^2 \zeta _3}{9}-272 \zeta _5\right)\\
                   &+&\displaystyle C_F^2 T_R n_f\left(\frac{5906}{27}-\frac{52 \pi
                       ^2}{9}-\frac{56 \pi^4}{27}+\frac{1024 \zeta
                       _3}{9}\right)\\
                   &+&\displaystyle C_F C_A T_R n_f\left(-\frac{34636}{729}+\frac{5188
                       \pi ^2}{243}+\frac{44 \pi^4}{45}-\frac{3856 \zeta
                       _3}{27}\right)\\
                   &+&\displaystyle C_F T_R^2 n_f^2\left(\frac{19336}{729}-\frac{80 \pi ^2}{27}-\frac{64 \zeta _3}{27}\right)\bigg]\,.
  \end{array}
\end{equation}
For the anomalous dimension $K$ we have
\begin{equation}
  \label{appen3-kerK}
  \small
  \begin{array}{rcl}
    K^{(0)} &=&\displaystyle 0\,,\\
    K^{(1)} &=&\displaystyle C_F\left[C_A\left(28 \zeta _3-\frac{808}{27}\right)+\frac{224 T_R n_f}{27}\right]\,,\\
    K^{(2)} &=&\displaystyle 2 C_F\bigg[\frac{1}{2}
                C_A^2\left(-\frac{176}{3}  \zeta _3 \zeta _2+\frac{6392
                \zeta _2}{81}+\frac{12328 \zeta _3}{27}+\frac{154 \zeta
                _4}{3}-192 \zeta _5-\frac{297029}{729}\right)\\
            &+&\displaystyle C_A T_R n_f\left(-\frac{824}{81} \zeta _2-\frac{904 \zeta _3}{27}+\frac{20 \zeta _4}{3}+\frac{62626}{729}\right)\\
            &+&\displaystyle 2 T_R^2 n_f^2\left(-\frac{32}{9} \zeta _3-\frac{1856}{729}\right)+C_F T_R n_f\left(-\frac{304 }{9} \zeta _3-16 \zeta _4+\frac{1711}{27}\right)\bigg]\,.
  \end{array} 
\end{equation}
By plugging the coefficients in Eqs.~(\ref{appen3-cusp}),
(\ref{appen3-gaF}), and (\ref{appen3-kerK}) into
Eq.~(\ref{eq:AnDimToAB}), we have checked that we find agreement with
the perturbative coefficients of the functions $A$ and $B$ as
implemented in the {\tt DYTURBO} code~\cite{Camarda:2019zyx}.

\section{Hysteresis in the Sudakov form factor}\label{app:hyst}

In this appendix, following Ref.~\cite{Bertone:2022sso}, 
we elaborate on the concept of perturbative hysteresis applied to the
analytically computed Sudakov form factor.

We start by noting that the scale $M$ in the logarithm that multiplies
the function $A$ in Eq.~(\ref{eq:SudFF2}) is to be identified with the
rapidity scale $\zeta$. However, the value of
$\zeta$ is fixed by the external kinematics and can be set to the
hard scale $M$ throughout. We thus write the Sudakov form factor,
controlling the evolution of a TMD from the scale $\mu_1$ to the scale
$\mu_2$ at a given hard scale $M$, as
\begin{equation}\label{eq:SudFF3}
  S(\mu_2,\mu_1;M) = - \int_{\mu_1^2}^{\mu_2^2}\frac{d\mu'^2}{\mu'^2}\left[A(a_s(\mu'))\ln\frac{M^2}{\mu'^2}+B(a_s(\mu'))\right]\,.
\end{equation}
Perturbative hysteresis is characterised by
$S(\mu_2,\mu_1;M)+S(\mu_1,\mu_2; M) \neq 0$.  Unlike the case of a
numerical evaluation, the analytic Sudakov factor computed in terms of
the $g$-functions leads to perturbative hysteresis. To assess its
magnitude, we observe that the $g$-functions are computed by assuming
that the upper integration limit in Eq.~(\ref{eq:SudFF3}), $\mu_2$, is
of the order of the hard scale $M$, and that this assumption allows
any displacement of $\mu_2$ from $M$ to be reabsorbed into the
function $B$ (see Eq.~(\ref{eq:Bdisplacement})). In order to compute
$S(\mu_2,\mu_1;M)+S(\mu_1,\mu_2; M)$, we release this assumption and
rewrite the Sudakov form factor as follows,
\begin{equation}\label{eq:SudFF4}
  S(\mu_2,\mu_1;M) = - \int_{\mu_1^2}^{\mu_2^2}\frac{d\mu'^2}{\mu'^2}\left[A(a_s(\mu'))\ln\frac{\mu_2^2}{\mu'^2}+B(a_s(\mu'))\right]- 2\ln\frac{M}{\mu_2}\int_{\mu_1^2}^{\mu_2^2}\frac{d\mu'^2}{\mu'^2}A(a_s(\mu'))\,.
\end{equation}
The first integral can now be computed analytically and expressed in
terms of the usual $g$-functions,
\begin{equation}
  -\int_{\mu_1^2}^{\mu_2^2}\frac{d\mu'^2}{\mu'^2}\left[A(a_s(\mu'))\ln\frac{\mu_2^2}{\mu'^2}+B(a_s(\mu'))\right]
  = Lg_1(\lambda)+g_2(\lambda)+a_s(\mu_2) g_3(\lambda)+\dots\,
\end{equation}
with
\begin{equation}
  L = \ln\frac{\mu_1}{\mu_2}\, \quad\mbox{and}\quad \lambda= a_s(\mu_2)\beta_0L\,.
\end{equation}
The integral in the second term of the r.h.s.~of Eq.~(\ref{eq:SudFF4})
can also be computed order by order in perturbation theory using the
analytic running of the strong coupling, yielding\footnote{Since the
  logarithm $\ln(M^2/\mu_2^2)$ is potentially large, the perturbative
  counting of the second term on the r.h.s. of Eq.~(\ref{eq:SudFF4})
  follows that of the function $A$ in the first term.}
\begin{equation}\label{eq:residualterm}
  2\int_{\mu_1^2}^{\mu_2^2}\frac{d\mu'^2}{\mu'^2}A(a_s(\mu')) =h_1(\lambda)+a_s(\mu_2) h_2(\lambda)+a_s^2(\mu_2) h_3(\lambda)+\dots\,,
\end{equation}
Up to NLL accuracy, and setting $\kappa=1$, we find
\begin{equation}\label{eq:residualterm-bis}
  \begin{array}{l}
    \displaystyle h_1(\lambda) = \frac{4A^{(1)}}{\beta_0}\ln\left(1-\lambda\right)\,,\\
    \\
    \displaystyle h_2(\lambda)=\frac{4A^{(1)}b_1}{\beta_0}\frac{\lambda+\ln\left(1-\lambda\right)}{1-\lambda}-\frac{4A^{(2)}}{\beta_0}\frac{\lambda}{1-\lambda}\,.
  \end{array}
\end{equation}
  
This generalisation of the Sudakov form factor enables us to compute
the quantity $S(\mu_2,\mu_1;M)+S(\mu_1,\mu_2; M)$. For simplicity, we
set $M=\mu_2$ so that the ``forward'' Sudakov form factor
$S(\mu_2,\mu_1;\mu_2)$ can be evaluated using the $g$-functions
only. The ``backward'' Sudakov form factor $S(\mu_1,\mu_2;\mu_2)$ is
instead computed using the generalisation in Eq.~(\ref{eq:SudFF4}),
that is,
\begin{equation}
  \begin{array}{rcl}
    \displaystyle S(\mu_1,\mu_2;\mu_2) &=&\displaystyle
                                           \widetilde{L}g_1(\widetilde{\lambda})+g_2(\widetilde{\lambda})+a_s(\mu_1)
                                           g_3(\widetilde{\lambda})+\dots
    \\
    \\
                                       &-&\displaystyle \widetilde{L}\left[h_1(\widetilde{\lambda})+a_s(\mu_1) h_2(\widetilde{\lambda})+a_s^2(\mu_1) h_3(\widetilde{\lambda})+\dots\right]\,,
  \end{array}
\end{equation}
where
\begin{equation}
  \widetilde{L} = \ln\frac{\mu_2}{\mu_1}\,\quad\mbox{and}\quad \widetilde{\lambda}= a_s(\mu_1)\beta_0\widetilde{L}\,.
\end{equation}

At LL, the Sudakov form factor does not produce any perturbative
hysteresis. To see this, we simply need to consider the functions
$g_1$ and $h_1$. Using the LL evolution of the strong coupling we find
\begin{equation}
  \widetilde{\lambda}= -\frac{ \lambda}{1-\lambda}\,,
\end{equation}
so that
\begin{equation}
  g_1(\widetilde{\lambda}) =
  \frac{4A^{(1)}}{\beta_0}\frac{\lambda+(1-\lambda)\ln\left(1-\lambda\right)}{\lambda}\,
  \quad\mbox{and}\quad h_1(\widetilde{\lambda}) = -\frac{4A^{(1)}}{\beta_0}\ln\left(1-\lambda\right)\,,
\end{equation}
which immediately implies that
$S(\mu_2,\mu_1;\mu_2)+S(\mu_1,\mu_2;\mu_2)=0$.

\begin{figure}[h]
  \begin{centering}
    \includegraphics[width=0.6\textwidth]{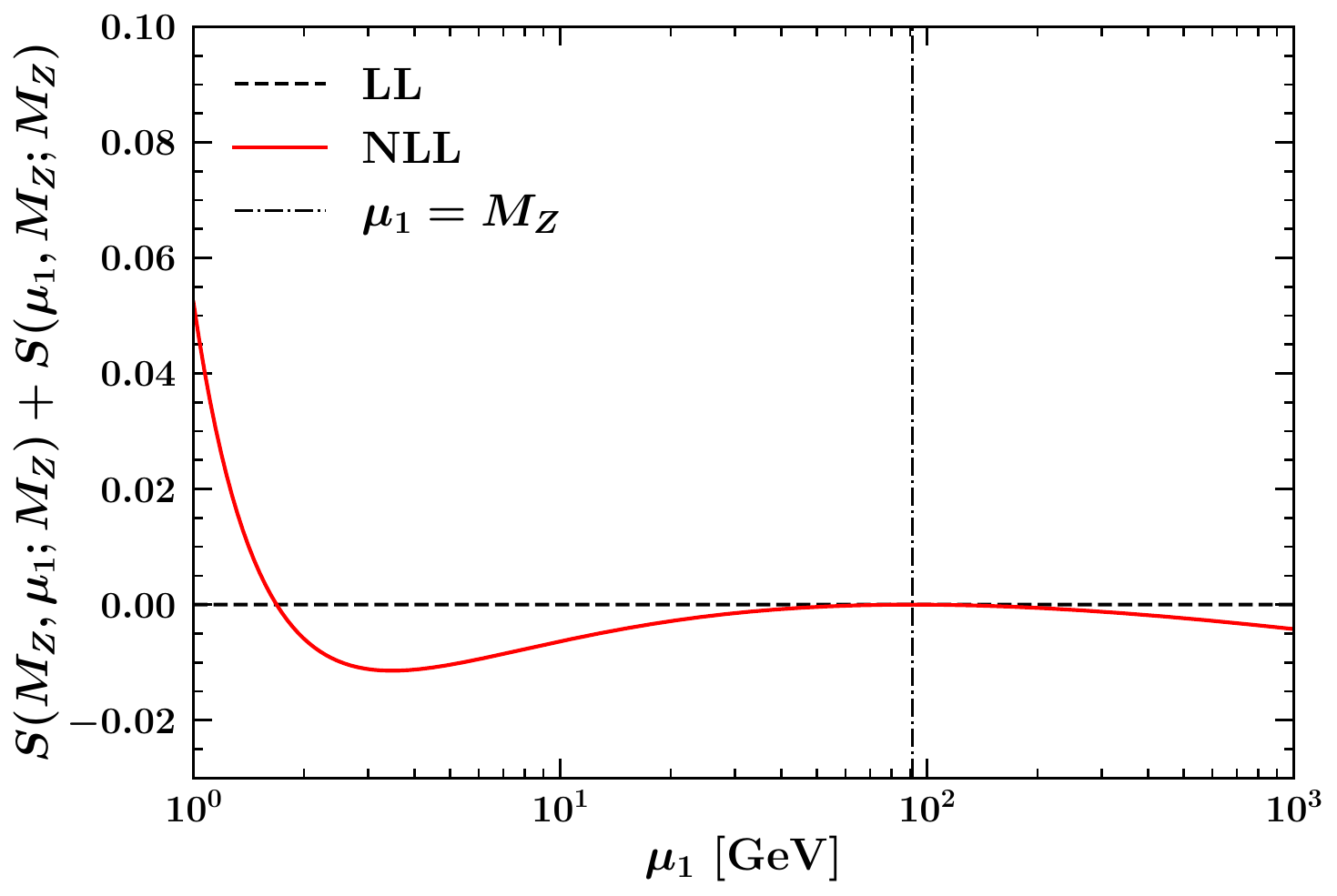}
    \caption{Perturbative hysteresis of the analytic Sudakov form
      factor at LL and NLL accuracies as a function of the scale
      $\mu_1$ in the range $\mu_1\in[1:1000]$~GeV. The strong-coupling
      boundary condition $\alpha_s(M_Z)=0.118$ is used. The vertical
      dot-dashed line indicates the scale $\mu_1=M_Z$ where the
      boundary condition on $\alpha_s$ is set. At this scale the NLL
      hysteresis touches zero.}\label{fig:SudakovHysteresis}
  \end{centering}
\end{figure}

Beyond LL, this equality is broken by the analytic solution. To show
this, in Fig.~\ref{fig:SudakovHysteresis} we plot the quantity
$S(\mu_2,\mu_1;\mu_2)+S(\mu_1,\mu_2;\mu_2)$ both at LL and NLL as a
function of $\mu_1$ setting $\mu_2=M_{Z}$ and $\alpha_s(M_Z)=0.118$.
As expected, the LL curve is identically zero over the entire range in
$\mu_1$ considered. At NLL, as a consequence of the perturbative
hysteresis, the curve deviates from zero. Deviations are enhanced at
small scales where the value of $\alpha_s$ is largest. At $\mu_1=M_Z$,
where no evolution takes place and both $S(M_Z,\mu_1;M_Z)$ and
$S(\mu_1,M_Z;M_Z)$ are separately zero, the NLL curve touches zero but
it slowly restarts departing from it at larger scales.

\bibliographystyle{ieeetr}
\bibliography{bibliography}

\end{document}